\documentclass[11pt,a4paper]{article}

\usepackage{slashed}
\usepackage{hyperref}
\usepackage{color}
\usepackage{amsthm}
\usepackage{mathtools}
\usepackage[normalem]{ulem} 
\usepackage{graphicx}
\usepackage{amssymb,latexsym,cite}
\usepackage{amsmath}
\usepackage{amsfonts}
\usepackage{mathrsfs}
\usepackage{bbm}
\usepackage{bm}
\usepackage{cleveref}
\usepackage[T1]{fontenc}
\usepackage{multicol}
\usepackage{multirow}
\usepackage{float}

\usepackage{tikz-cd}
\usetikzlibrary{decorations.markings}
\usetikzlibrary{intersections}
\usetikzlibrary{calc}
\tikzset{
    scalplusnoarrow/.style={postaction={decorate}, line width= 0.8pt},
}

\usepackage{simplewick}

\usepackage[matrix,arrow,color]{xy}

\usepackage{enumitem}

\definecolor{dgreen}{rgb}{.2,.6,.2}

\definecolor{dred}{RGB}{250,0,100}

\def\slasha#1{\setbox0=\hbox{$#1$}#1\hskip-\wd0\hbox to\wd0{\hss\sl/\/\hss}}

\def\periodb#1{\setbox0=\hbox{$#1$}#1\hskip-\wd0\hbox to\wd0{-}}

\newcommand{\e}{{\mathrm{e}}}               
\newcommand{\ii}{{\mathrm{i}}}              
\newcommand{\id}{\mathrm{id}}               

\newcommand{\CF}{\mathcal{F}}

\newcommand{\CS}{\mathcal{S}}

\newcommand{\mbf}[1]{{\boldsymbol {#1} }}

\newcommand{\Sym}{{\mathrm{Sym}}}

\newcommand{\BV}{{\textrm{\tiny BV}}}
\newcommand{\BVL}{{\mathsf\Delta}_{\textrm{\tiny BV}}}

\newcommand{\FR}{\mathbbm{R}}               
\newcommand{\FC}{\mathbbm{C}}               
\newcommand{\RZ}{\mathbbm{Z}}               

\newcommand{\dd}{\mathrm{d}}                

\newcommand{\sH}{\mathsf{H}}

\newcommand{\sP}{\mathsf{P}}

\newcommand{\sI}{\mathsf{I}}

\newcommand{\comment}[1]{}                  
\def\tyng(#1){\hbox{\tiny$\yng(#1)$}}           
\def\tyoung(#1){\hbox{\tiny$\young(#1)$}}           

\newcommand{\beq}{\begin{eqnarray}}
\newcommand{\eeq}{\end{eqnarray}}

\newcommand{\sgreen}{\mathsf{G}}

\newcommand{\sfp}{{\sf p}}

\newcommand{\sfh}{{\sf h}}

\newcommand{\tte}{\mathtt{e}}

\newcommand{\sfR}{\mathsf{R}}

\definecolor{outrageousorange}{rgb}{1.0, 0.43, 0.29}

\newenvironment{myitemize}{\begin{itemize}[itemsep=-0.05cm, leftmargin=*, topsep=0.1cm]}{\end{itemize}}

\newcommand{\nn}{\nonumber}

\def\RR{{\mathcal R}}

\def\d{{\rm d}}
\def\swone{{\textrm{\tiny $(1)$}}}
\def\swtwo{{\textrm{\tiny $(2)$}}}

\def\swzero{{\textrm{\tiny $(0)$}}}

\numberwithin{equation}{section}
\catcode`@=12

\theoremstyle{plain}

\theoremstyle{definition}

\newtheorem{example}[equation]{Example}

\begin{document}


\renewcommand{\thefootnote}{\fnsymbol{footnote}}
    
\vspace{1cm}

\begin{center}

{\LARGE{\bf Braided scalar quantum field theory}}

\vspace*{1.5cm}

\baselineskip=14pt
        
{\large\bf Djordje Bogdanovi\'c}${}^{\,(a)\,,\,}$\footnote{Email: \ {\tt
djbogdan@@ipb.ac.rs}} \ \ \ \ \ {\large\bf Marija Dimitrijevi\'c \'Ciri\'c}${}^{\,(a)\,,\,}$\footnote{Email: \ {\tt
dmarija@ipb.ac.rs}} \\[2mm] {\large\bf Voja Radovanovi\'c}${}^{\,(a)\,,\,}$\footnote{Email: \ {\tt
rvoja@ipb.ac.rs}} \ \ \ \ \ {\large\bf Richard J. Szabo}${}^{\,(b)\,,\,}$\footnote{Email: \ {\tt R.J.Szabo@hw.ac.uk}} \ \ \ \ \ {\large\bf Guillaume Trojani}${}^{\,(b)\,,\,}$\footnote{Email: \ {\tt gt43@hw.ac.uk}}
\\[6mm]

\noindent  ${}^{(a)}$ {\it Faculty of Physics, University of
Belgrade}\\ {\it Studentski trg 12, 11000 Beograd, Serbia}
\\[3mm]

\noindent  ${}^{(b)}$ {\it Department of Mathematics, Heriot-Watt University\\ Colin Maclaurin Building,
Riccarton, Edinburgh EH14 4AS, U.K.}\\ and {\it Maxwell Institute for
Mathematical Sciences, Edinburgh, U.K.}
\\[30mm]

\end{center}

\begin{abstract}
\noindent
We formulate scalar field theories in a curved braided $L_\infty$-algebra formalism and analyse their correlation functions using Batalin--Vilkovisky quantization. We perform detailed calculations in cubic braided scalar field theory up to two-loop order and three-point multiplicity. The divergent tadpole contributions are eliminated by a suitable choice of central curvature for the $L_\infty$-structure, and we confirm the absence of UV/IR mixing. The calculations of higher loop and higher multiplicity correlators in homological perturbation theory are facilitated by the introduction of a novel diagrammatic calculus. We derive an algebraic version of the Schwinger--Dyson equations based on the homological perturbation lemma, and use them to prove the braided Wick theorem.
\end{abstract}

\vspace*{2cm}



\newpage

{\baselineskip=14pt
\tableofcontents
}

\setcounter{footnote}{0}
\renewcommand{\thefootnote}{\arabic{footnote}}

\bigskip

\setcounter{page}{1}
\newcommand{\Section}[1]{\setcounter{equation}{0}\section{#1}}
\renewcommand{\theequation}{\arabic{section}.\arabic{equation}}

\section{Introduction}

Noncommutative field theories appear as effective theories in many physical situations (see e.g.~\cite{Douglas:2001ba,Szabo:2001kg,Szabo:2004ic,Szabo:2009tn,Hersent:2022gry} for reviews). They are also believed to provide frameworks for models of quantized spacetime and quantum gravity. In order for them to be viable candidate descriptions of any physics, one should first establish whether they define sensible quantum theories. The standard approach to quantization of Lagrangian field theories proceeds through the path integral formalism, but this is limited to free theories and perturbation theory, while being subjected to various layers of technical subtleties and difficulties. On the other hand, methods based on the intrinsic homotopy algebraic structure of a quantum field theory have sharpened our understanding of the algebraic structures of correlation functions and dynamical processes~\cite{Hohm:2017pnh,Doubek:2017naz,Jurco:2018sby,Nutzi:2018vkl,Arvanitakis:2019ald,Jurco:2019yfd,Masuda:2020tfa,Arvanitakis:2020rrk,Saemann:2020oyz,Chiaffrino:2021pob,Okawa:2022sjf,Konosu2023,Konosu:2023rkm,Bonezzi:2023xhn,Konosu:2024dpo}. The quantum Batalin--Vilkovisky (BV) formalism gives explicit homological constructions in a  purely algebraic fashion without resorting to path integral techniques. The $L_\infty$-structures of the standard noncommutative field theories are discussed in~\cite{Nguyen:2021rsa,Giotopoulos:2021ieg,Szabo:2023cmv}.

This paper is part of the ongoing exploration of a new class of noncommutative field theories, called \emph{braided field theories}. These theories were introduced in~\cite{DimitrijevicCiric:2021jea} by twisting the classical $L_\infty$-algebra formulation of field theories~\cite{Hohm:2017pnh,Jurco:2018sby} to theories organised by a new homotopy algebraic structure called a \emph{braided $L_\infty$-algebra}; see~\cite{Giotopoulos:2021ieg} for a detailed exposition with further developments. The classical dynamics of these theories are governed by a Maurer--Cartan theory of their underlying braided $L_\infty$-algebras. Their quantum dynamics are formulated through a braided version of BV quantization, introduced in~\cite{Nguyen:2021rsa} and further developed in~\cite{DimitrijevicCiric:2023hua}. In this instance the homotopy algebraic approach is a necessity: braided field theories lack a path integral formulation or even a good notion of a functional differential calculus, in marked contrast to the standard noncommutative field theories.

In this paper we undertake a detailed study of real braided scalar field theories with polynomial interactions and the Moyal--Weyl twist deformation, whose braided $L_\infty$-algebra formulation and BV quantization were briefly addressed by~\cite{Giotopoulos:2021ieg,DimitrijevicCiric:2023hua,Bogdanovic:2023izt}. As we review in \underline{Section~\ref{sub:braidedphi3}}\,, the Maurer--Cartan theories for both the braided and conventional $L_\infty$-algebra formulations in this case coincide, and so classically braided scalar field theory is the same as the standard noncommutative scalar field theory. However, their quantizations differ: while the braided BV formalism leads to the same interaction vertex as in the standard theory, it produces a deformation of the usual Wick expansion of correlation functions. 

The additional braiding features which appear upon quantization using homological perturbation theory seem to cancel out the noncommutative effects of the deformed interactions. In all perturbative calculations performed so far, one finds that the braided correlation functions match exactly with their commutative counterparts. This has sparked excitement over the prospect of obtaining an interacting noncommutative field theory which is free from the notorious problem of UV/IR mixing~\cite{Minwalla:1999px,Szabo:2001kg}, and hence may be perturbatively renormalizable. 

In \underline{Section~\ref{sec:phi3corrs}} we corroborate these findings with detailed computations of the correlation functions in cubic braided scalar field theory up to two-loop order and multiplicity three; some preliminary results of this investigation were reported in~\cite{Bogdanovic:2023izt}. We find that noncommutative effects appear at most in phase factors dependent solely on the external momenta of correlators, exactly like the planar Feynman diagrams of conventional noncommutative quantum field theory~\cite{Minwalla:1999px,Szabo:2001kg}. This suggests that the perturbative expansion of braided scalar field theory does not involve any non-planar graphs, which are the source of UV/IR mixing. 

Explicit calculations of correlation functions in homological perturbation theory at higher-loop orders and higher multiplicities become overwhelmingly complicated. One of the new developments of this paper is a diagrammatic calculus that serves to simplify and control these computations, similarly to the graphical methods introduced in~\cite{Nguyen:2021rsa}. The particular diagrams used in this paper are summarised in \underline{Appendix~\ref{subsub:pictograms}}\,.

Another novelty we introduce in this paper is the necessity of a \emph{curved} braided $L_\infty$-structure in the effective theory underlying braided scalar field theories in the presence of external sources. For the cubic scalar field theory, this is used to cancel tadpole diagrams. A constant central curvature is introduced and used in a systematic algebraic way to eliminate the uncontrollable tadpole divergences in homological perturbation theory. In \underline{Appendix~\ref{app:BV}} we generalize the definition of braided $L_\infty$-algebra to include the curved case, and describe the modifications incurred in braided BV quantization.

In \underline{Section~\ref{sec:SD}} we develop a new avenue for analysing the noncommutative features of braided quantum field theory. We derive an algebraic version of the Schwinger--Dyson equations using the braided homological perturbation lemma, which lead to recursion relations among the full interacting correlation functions of any braided scalar field theory. While these equations reduce to the anticipated recursive equations, they take a somewhat unconventional form in the $L_\infty$-algebra formalism as a sort of braided symmetrization of the standard identities. For the free scalar field, we apply the recursion to rigorously prove the braided Wick theorem which was proposed in~\cite{DimitrijevicCiric:2023hua} based on heuristic arguments and low multiplicity checks in homological perturbation theory. For the cubic braided scalar field theory we obtain a recursion between the full two-point and three-point functions wherein noncommutativity again only appears in external momentum phase factors.

While the primary purpose of the present work is to further advance the development of braided quantum field theory, the techniques we introduce (e.g. our diagrammatic calculus and our algebraic Schwinger--Dyson equations) are more broadly applicable. They should be of interest to practitioners working on the interactions between homotopical algebra and quantum field theory in general. Our main results and future perspectives are  summarised in \underline{Section~\ref{sec:discussion}}\,.

\section{Correlation functions of braided scalar field theories}
\label{sub:braidedphi3}

In this paper we study the simplest class of examples of a braided noncommutative field theory, introduced in~\cite{Giotopoulos:2021ieg,DimitrijevicCiric:2023hua,Bogdanovic:2023izt}, provided by the model of a real scalar field $\phi$ of mass $m$ in $d$ dimensions with polynomial interactions. The noncommutative deformation is introduced via the Drinfel'd twist formalism. In this paper we will use the Moyal--Weyl twist
\begin{align}\label{eq:MWtwist}
\CF =\exp\big(-\tfrac{\mathrm{i}}2\,\theta^{\mu\nu}\,\partial_\mu\otimes\partial_\nu\big)  \qquad \mbox{with} \quad \CF^{-1} = \sf{f}_\alpha\otimes \sf{f}^\alpha
  \ ,
\end{align}
where $(\theta^{\mu\nu})$ is a $d{\times}d$ antisymmetric real-valued
matrix, and $\partial_\mu=\frac\partial{\partial x^\mu}\in\Gamma(T\FR^{1,d-1})$ for $\mu=1,\dots,d$ are vector fields generating spacetime translations in \smash{$\FR^{1,d-1}$}.  The last equality represents the formal power series expansion of  the inverse twist $\CF^{-1}$ into bidifferential operators.

This twist deforms the pointwise product $f\cdot g$ of functions  \smash{$f,g\in C^\infty(\FR^{1,d-1})$} to the noncommutative star-product 
\begin{equation}
f\star g = {\sf{f}}_\alpha(f)\cdot {\sf{f}}^\alpha(g) = f\cdot g + \sum_{n=1}^\infty\,\frac1{n!}\,\bigg(\frac{\ii}{2}\bigg)^n\,\theta^{\mu_1\nu_1}\cdots\theta^{\mu_n\nu_n} \, \partial_{\mu_1}\cdots\partial_{\mu_n} f\cdot\partial_{\nu_1}\cdots\partial_{\nu_n} g  \ . \label{MWStar}
\end{equation}
The star-product is braided commutative,
\begin{align*}
f\star g = \sfR_\alpha( g)\star \sfR^\alpha( f) \ ,
\end{align*}
where the braiding is encoded by the $\RR$-matrix $\RR=\CF^{-2}$ whose inverse expands into bidifferential operators as
\begin{align*}
\RR^{-1} = \CF^2 = \sfR_\alpha\otimes\sfR^\alpha \ .
\end{align*}

In this section we introduce the techniques of braided homological perturbation theory that we will use to study the quantization of these models in subsequent sections. See Appendix~\ref{app:BV} for our notational conventions and for background on the general formalism.

\subsection{Classical braided scalar field theory}

A scalar field theory has no gauge symmetries and therefore its flat $L_\infty$-structure has underlying graded vector space $V= V_{1}\oplus V_{2}$ concentrated in degrees 1 and 2, where $V_1=V_2=C^\infty(\FR^{1,d-1})$ respectively contain the physical fields $\phi\in V_1$ and their antifields $\phi^+\in V_2$. The  non-vanishing brackets are valued in $V_2$ and given by
\begin{align}
\ell^\star_1(\phi)  &= \big(-\square - m^2\big)\,\phi  \ , \label{eq:l1} \\[4pt]
\ell^\star_{n}(\phi_1,\dots, \phi_{n}) &= -(-1)^{{n\choose 2}} \ \lambda_n \ \phi_1\star\cdots\star \phi_{n} \ , \label{eq:intbrackets}
\end{align}
for $\phi,\phi_1,\dots,\phi_{n}\in V_1$. We assume that $\lambda_n\in\FR$ are non-zero for only finitely many $n\geqslant 2$.
The cyclic pairing of degree $-3$ is taken to be
\begin{align}
\langle\phi, \phi ^+\rangle_\star & = \int\d^d x \ \phi \star \phi^+ = \int\dd^dx \ \phi \cdot \phi^+  \label{eq:scalarpairing}
\end{align}
where $\phi\in V_1$ and $\phi^+\in V_2$. The braided homotopy Jacobi identities as well as cyclicity all follow trivially for degree reasons in this theory.

The Maurer--Cartan theory for this braided $L_\infty$-algebra leads to the action functional
\begin{align}
\begin{split}
S(\phi) & = \frac{1}{2!} \, \langle \phi, \ell^\star_1(\phi)\rangle_\star 
+\sum_{n\geqslant 2}\, \frac{(-1)^{{n\choose 2}}}{(n+1)!} \, \langle\phi, \ell^\star_n(\phi^{\otimes n})\rangle_\star \\[4pt]
& = \int\d^d x \ \frac{1}{2}\,\phi\,\big(-\square - m^2\big)\,\phi - \sum_{n\geqslant 3}\, \frac{\lambda_{n-1}}{n!}\,\phi^{\star n} \ , \label{eq:S0scalar}
\end{split}
\end{align}
and the equation of motion
\begin{align} \label{eq:eom}
F_\phi & = \ell^\star_1(\phi) + \sum_{n\geqslant 2}\, \frac{(-1)^{{n\choose 2}}}{n!}\,\ell^\star_n(\phi^{\otimes n}) = \big(-\square - m^2\big)\,\phi - \sum_{n\geqslant 2}\,\frac{\lambda_n}{n!}\,\phi^{\star n} = 0 \ .
\end{align}
These are simply the classical dynamical equations for standard noncommutative scalar field theory.

\subsection{Batalin--Vilkovisky quantization}
\label{sub:BVquant}

Let us now explain how to compute correlation functions of the interacting braided scalar field theory using the braided Batalin--Vilkovisky (BV) formalism developed in \cite{Nguyen:2021rsa}. Further details can be found in \cite{DimitrijevicCiric:2023hua} and in Appendix~\ref{app:BV}.

We start from the cohomology $H^\bullet(V)$ of the underlying cochain complex $(V,\ell_1^\star)$, which describes the classical vacua of the free (braided) scalar field theory on $\FR^{1,d-1}$. This is also a cochain complex concentrated in degrees~$1$ and~$2$, given by the solution space $H^1(V)=\ker(\ell_1^\star)$ of the massive Klein--Gordon equation $(\square + m^2)\,\phi=0$ and the space $H^2(V)={\rm coker}(\ell_1^\star)$, with the trivial differential~$0$.

Following the discussion in Appendix~\ref{app:BV}, we need to define a translation-equivariant projection $\sfp: V\longrightarrow H^\bullet(V)$ of degree $0$ and a translation-invariant contracting homotopy $\sfh: V_2\longrightarrow V_1$. For this, let $\sgreen:C^\infty(\FR^{1,3})\longrightarrow C^\infty(\FR^{1,3})$ denote the scalar Feynman propagator
\begin{align}
\sgreen = -\frac1{\square+m^2} \quad , \quad \tilde \sgreen(k) = \frac1{k^2-m^2} \ ,\label{TildeG}
\end{align}
where $\tilde \sgreen(k)$ are the eigenvalues of the Green operator $\sgreen$ when acting on plane wave eigenfunctions of the form $\e^{\pm\,\ii\, k\cdot x}$. It satisfies
\begin{align*}
\ell_1^\star\circ \sgreen = -\big(\square+m^2\big)\circ \sgreen = \id_{C^\infty(\FR^{1,d-1})} \ .
\end{align*}

We are interested in calculating vacuum correlation functions. For this, we take the trivial projection map $\sfp=0$~\cite{Masuda:2020tfa,Okawa:2022sjf},
or more accurately we restrict the cochain complex of $H^\bullet(V)$ to its trivial subspaces.
With these choices, the contracting homotopy $\sfh:V_2\longrightarrow V_1$ is given through the propagator as $\sfh = -\sgreen$. Explicitly
\begin{align}\label{eq:htwo}
\sfh(\phi^+)(x) = \frac1{\square+m^2}\, \phi^+(x) = - \int \d^d y \ \int_k \, \frac{\e^{-\ii\,k\cdot(x-y)}}{k^2-m^2 - \ii\,\epsilon} \ \phi^+(y) \ ,
\end{align}
for $\phi^+\in V_2$ and $\epsilon > 0$, where we use the shorthand \smash{$\int_k=\int\frac{\d^dk}{(2\pi)^d}$}. In the sequel we will drop the Feynman $\ii\,\epsilon$-prescription for brevity, and implicitly assume that all integrals are suitably regularised where needed to make sense of expressions. By Fourier transforming \eqref{TildeG} to momentum space representation, the contracting homotopy acts as
\begin{equation}
\sfh(\tilde\phi^+)(k) = - \frac{\tilde\phi^+(k)}{k^2-m^2} \ .\nn
\end{equation}

Now we apply braided homological perturbation theory. We extend the maps $\sfp$ and $\sfh$ to the braided space of functionals $\Sym_\RR (V[2])$ on $V$, defined on generators $\varphi_i \in V[2]$ by the twisted symmetric product $\odot_\star$, which is braided graded commutative:
\begin{align*}
\varphi_i\odot_\star\varphi_j = (-1)^{|\varphi_i|\,|\varphi_j|} \ \sfR_\alpha(\varphi_j)\odot_\star\sfR^\alpha(\varphi_i) \ ,
\end{align*}
where $|\varphi|$ denotes the degree of a homogeneous element $\varphi\in V[2]$.
We will usually abbreviate such Koszul sign factors as $\pm$. 

The data above induce a projection map $\sP:\Sym_\RR( V[2])\longrightarrow (\Sym_\RR( H^\bullet(V[2])))_0$ to the trivial subspace $\FR\subset\Sym_\RR( H^\bullet(V[2]))$ in degree~$0$ given by
\begin{align}\label{eq:trivial project}
\sP(1)=1 \qquad \mbox{and} \qquad \sP(\varphi_1\odot_\star\cdots\odot_\star\varphi_n) = 0 \ .
\end{align}
The extended contracting homotopy $\sH:\Sym_\RR( V[2])\longrightarrow \Sym_\RR( V[2])[-1]$ of symmetric and total degree~$0$ is defined as
\begin{align}\label{eq:sfH}
\begin{split}
\sH(1)&=0 \ , \\[4pt]
\sH(\varphi_1\odot_\star\cdots\odot_\star\varphi_n) &= \frac1n \, \sum_{i=1}^n \, \pm \ \varphi_1\odot_\star\cdots\odot_\star\varphi_{i-1}\odot_\star\sfh(\varphi_i) \odot_\star\varphi_{i+1} \odot_\star\cdots\odot_\star \varphi_n \ ,
\end{split}
\end{align}
for all $\varphi_1,\dots,\varphi_n\in V[2]$. We used the translation-invariance of $\sfh$ in (\ref{eq:sfH}) which trivializes the actions of $\RR$-matrices.

We perturb the free differential $\ell_1^\star$, extended using translation-invariance as a strict graded derivation to all of $\Sym_\RR (V[2])$, to the `quantum' differential
\begin{align*}
Q_{\mbf\delta} = \ell_1^\star + \mbf\delta
\end{align*}
on $\Sym_\RR(V[2])$, with a formal translation-invariant perturbation $\mbf\delta$. The braided extension of the homological perturbation lemma \cite{Nguyen:2021rsa} then constructs the perturbed  projection $\tilde{\sP}=\sP+\sP_{\mbf\delta}$  of \eqref{SDR3}, where 
\begin{align*}
\sP_{\mbf\delta} = \sP\,\big(\id_{\Sym_\RR(V[2])} - {\mbf\delta}\,\sH\big)^{-1} \, \mbf\delta \, \sH \ .
\end{align*}
The interacting quantum field theory is defined by the small perturbation 
\begin{align*}
\mbf\delta= -\ii\,\hbar\,\BVL - \{\CS _{\rm int},-\}_\star \ .
\end{align*}
The definitions as well as the basic properties of the braided BV Laplacian $\BVL$ and the braided antibracket $\{-,-\}_\star$ are found in Appendix~\ref{app:BV}.

The interaction functional $\CS _{\rm int}\in \Sym_\RR(V[2])$ is defined by
\begin{align}\label{eq:Sintxi}
\CS _{\rm int} := \sum_{n\geqslant 2} \, \frac{(-1)^{{n\choose 2}}}{(n+1)!} \, \langle \xi \,,\,\ell^{\star \, {\rm ext}}_n(\xi^{\otimes n})\rangle^{\rm ext}_\star \ .
\end{align}
The extended brackets and pairing are defined in Appendix~\ref{app:BV}, while the contracted coordinate functions $\xi\in\Sym_\RR (V[2])\otimes V$ are given by
\begin{align*}
\xi = \int_k \, \big(\tte_k\otimes \tte^k + \tte^k\otimes \tte_k\big) \ ,
\end{align*}
where $\tte_k(x)=\e^{-\ii\, k\cdot x}$ is the basis of plane waves for $V_1$ with dual basis 
$
\tte^k(x)=\tte_k^*(x)=\e^{\,\ii\,k\cdot x}
$
for $V_2$. These bases are dual with respect to the inner product \eqref{eq:scalarpairing}, 
\begin{align*}
\langle\tte_k,\tte^p\rangle_\star = (2\pi)^d \, \delta(k-p) = \langle\tte^p,\tte_k\rangle_\star \ .
\end{align*}
where we use
\begin{align*}
\int \dd^dx \ \e^{\pm\,\ii\,k\cdot x} = (2\pi)^d \, \delta(k) \ .
\end{align*}
The star-products among basis fields are
\begin{align}\label{eq:ekstarep}
\tte_k\star \tte_p = \e^{\, -\frac\ii2 \, k\cdot\theta\, p} \ \tte_{k+p} \ ,
\end{align}
where $k\cdot\theta\, p:=k_\mu\,\theta^{\mu\nu}\,p_\nu=-p\cdot\theta\, k$, while the action of the inverse $\RR$-matrix on them is given by
\begin{align}\label{eq:Rekotimesep}
\RR^{-1}(\tte_k\otimes \tte_p) = \sfR_\alpha(\tte_k)\otimes\sfR^\alpha(\tte_p) = \e^{\,\ii\,k\cdot\theta\,p} \ \tte_k\otimes \tte_p \ .
\end{align}

Using these definitions, we obtain
\begin{align}
\begin{split}
\CS _{\rm int} & = \sum_{n\geqslant 2} \, \frac{(-1)^{{n\choose 2}}}{(n+1)!} \ \int_{k_1,\dots,k_{n+1}} \, \langle \tte^{k_1}\otimes \tte_{k_1}\,,\,\ell_n^{\star\, {\rm ext}}(\tte^{k_2}\otimes \tte_{k_2},\dots,\tte^{k_{n+1}}\otimes \tte_{k_{n+1}})\rangle^{\rm ext}_\star \\[4pt]
& =: \sum_{n\geqslant 3} \ \int_{k_1,\dots,k_n} \, V_n(k_1,\dots,k_n) \ \tte^{k_1}\odot_\star \cdots\odot_\star \tte^{k_n} \ , \label{IntVert3}
\end{split}
\end{align}
where the interaction vertices are
\begin{align}\label{eq:Vint3}
V_n(k_1,\dots,k_n) = -\frac{\lambda_{n-1}}{n!} \ \e^{\,\frac\ii2\,\sum\limits_{i<j} \, k_i\cdot\theta\, k_j} \ (2\pi)^d \, \delta(k_1+\cdots+k_n) \ ,
\end{align}
with $i,j =1,\dots,n$. These are just the usual vertices of the standard noncommutative scalar field theories. The vertex functions (\ref{eq:Vint3}) have the braided symmetry
\begin{align}\label{eq:V3braidedsym}
V_n(\ \  k_{i+1},k_i\ \ ) = \e^{-\ii\,k_i\cdot\theta\, k_{i+1}} \ V_n(k_1,\dots,k_n)
\end{align}
under the interchange of any pair of neighbouring momenta, and also the cyclic symmetry
\begin{align}\label{eq:V3cyclicsym}
V_n(k_1,\dots,k_{n-1},k_n) = V_n(k_n,k_1,\dots,k_{n-1}) \ .
\end{align}

Following~\cite{DimitrijevicCiric:2023hua,Bogdanovic:2023izt}, by formally expanding the perturbation $\sP_{\mbf\delta}$, the interacting correlation functions of the braided quantum field theory are given by \cite{Nguyen:2021rsa}
\begin{align}\label{eq:intcorrelationfn}
\begin{split}
G_n^\star(x_1,\dots,x_n)
:= \sum_{m=1}^\infty \, \sP\,\big((-\ii\,\hbar\,\BVL\,\sH - \{\CS _{\rm int},-\}_\star\,\sH)^m\,(\delta_{x_1}\odot_\star\cdots\odot_\star\delta_{x_n})\big) \ ,
\end{split}
\end{align}
where $\delta_{x_i}(x) = \delta(x-x_i)$. By Fourier transformation
\begin{align*}
G_n^\star(x_1,\dots,x_n) = \int_{p_1,\dots,p_n} \, \e^{-\ii\,p_1\cdot x_1 - \cdots - \ii\,p_n\cdot x_n} \ \tilde G_n^\star(p_1,\dots,p_n)
\end{align*}
this has momentum space representation
\begin{equation}\label{eq:intcorrelationfnMomentum}
\tilde G_n^\star(p_1,\dots,p_n) = \sum_{m=1}^\infty \, \sP\,\big((-\ii\,\hbar\,\BVL\,\sH - \{\CS _{\rm int},-\}_\star\,\sH)^m\, (\tte^{p_1}\odot_\star\cdots\odot_\star\tte^{p_n})\big) \ ,
\end{equation}
where $\tte^{p_i}(x)=\e^{\,\ii\,p_i\cdot x}$ and all momenta are treated as incoming in propagators.
Since only $\sP(1) = 1$ is non-zero, this is a formal power series expansion over $\FC$ in $\hbar$ and the coupling constants $\lambda_r$. 

\subsection{External sources and curved $L_\infty$-structures}
\label{sub:sources}

Braided scalar field theories in the presence of source terms or external fields $J\in C^\infty(\FR^{1,d-1})$ lead to an effective theory based on a \emph{curved} braided $L_\infty$-algebra, see Appendix~\ref{app:BV}, with curvature $J$. Regarding the ground field $\FR$ as sitting in degree~$0$, we introduce a new linear map $\ell_0^\star:\FR\longrightarrow V_2$ of degree~$2$ defined by
\begin{align*}
 \ \quad \ell_0^\star(1) = J \ ,
\end{align*}
while keeping the same non-zero higher brackets $\ell_n^\star$ for $n\geqslant1$.  For degree reasons, the curvature $\ell_0^\star(1)\in V_2$ is central, so that again $(\ell_1^\star){}^2=0$, i.e. the curved $L_\infty$-algebra in the present case is based on the same cochain complex $(V,\ell_1^\star)$ of free fields.

The curvature of the braided $L_\infty$-algebra modifies  (\ref{eq:S0scalar}) to the curved Maurer--Cartan functional
\begin{equation*}
S_J(\phi) = S(\phi) + \langle\phi,\ell_0^\star(1)\rangle_\star   = S(\phi) + \int\dd^dx \ J \cdot \phi \ ,
\end{equation*}
and the equation of motion \eqref{eq:eom} to the curved Maurer--Cartan equation
\begin{align*}
F_\phi = -\ell_0^\star(1) = -J \ .
\end{align*}

In braided homological perturbation theory, the modification of the interaction functional \eqref{IntVert3} is given by the addition of a linear interaction term:
\begin{align*}
\CS_{\rm int}^J = \CS_{\rm int} + \langle\xi,\ell_0^{\star\,{\rm ext}}(1)\rangle_\star^{\rm ext}  = \CS_{\rm int} + \int_{k_0} \, \tilde J(k_0) \ \tte^{k_0}  \ ,
\end{align*}
where $\tilde J(k) = \int\dd^dx \ \e^{-\ii\,k\cdot x} \, J(x)$ is the Fourier transform of the external source $J$. In Section~\ref{sec:phi3corrs} we will use an effective theory of this type to cancel certain vacuum diagrams by linear counterterms. 

\section{Correlation functions of braided $\lambda\,\phi^3$-theory}
\label{sec:phi3corrs}

In this section we present a detailed perturbative analysis of the simplest member of the family of braided scalar field theories considered in Section~\ref{sub:braidedphi3}. This is the model with $\lambda\,\phi^3$-interaction, obtained by taking only $\lambda:=\lambda_2$ to be non-zero. The interaction vertex in braided $\lambda\,\phi^3$-theory is given by 
\begin{equation}\label{interaction-vertex}
V_3(k_1,k_2,k_3)=-
\frac{\lambda}{3!} \, \e^{\,\frac{\mathrm{i}}{2}\,k_1\cdot\theta\, k_2} \ (2\pi)^d\,\delta(k_1+k_2+k_3) \ .
\end{equation}

In the commutative case, this model presents several problems. Firstly, the vacuum expectation value of $\phi$ is different from zero, which manifests itself in a non-zero tadpole diagram. Secondly, the Hamiltonian does not have a ground state, that is, the energy is not bounded from below. In the braided case, the non-trivial tadpole issue will be dealt with in Section~\ref{sub:moretadpole} while the problem of arbitrarily large negative energy will be ignored, as our main goal in this section is to illustrate the techniques of braided homological perturbation theory within the braided BV formalism. Since we do not address any renormalizability issues in this paper, we continue to work in general spacetime dimension $d$ in the following.

\subsection{One-point function at one-loop}
\label{sub:1ptfunction}

The one-point function follows from (\ref{eq:intcorrelationfnMomentum}) as
\begin{align}\label{eq:1PointPhi3}
\tilde G_1^\star(p) = \sum_{m=1}^\infty \, \sP\,\big((-\ii\,\hbar\,\BVL\,\sH - \{\CS _{\rm int},-\}_\star\,\sH)^m\, (\tte^{p})\big)\nn \ . 
\end{align}
The one-loop contribution is 
\begin{align}
\tilde G_1^\star(p)^{\swone}= \ii\,\hbar\,\BVL\,\sH\, \{\CS _{\rm int}, \sH\, (\tte^p) \}_\star = -\frac{\lambda}{2}\,\ii\,\hbar\,\frac{(2\pi)^d\,\delta(p)}{p^2-m^2}\,\int_{k}\, \frac{1}{k^2 - m^2} \ .\nn
\end{align}
This is the same as the tadpole contribution in commutative $\phi^3$-theory and is represented by the diagram defined in Appendix~\ref{subsub:pictograms} as
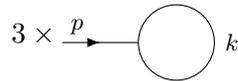
\begin{equation*}\label{fig:Phi31}
3\times {\footnotesize
\begin{tikzpicture}[baseline]
        \coordinate (k) at (0,0);
        \coordinate[label=above right: $p$] (p) at (180:1);

        \draw[decoration={markings, mark=at position 0.5 with {\arrow{Latex}} }, postaction={decorate}] (p) -- (k);
        \draw[decoration={markings }, postaction={decorate}] ($(k)+(0:0.5)$) circle (0.5) node[right=15]{$k$};
    \end{tikzpicture}
    } \normalsize
\end{equation*}
It is non-zero and divergent in any dimension. We will discuss how to remove it in Section~\ref{sub:moretadpole}.

\subsection{Two-point function at one-loop}

The two-point function is defined similarly as
\begin{equation}\label{eq:2PointPhi3}
\tilde G_2^\star(p_1,p_2) = \sum_{m=1}^\infty \, \sP\,\big((-\ii\,\hbar\,\BVL\,\sH - \{\CS _{\rm int},-\}_\star\,\sH)^m\, (\tte^{p_1}\odot_\star\tte^{p_2})\big) \ .
\end{equation}

\subsubsection{Tree-level contribution}

The tree-level contribution is just the free propagator 
\begin{equation}
\tilde G_2^{\star}(p_1,p_2)^{\swzero} = -\ii\,\hbar\,\BVL\,\sH\,(\tte^{p_1}\odot_\star\tte^{p_2}) = \ii\,\hbar\,\frac{(2\pi)^d\,\delta(p_1+p_2)}{p_1^2-m^2} \ .\label{G20}
\end{equation}
This is represented diagrammatically as
\begin{equation*}
\footnotesize
    \begin{tikzpicture}[baseline]
        \coordinate[label=above left: $p_1$] (p1) at (-1,0);
        \coordinate[label=above right: $p_2$] (p2) at (1,0);
        \draw[decoration={markings}, postaction={decorate}] (p1) -- (p2);
    \end{tikzpicture}
    \normalsize
\end{equation*}

\subsubsection{One-loop corrections}

The one-loop corrections to the free two-point function are given by the sum of the two terms
\begin{align}\label{eq:2pt2loop}
\begin{split}
& \tilde G_2^\star(p_1,p_2)_1^{\swone} = \ii\,\hbar\,\BVL\,\sH \,\bigl\{ \CS _{\rm int},\sH\, \big((\ii\,\hbar\,\BVL\,\sH)\,\bigl\{\CS _{\rm int},\sH\,(\tte^{p_1}\odot_\star \tte^{p_2})\bigr\}_\star \big)\bigr\}_\star \ , \\[4pt]
& \tilde G_2^\star(p_1,p_2)_2^{\swone} = (\ii\,\hbar\,\BVL\,\sH)^2 \, \bigl\{ \CS _{\rm int},\sH\, \bigl\{\CS _{\rm int},\sH\,(\tte^{p_1}\odot_\star \tte^{p_2})\bigr\}_\star \bigr\}_\star \ .
\end{split}
\end{align}

We start with 
\begin{align}\label{S2}
\begin{split}
& \bigl\{\CS _{\rm int},\sH\,(\tte^{p_1}\odot_\star \tte^{p_2})\bigr\}_\star = -\frac{3}{2} \, \int_{k_1,k_2,k_3} \, V_3(k_1,k_2,k_3) \, \bigl( \langle \tte^{k_3}, \sgreen(\tte^{p_1})\rangle_\star \,  \tte^{k_1}\odot_\star \tte^{k_2}\odot_\star \tte^{p_2} \\
& \hspace{9cm} + \langle \tte^{k_3},  \sgreen(\tte^{p_2})\rangle_\star \,  \tte^{p_1}\odot_\star \tte^{k_1}\odot_\star \tte^{k_2} \bigr) \ ,
\end{split}
\end{align}
where the non-zero pairings are of the form
\begin{align}\nn
\begin{split}
& \langle \tte^{k} , \sgreen (\tte^{p})\rangle_\star = \langle \sgreen(\tte^{p}) , \tte^{k}\rangle_\star = \langle \tte^{p} , \sgreen(\tte^{k})\rangle_\star = \tilde \sgreen(p) \, (2\pi)^d \, \delta(k+p) \ .
\end{split}
\end{align}

To calculate $\tilde G_2^\star(p_1,p_2)_1^{\swone}$, we apply $\ii\,\hbar\,\BVL\,\sH$ to (\ref{S2}) and obtain
\begin{align}\nn
&(\ii\,\hbar\,\BVL\,\sH)\,\bigl\{\CS _{\rm int},\sH\,(\tte^{p_1}\odot_\star \tte^{p_2})\bigr\}_\star \\[4pt]
& \qquad = -\frac{3}{2} \, \int_{k_1,k_2,k_3} \, V_3(k_1,k_2,k_3) \,
\bigl( \langle \tte^{k_3},\sgreen(\tte^{p_1})\rangle_\star \,  (\ii\,\hbar\,\BVL\,\sH) \, (\tte^{k_1}\odot_\star \tte^{k_2}\odot_\star \tte^{p_2}) \nn\\
& \hspace{6cm} + \langle \tte^{k_3}, \sgreen(\tte^{p_2})\rangle_\star \, (\ii\,\hbar\,\BVL\,\sH) \, (\tte^{p_1}\odot_\star \tte^{k_1}\odot_\star \tte^{k_2}) \bigr) \ .\nn
\end{align}
These two terms result in
\begin{align} \label{H31}
\begin{split}
(\ii\,\hbar\,\BVL\,\sH) \, (\tte^{k_1}\odot_\star \tte^{k_2}\odot_\star \tte^{p_2}) &= -\tfrac{2}{3} \, \ii\,\hbar \, \bigl( \langle \tte^{k_1} , \sgreen(\tte^{k_2})\rangle_\star\, \tte^{p_2}\\ 
&\hspace{15mm} + \langle \tte^{k_2} , \sgreen(\tte^{p_2})\rangle_\star\, \tte^{k_1} \, + \, \e^{\,-\ii\,k_1\cdot\theta\,k_2} \, \langle \tte^{k_1} , \sgreen(\tte^{p_2})\rangle_\star\, \tte^{k_2}\bigr)
\end{split}
\end{align}
and
\begin{align} \label{H32}
\begin{split}
(\ii\,\hbar\,\BVL\,\sH) \, (\tte^{p_1}\odot_\star \tte^{k_1}\odot_\star \tte^{k_2}) &= -\tfrac{2}{3} \, \ii\,\hbar \, \bigl( \langle \tte^{k_1} , \sgreen(\tte^{p_1})\rangle_\star\, \tte^{k_2}\\ 
&\hspace{15mm} +\langle \tte^{k_1} , \sgreen(\tte^{k_2})\rangle_\star\, \tte^{p_1} \, + \, \e^{\,\ii\,k_2\cdot\theta\,k_1} \, \langle \tte^{p_1} , \sgreen(\tte^{k_2})\rangle_\star\, \tte^{k_1}\bigr) \ . 
\end{split}
\end{align}

Adding another vertex insertion introduces three more internal momenta, two of which will be contracted via another application of $\ii\,\hbar\,\BVL\,\sH$, resulting in
\begin{align} \label{DeltaSH1}
\begin{split}
& (\ii\,\hbar\,\BVL\,\sH)\, \{\CS _{\rm int},\sH(\tte^{r})\}_\star = 3 \, \int_{q_1,q_2,q_3} \, V_3(q_1,q_2,q_3) \,  \langle \tte^{q_3}, \sgreen(\tte^{r})\rangle_\star \,  \langle \sgreen(\tte^{q_1}),\tte^{q_2}\rangle_\star \ ,
\end{split}
\end{align}
where we used the symmetry properties (\ref{eq:V3braidedsym}) and (\ref{eq:V3cyclicsym}). Applying (\ref{DeltaSH1}) to (\ref{H31}) and (\ref{H32}), and adding all terms, we get the first correction term
\begin{align} \label{PrvaKorekcija}
\begin{split}
\tilde G_2^\star(p_1,p_2)_1^{\swone} &=\frac{(\ii\,\hbar\,\lambda)^2}{3}\, \frac{(2\pi)^d\, \delta(p_1+p_2)}{(p_1^2 - m^2)\,(p_2^2 - m^2)} \ \int_k\, \frac{1}{(0-m^2)\,(k^2 - m^2)} \\[4pt]
& \quad\, +\frac{(\ii\,\hbar\,\lambda)^2}{6}\, \frac{(2\pi)^d\, \delta(p_1)}{p_1^2-m^2} \ \left[\int_{k}\, \frac{1}{k^2-m^2} \right]^2 \ \frac{(2\pi)^d\, \delta(p_2)}{p_2^2-m^2} \ .
\end{split}
\end{align}

We recognize this as the tadpole contributions to the two-point function at one-loop: the first term in $\tilde G_2^\star(p_1,p_2)_1^{\swone}$ is the contribution from the connected tadpole diagram 
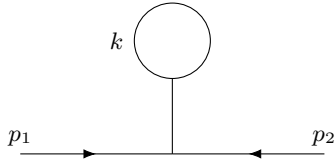
\begin{equation*}
\footnotesize
    \begin{tikzpicture}[baseline]
        \coordinate (k) at (0,0);
        \coordinate (l) at ($(k)+(90:1)$);
        \coordinate[label=above: $p_1$] (p1) at ($(k) +(180:2)$);
        \coordinate[label=above: $p_2$] (p2) at ($(k) +(0:2)$);

        \draw[decoration={markings, mark=at position 0.5 with {\arrow{Latex}} }, postaction={decorate}] (p1) -- (k);
        \draw[decoration={markings, mark=at position 0.5 with {\arrow{Latex}} }, postaction={decorate}] (p2) -- (k);
        \draw[decoration={markings }, postaction={decorate}] (k) -- (l);
        \draw[decoration={markings }, postaction={decorate}] ($(l)+(90:0.5)$) circle (0.5) node[left=15]{$k$};
    \end{tikzpicture}
\normalsize
\end{equation*}
while the second term is the contribution from the disconnected pair of tadpole diagrams
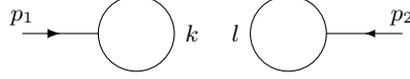
\begin{equation*}
\footnotesize
    \begin{tikzpicture}[baseline]
        \coordinate (k) at (-1.5,0);
        \coordinate (l) at (1.5,0);
        \coordinate[label=above: $p_1$] (p1) at (180:2.5);
        \coordinate[label=above: $p_2$] (p2) at (0:2.5);

        \draw[decoration={markings, mark=at position 0.5 with {\arrow{Latex}} }, postaction={decorate}] (p1) -- (k);
        \draw[decoration={markings, mark=at position 0.5 with {\arrow{latex}} }, postaction={decorate}] (p2) -- (l);
        \draw[decoration={markings }, postaction={decorate}] ($(k)+(0:0.5)$) circle (0.5) node[right=15]{$k$};
        \draw[decoration={markings }, postaction={decorate}] ($(l)-(0:0.5)$) circle (0.5) node[left=15]{$l$};
    \end{tikzpicture}
\label{fig:Phi321a}
\normalsize
\end{equation*}

To calculate $\tilde G_2^\star(p_1,p_2)_2^{\swone}$, we first have to add one more vertex to (\ref{S2}):
\begin{align} \label{eq:SintHSintH3e}
\begin{split}
& \bigl\{ \CS _{\rm int},\sH\, \{\CS _{\rm int},\sH\,(\tte^{p_1}\odot_\star \tte^{p_2})\}_\star \bigr\}_\star  \\[4pt]
& \hspace{1cm} = -\frac{3}{2} \, \int_{k_1,k_2,k_3} \, V_3(k_1,k_2,k_3) \, \bigl( \langle \tte^{k_3}, \sgreen(\tte^{p_1})\rangle_\star \,  \bigl\{\CS_{\rm int},\sH\, (\tte^{k_1}\odot_\star \tte^{k_2}\odot_\star \tte^{p_2}) \bigr\}_\star \, \\
& \hspace{7cm} + \langle \tte^{k_3}, \sgreen(\tte^{p_2})\rangle_\star \,\bigl\{\CS_{\rm int},\sH\, (\tte^{p_1}\odot_\star \tte^{k_1}\odot_\star \tte^{k_2}) \bigr\}_\star \bigr) \ .
\end{split}
\end{align}
Using braided symmetry \eqref{eq:V3braidedsym} of the interaction vertex, the first term is given explicitly by
\begin{align}\label{eq:SintH3e}
\begin{split}
& \bigl\{\CS_{\rm int},\sH\, (\tte^{k_1}\odot_\star \tte^{k_2}\odot_\star \tte^{p_2}) \bigr\}_\star  \\[4pt]
& \hspace{1cm} = -\frac{3}{3} \, \int_{q_1,q_2,q_3} \, V_3(q_1,q_2,q_3) \, \bigl( \langle \tte^{q_3}, \sgreen(\tte^{k_1})\rangle_\star \, \tte^{q_1}\odot_\star \tte^{q_2}\odot_\star \tte^{k_2} \odot_\star \tte^{p_2} \, \\
& \hspace{6cm} + \langle \tte^{q_3}, \sgreen(\tte^{k_2})\rangle_\star \, \tte^{k_1}\odot_\star \tte^{q_1}\odot_\star \tte^{q_2} \odot_\star \tte^{p_2} \, \\
& \hspace{7cm} + \langle \tte^{q_3}, \sgreen(\tte^{p_2})\rangle_\star \, \tte^{k_1}\odot_\star \tte^{k_2}\odot_\star \tte^{q_1} \odot_\star \tte^{q_2} \bigr) \ , 
\end{split}
\end{align}
and similarly for the second term.

The next step is to apply the operator $\ii\,\hbar\,\BVL\,\sH$ twice in a row, which is equivalent to the braided Wick theorem applied to any product of four fields found under the integral \cite{DimitrijevicCiric:2023hua}. In Section~\ref{sub:free theory sde} we will offer a new perspective and a rigorous derivation of the Wick theorem for braided scalar field theory; in the present calculation we use \eqref{eq:4pointfree}. The result is given by
\begin{align}
&(\ii\,\hbar\,\BVL\,\sH)^2 \, \bigl\{ \CS _{\rm int},\sH\, \bigl\{\CS _{\rm int},\sH\,(\tte^{p_1}\odot_\star \tte^{p_2})\bigr\}_\star \bigr\}_\star \nn \\[4pt]
& \quad = \frac{3}{2} \, \int_{k_1, k_2, k_3} \ \int_{q_1, q_2, q_3} \, V_3(k_1,k_2,k_3) \, V_3(q_1,q_2,q_3) \, \Bigl( \langle \tte^{k_3}, \sgreen(\tte^{p_1})\rangle_\star \, \langle \tte^{q_3}, \sgreen(\tte^{k_1})\rangle_\star\nn \\
&\hspace{3cm} \times \bigl( \langle \tte^{q_1}, \sgreen(\tte^{q_2})\rangle_\star\,\langle \tte^{k_2}, \sgreen(\tte^{p_2})\rangle_\star + \e^{\,\ii\,k_2\cdot\theta\,p_2}\, \langle \tte^{q_1}, \sgreen(\tte^{k_2})\rangle_\star\,\langle \tte^{q_2}, \sgreen(\tte^{p_2})\rangle_\star \nn \\
& \hspace{8cm} + \langle \tte^{q_1}, \sgreen(\tte^{p_2})\rangle_\star\, \langle \tte^{q_2}, \sgreen(\tte^{k_2})\rangle_\star\bigr) + \cdots \Bigr) \ ,\nn
\end{align}
where for brevity we display explicitly only the braided Wick expansion of the first product arising.

Unraveling everything, one gets the second correction term 
\begin{align} \label{DrugaKorekcija}
\begin{split}
\tilde G_2^\star(p_1,p_2)_2^{\swone} &= \frac{(\ii\,\hbar\,\lambda)^2}{12}\, \frac{(2\pi)^d\, \delta(p_1)}{p_1^2-m^2} \ \left[\int_{k}\, \frac{1}{k^2-m^2} \right]^2 \ \frac{(2\pi)^d\, \delta(p_2)}{p_2^2-m^2} \\
& \quad\, +\frac{(\ii\,\hbar\,\lambda)^2}{6}\, \frac{(2\pi)^d\, \delta(p_1+p_2)}{(p_1^2 - m^2)\,(p_2^2 - m^2)} \ \int_k\, \frac{1}{(0-m^2)\,(k^2 - m^2)}\\
& \quad\, +\frac{(\ii\,\hbar\,\lambda)^2}{2}\, \frac{(2\pi)^d\, \delta(p_1+p_2)}{(p_1^2 - m^2)\,(p_2^2 - m^2)} \ \int_k\, \frac{1}{\big(k^2-m^2\big)\,\big((p_1-k)^2 - m^2\big)}  \ .
\end{split}
\end{align}
The first and second terms are the same as the two tadpole contributions we found in $\tilde G_2^\star(p_1,p_2)^{\swone}_1$. The third term in \smash{$\tilde G_2^\star(p_1,p_2)^{\swone}_2$} comes from the vacuum polarization diagram
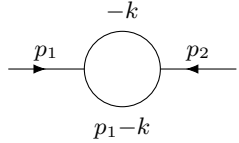
\begin{equation*}
\footnotesize
\begin{tikzpicture}
    \coordinate (k) at (-0.5,0);
    \coordinate (l) at (0.5,0);

    \draw[decoration={markings, mark=at position 0.5 with {\arrow{Latex}} }, postaction={decorate}] ($(k) + (180:1)$) -- (k) node[above, pos=0.5]{$p_1$};
    \draw[decoration={markings, mark=at position 0.5 with {\arrow{Latex[reversed]}} }, postaction={decorate}] (l) -- ($(l) + (0:1)$) node[above, pos=0.5]{$p_2$};
    \draw[decoration={markings }, postaction={decorate}] ($(k)+(0:0.5)$) circle (0.5) node[above=15]{$-k$} node[below=15]{$p_1{-}k$};    
    
\end{tikzpicture}
\normalsize
\end{equation*}

Finally, by adding the two corrections one finds that the one-loop contribution to the two-point function in momentum space is given by
\begin{align} \label{KorekcijaPhi3}
\begin{split}
\tilde G_2^\star(p_1,p_2)^{\swone} &= \frac{(\ii\, \hbar\,\lambda)^2}{2} \, \frac{(2\pi)^d\, \delta(p_1+p_2)}{(p_1^2 - m^2)\,(p_2^2 - m^2)} \ \int_k\, \frac{1}{\big(k^2-m^2\big)\,\big((p_1-k)^2 - m^2\big)}\\
& \quad \, + \frac{(\ii\, \hbar\,\lambda)^2}{2} \, \frac{(2\pi)^d\, \delta(p_1+p_2)}{(p_1^2 - m^2)\,(p_2^2 - m^2)} \ \int_k\, \frac{1}{(0-m^2)\,(k^2 - m^2)} \\
& \quad \, + \frac{(\ii\, \hbar\,\lambda)^2}{4} \, \frac{(2\pi)^d\, \delta(p_1)}{p_1^2-m^2} \ \left[\int_{k}\, \frac{1}{k^2-m^2} \right]^2 \ \frac{(2\pi)^d\, \delta(p_2)}{p_2^2-m^2} \ .
\end{split}
\end{align}
This is exactly the same result as in the commutative case because braided features that come from different sources, such as the vertex and the braided Wick theorem, have cancelled each other out. Diagrammatically, the result (\ref{KorekcijaPhi3}) is represented  with corresponding diagrams defined in Appendix~\ref{subsub:pictograms} as
\begin{equation}
\tilde G_2^\star(p_1,p_2)^{\swone} = { \footnotesize 18 \times \begin{tikzpicture}[scale=0.5, baseline]
        \coordinate (k) at (-0.5,0);
        \coordinate (l) at (0.5,0);
        \draw[decoration={markings, mark=at position 0.5 with {\arrow{Latex[reversed]}} }, postaction={decorate}] (k) -- ($(k) + (180:2)$) node[above]{$p_1$};
        \draw[decoration={markings, mark=at position 0.5 with {\arrow{Latex[reversed]}} }, postaction={decorate}] (l) -- ($(l) + (0:2)$) node[above]{$p_2$};
        \draw[decoration={markings }, postaction={decorate}] ($(k)+(0:0.5)$) circle (0.5);    
    \end{tikzpicture}
+18 \times  \begin{tikzpicture}[scale=0.5, baseline]
        \coordinate (k) at (0,0);
        \coordinate (l) at ($(k)+(90:1)$);
        \coordinate[label=above: $p_1$] (p1) at ($(k) +(180:2.5)$);
        \coordinate[label=above: $p_2$] (p2) at ($(k) +(0:2.5)$);

        \draw[decoration={markings, mark=at position 0.5 with {\arrow{Latex[reversed]}} }, postaction={decorate}] (k) -- (p1);
        \draw[decoration={markings, mark=at position 0.5 with {\arrow{Latex[reversed]}} }, postaction={decorate}] (k) -- (p2);
        \draw[decoration={markings}, postaction={decorate}] (k) -- (l);
        \draw[decoration={markings}, postaction={decorate}] ($(l)+(90:0.5)$) circle (0.5);
    \end{tikzpicture}
+9 \times  \begin{tikzpicture}[scale=0.5, baseline]
        \coordinate (k) at (-1.5,0);
        \coordinate (l) at (1.5,0);
        \coordinate[label=above: $p_1$] (p1) at (180:2.5);
        \coordinate[label=above: $p_2$] (p2) at (0:2.5);

        \draw[decoration={markings, mark=at position 0.5 with {\arrow{Latex[reversed]}} }, postaction={decorate}] (k) -- (p1);
        \draw[decoration={markings, mark=at position 0.5 with {\arrow{latex[reversed]}} }, postaction={decorate}] (l) -- (p2);
        \draw[decoration={markings}, postaction={decorate}] ($(k)+(0:0.5)$) circle (0.5);
        \draw[decoration={markings}, postaction={decorate}] ($(l)-(0:0.5)$) circle (0.5);
    \end{tikzpicture}  } \normalsize \ . \label{KorekcijaPhi3Diagrams}
\end{equation}


\subsection{Diagrammatic calculus}
\label{sub:diagramcalculus}

Calculations performed here using braided homological perturbation theory become increasingly more and more cumbersome as one goes to higher point correlation functions and higher loop orders. Therefore we now present a simplifying diagrammatic way to calculate the $n$-point correlation functions at one-loop. The method can be easily generalized to higher loop orders.

We start by introducing the relevant notation:
\begin{myitemize}

\item[$\bullet$] Black lines represent antifields $\tte^k$ with the corresponding momentum $k$.

\item[$\bullet$] Coloured lines denote propagator factors \smash{$G_{pk} := \langle\tte^p,\sgreen(\tte^k)\rangle_\star = \tilde \sgreen(k) \, (2\pi)^d\,\delta(k+p) = G_{kp}$} that appear after a contraction.  
Large coloured dots correspond to the vertices and are labelled with the arbitrary momentum introduced. Each one carries a numerical factor from (\ref{interaction-vertex}) with the corresponding momentum. Red colouring corresponds to the previous steps, while the green colouring represents the current step.

\item[$\bullet$] Acting with $\BVL\,\sH$ turns two black lines into one green line, creating a propagator.

\item[$\bullet$] Remembering that $\{\CS _{\rm int},- \}\,\sH$ is invariant under the action of the twist, acting with $\{\CS _{\rm int},- \}\,\sH$ turns one black line into one green line, creating a propagator, while at the same time introducing two more black lines (antifields) and a vertex.

\item[$\bullet$] To implement braiding, we have to pay attention to the following. Labelling momenta clockwise in a vertex corresponds to their order in the sense of antifields in the symmetric algebra. Cyclic permutations are allowed because of the vertex property (\ref{eq:V3cyclicsym}). Any other permutation introduces a phase factor. Distinct disconnected diagrams are drawn in an ordered manner. In order to pass a black line from one diagram to another, one should first permute it with other black lines inside one diagram and set it next to the other, and then permute that black line with black lines of the other diagram.

\end{myitemize}

With these diagrammatic rules we will now recalculate the two terms in (\ref{eq:2pt2loop}) separately. For the first term we start with
\begin{eqnarray*}
\tilde G_2^\star(p_1,p_2)^{\swone}_1 &=& \ii\,\hbar\,\BVL\,\sH\, \Big( \big\{\CS _{\rm int}, \sH\, \big(\ii\,\hbar\,\BVL\,\sH\, \{\CS _{\rm int}, \sH\, \Bigg( { \footnotesize \begin{tikzpicture}[scale=0.5, baseline]
        \coordinate (p1) at (90:1);
        \coordinate[label=below: {$p_1$}] (p1in) at (270:1);
        \coordinate (p2) at ($(p1) +(0:2)$);
        \coordinate[label=below: {$p_2$}] (p2in) at ($(p1in) +(0:2)$);

        \draw[decoration={markings, mark=at position 0.5 with {\arrow{Latex}}}, postaction={decorate}] (p1in) -- (p1);
        \draw[decoration={markings, mark=at position 0.5 with {\arrow{Latex}}}, postaction={decorate}] (p2in) -- (p2);
    \end{tikzpicture} } \normalsize
\Bigg) \}_\star \big) \big\}_\star \Big)\\[4pt]
&=& -\frac{3}{2}\,\ii\,\hbar\,\BVL\,\sH\, \Big( \big\{\CS _{\rm int}, \sH\, \big(\ii\,\hbar\,\BVL\,\sH\, \Bigg(
   { \footnotesize  \begin{tikzpicture}[scale=0.5, baseline]
        \coordinate[label=left: \textcolor{dgreen}{$k$}] (k) at (0,0);
        \coordinate (k1) at (270:2);
        \coordinate[label=right: {$k_3$}] (k3) at (45:2);
        \coordinate[label=left: {$k_2$}] (k2) at (135:2);
        \coordinate[label=below: {$p_2$}] (p2in) at ($(k1) +(0:2)$);
        \coordinate (p2) at ($(k) +(0:2)$);        
        \draw[fill, dgreen] (k) circle[radius=0.1];
        
        \draw[decoration={markings, mark=at position 0.5 with {\arrow{Latex}}}, postaction={decorate}] (k3) -- (k);
        \draw[decoration={markings, mark=at position 0.5 with {\arrow{Latex}}}, postaction={decorate}] (k2) -- (k);
        \draw[decoration={markings, mark=at position 0.5 with {\arrow{Latex}}}, postaction={decorate}] (p2in) -- (p2);

        \draw[color=dgreen, decoration={markings}, postaction={decorate}] (k1) -- (k) node[left, pos=0.5]{$G_{k_1 p_1}$};
    \end{tikzpicture} } \normalsize
+  { \footnotesize \begin{tikzpicture}[scale=0.5, baseline]
        \coordinate[label=right: \textcolor{dgreen}{$k$}] (k) at (0,0);
        \coordinate (k1) at (270:2);
        \coordinate[label=right: {$k_3$}] (k3) at (45:2);
        \coordinate[label=left: {$k_2$}] (k2) at (135:2);
        \coordinate[label=below: {$p_1$}] (p1in) at ($(k1) +(180:2)$);
        \coordinate (p1) at ($(k) +(180:2)$);
        \draw[fill, dgreen] (k) circle[radius=0.1];
        
        \draw[decoration={markings, mark=at position 0.5 with {\arrow{Latex}}}, postaction={decorate}] (k3) -- (k);
        \draw[decoration={markings, mark=at position 0.5 with {\arrow{Latex}}}, postaction={decorate}] (k2) -- (k);
        \draw[decoration={markings, mark=at position 0.5 with {\arrow{Latex}}}, postaction={decorate}] (p1in) -- (p1);
    
        \draw[color=dgreen, decoration={markings}, postaction={decorate}] (k1) -- (k) node[right, pos=0.5]{$G_{k_1 p_2}$};
    \end{tikzpicture}   } \normalsize
\Bigg) \big) \big\}_\star \Big)\\[4pt]
&=& \frac{3\cdot 2}{2\cdot 3}\,\ii\,\hbar\,\BVL\,\sH\, \Big( \big\{\CS _{\rm int}, \sH\, \Bigg(
   { \footnotesize \begin{tikzpicture}[scale=0.5, baseline]
        \coordinate[label=left: \textcolor{red}{$k$}] (k) at (0,0);
        \coordinate (k1) at (270:2);
        \coordinate (k3) at (45:2);
        \coordinate (k2) at (135:2);
        \coordinate[label=below: {$p_2$}] (p2in) at ($(k1) +(0:2)$);
        \coordinate (p2) at ($(k) +(0:2)$);
        \draw[fill, red] (k) circle[radius=0.1];

        \draw[color=dgreen, decoration={markings}, postaction={decorate}] ($(k) + (90:0.5)$) circle (0.5) node[above=10]{$G_{k_2 k_3}$};
        \draw[color=red, decoration={markings}, postaction={decorate}] (k1) -- (k) node[left, pos=0.5]{$G_{k_1 p_1}$};
        \draw[decoration={markings, mark=at position 0.5 with {\arrow{Latex}}}, postaction={decorate}] (p2in) -- (p2);
    \end{tikzpicture} } \normalsize
+ \e^{\,\ii\, p_2 \cdot \theta\, k_3} \times  {\footnotesize \begin{tikzpicture}[scale=0.5, baseline]
        \coordinate[label=left: \textcolor{red}{$k$}] (k) at (0,0);
        \coordinate (k1) at (270:2);
        \coordinate[label=right: {$k_3$}] (k3) at (45:2);
        \coordinate (k2) at (135:2);
        \coordinate (p2in) at ($(k1) +(0:2)$);
        \coordinate (p2) at ($(k) +(0:2)$);
        \draw[fill, red] (k) circle[radius=0.1];
        
        \draw[decoration={markings, mark=at position 0.5 with {\arrow{Latex}}}, postaction={decorate}] (k3) -- (k);
        \draw[color=dgreen, decoration={markings}, postaction={decorate}] (k2) -- (k) node[below left, pos=0.5]{$G_{k_2 p_2}$};

        \draw[color=red, decoration={markings}, postaction={decorate}] (k1) -- (k) node[left, pos=0.5]{$G_{k_1 p_1}$};
    \end{tikzpicture} } \normalsize \\
&& \hspace{1cm} + { \footnotesize \begin{tikzpicture}[scale=0.5, baseline]
        \coordinate[label=left: \textcolor{red}{$k$}] (k) at (0,0);
        \coordinate (k1) at (270:2);
        \coordinate (k3) at (45:2);
        \coordinate[label=left: {$k_2$}] (k2) at (135:2);
        \coordinate (p2in) at ($(k1) +(0:2)$);
        \coordinate (p2) at ($(k) +(0:2)$);
        \draw[fill, red] (k) circle[radius=0.1];
        
        \draw[decoration={markings, mark=at position 0.5 with {\arrow{Latex}}}, postaction={decorate}] (k2) -- (k);
        \draw[color=dgreen, decoration={markings}, postaction={decorate}] (k3) -- (k) node[below right, pos=0.5]{$G_{k_3 p_2}$};

        \draw[color=red, decoration={markings}, postaction={decorate}] (k1) -- (k) node[left, pos=0.5]{$G_{k_1 p_1}$};
    \end{tikzpicture} } \normalsize
+ { \footnotesize \begin{tikzpicture}[scale=0.5, baseline]
        \coordinate[label=right: \textcolor{red}{$k$}] (k) at (0,0);
        \coordinate (k1) at (270:2);
        \coordinate[label=right: {$k_3$}] (k3) at (45:2);
        \coordinate (k2) at (135:2);
        \coordinate (p2in) at ($(k1) +(0:2)$);
        \coordinate (p2) at ($(k) +(0:2)$);
        \draw[fill, red] (k) circle[radius=0.1];
        
        \draw[decoration={markings, mark=at position 0.5 with {\arrow{Latex}}}, postaction={decorate}] (k3) -- (k);
        \draw[color=dgreen, decoration={markings}, postaction={decorate}] (k2) -- (k) node[below left, pos=0.5]{$G_{p_1 k_2}$};

        \draw[color=red, decoration={markings}, postaction={decorate}] (k1) -- (k) node[right, pos=0.5]{$G_{k_1 p_2}$};
    \end{tikzpicture} } \normalsize \\
&& \hspace{2cm} +\   \e^{\,\ii\, k_2 \cdot \theta\, p_1} \times {\footnotesize \begin{tikzpicture}[scale=0.5, baseline]
        \coordinate[label=right: \textcolor{red}{$k$}] (k) at (0,0);
        \coordinate (k1) at (270:2);
        \coordinate (k3) at (45:2);
        \coordinate[label=left: {$k_2$}] (k2) at (135:2);
        \coordinate (p2in) at ($(k1) +(0:2)$);
        \coordinate (p2) at ($(k) +(0:2)$);
        \draw[fill, red] (k) circle[radius=0.1];
        
        \draw[decoration={markings, mark=at position 0.5 with {\arrow{Latex}}}, postaction={decorate}] (k2) -- (k);
        \draw[color=dgreen, decoration={markings}, postaction={decorate}] (k3) -- (k) node[below right, pos=0.5]{$G_{p_1 k_3}$};

        \draw[color=red, decoration={markings}, postaction={decorate}] (k1) -- (k) node[right, pos=0.5]{$G_{k_1 p_2}$};
    \end{tikzpicture} } \normalsize
+  { \footnotesize \begin{tikzpicture}[scale=0.5, baseline]
        \coordinate[label=right: \textcolor{red}{$k$}] (k) at (0,0);
        \coordinate (k1) at (270:2);
        \coordinate (k3) at (45:2);
        \coordinate (k2) at (135:2);
        \coordinate[label=below: {$p_1$}] (p1in) at ($(k1) +(180:2)$);
        \coordinate (p1) at ($(k) +(180:2)$);
        \draw[fill, red] (k) circle[radius=0.1];
        
        \draw[color=dgreen, decoration={markings}, postaction={decorate}] ($(k) + (90:0.5)$) circle (0.5) node[right=10]{$G_{k_2 k_3}$};
        \draw[decoration={markings, mark=at position 0.5 with {\arrow{Latex}}}, postaction={decorate}] (p1in) -- (p1);
    
        \draw[color=red, decoration={markings}, postaction={decorate}] (k1) -- (k) node[right, pos=0.5]{$G_{k_1 p_2}$};
    \end{tikzpicture} } \normalsize \Bigg) \big\}_\star \Big) \ .
\end{eqnarray*}

Adding another interaction vertex results in
\begin{eqnarray*}
\tilde G_2^\star(p_1,p_2)^{\swone}_1 &=& -\frac{3\cdot 2 \cdot 3}{2\cdot 3 \cdot 1}\,\ii\,\hbar\,\BVL\,\sH\, \Bigg(
    {\footnotesize \begin{tikzpicture}[scale=0.5, baseline]
        \coordinate[label=left: \textcolor{red}{$k$}] (k) at (0,0);
        \coordinate (k1) at (270:2);
        \coordinate (k3) at (45:2);
        \coordinate (k2) at (135:2);
        \coordinate[label=right: \textcolor{dgreen}{$l$}] (l) at ($(k) +(0:2)$);
        \coordinate (l1) at ($(l) +(270:2)$);
        \coordinate[label=left: $l_2$] (l2) at ($(l) +(135:2)$);
        \coordinate[label=right: $l_3$] (l3) at ($(l) +(45:2)$);
        \draw[fill, red] (k) circle[radius=0.1];
        \draw[fill, dgreen] (l) circle[radius=0.1];

        \draw[color=red, decoration={markings}, postaction={decorate}] ($(k) + (90:0.5)$) circle (0.5) node[left=10]{$G_{k_2 k_3}$};
        \draw[color=red, decoration={markings}, postaction={decorate}] (k1) -- (k) node[left, pos=0.5]{$G_{k_1 p_1}$};
        
        \draw[color=dgreen, decoration={markings}, postaction={decorate}] (l1) -- (l) node[right, pos=0.5]{$G_{l_1 p_2}$};        
        \draw[decoration={markings, mark=at position 0.5 with {\arrow{Latex}}}, postaction={decorate}] (l2) -- (l);
        \draw[decoration={markings, mark=at position 0.5 with {\arrow{Latex}}}, postaction={decorate}] (l3) -- (l);
    \end{tikzpicture} } \normalsize
+ \e^{\,\ii\, p_2 \cdot \theta\, k_3} \times { \footnotesize \begin{tikzpicture}[scale=0.5, baseline]
        \coordinate[label=left: \textcolor{red}{$k$}] (k) at (0,0);
        \coordinate (k1) at (270:2);
        \coordinate[label=left: \textcolor{red}{$l$}] (l) at (45:2);
        \coordinate (k2) at (135:2);
        \coordinate[label=left: $l_2$] (l2) at ($(l) + (90:2)$);
        \coordinate[label=right: $l_3$] (l3) at ($(l) + (0:2)$);
        \draw[fill, red] (k) circle[radius=0.1];
        \draw[fill, dgreen] (l) circle[radius=0.1];
        
        \draw[color=red, decoration={markings}, postaction={decorate}] (k2) -- (k) node[below left, pos=0.5]{$G_{k_2 p_2}$};
        \draw[color=red, decoration={markings}, postaction={decorate}] (k1) -- (k) node[left, pos=0.5]{$G_{k_1 p_1}$};
        
        \draw[color=dgreen, decoration={markings}, postaction={decorate}] (l) -- (k) node[below right, pos=0.5]{$G_{k_3 l_1}$};
        \draw[decoration={markings, mark=at position 0.5 with {\arrow{Latex}}}, postaction={decorate}] (l2) -- (l);
        \draw[decoration={markings, mark=at position 0.5 with {\arrow{Latex}}}, postaction={decorate}] (l3) -- (l);
    \end{tikzpicture} } \normalsize \\
&& \hspace{1cm} + { \footnotesize  \begin{tikzpicture}[scale=0.5, baseline]
        \coordinate[label=left: \textcolor{red}{$k$}] (k) at (0,0);
        \coordinate (k1) at (270:2);
        \coordinate[label=right: \textcolor{dgreen}{$l$}] (l) at (135:2);
        \coordinate (k3) at (45:2);
        \coordinate[label=left: $l_2$] (l2) at ($(l) + (180:2)$);
        \coordinate[label=right: $l_3$] (l3) at ($(l) + (90:2)$);
        \draw[fill, red] (k) circle[radius=0.1];
        \draw[fill, dgreen] (l) circle[radius=0.1];
        
        \draw[color=red, postaction={decorate}] (k3) -- (k) node[below right, pos=0.5]{$G_{k_3 p_2}$};
        \draw[color=red, decoration={markings}, postaction={decorate}] (k1) -- (k) node[left, pos=0.5]{$G_{k_1 p_1}$};
        
        \draw[color=dgreen, decoration={markings}, postaction={decorate}] (l) -- (k) node[below left, pos=0.5]{$G_{k_2 l_1}$};
        \draw[decoration={markings, mark=at position 0.5 with {\arrow{Latex}}}, postaction={decorate}] (l2) -- (l);
        \draw[decoration={markings, mark=at position 0.5 with {\arrow{Latex}}}, postaction={decorate}] (l3) -- (l);
    \end{tikzpicture} } \normalsize
+ {\footnotesize  \begin{tikzpicture}[scale=0.5, baseline]
        \coordinate[label=right: \textcolor{red}{$k$}] (k) at (0,0);
        \coordinate (k1) at (270:2);
        \coordinate[label=left: \textcolor{dgreen}{$l$}] (l) at (45:2);
        \coordinate (k2) at (135:2);
        \coordinate[label=left: $l_2$] (l2) at ($(l) + (90:2)$);
        \coordinate[label=right: $l_3$] (l3) at ($(l) + (0:2)$);
        \draw[fill, red] (k) circle[radius=0.1];
        \draw[fill, dgreen] (l) circle[radius=0.1];
        
        \draw[color=red, decoration={markings}, postaction={decorate}] (k2) -- (k) node[below left, pos=0.5]{$G_{k_2 p_1}$};
        \draw[color=red, decoration={markings}, postaction={decorate}] (k1) -- (k) node[right, pos=0.5]{$G_{k_1 p_2}$};
        
        \draw[color=dgreen, decoration={markings}, postaction={decorate}] (l) -- (k) node[below right, pos=0.5]{$G_{k_3 l_1}$};
        \draw[decoration={markings, mark=at position 0.5 with {\arrow{Latex}}}, postaction={decorate}] (l2) -- (l);
        \draw[decoration={markings, mark=at position 0.5 with {\arrow{Latex}}}, postaction={decorate}] (l3) -- (l);
    \end{tikzpicture} } \normalsize \\
&&  \hspace{2cm} +\  \e^{\,\ii\, k_2 \cdot \theta\, p_1} \times {\footnotesize \begin{tikzpicture}[scale=0.5, baseline]
        \coordinate[label=right: \textcolor{red}{$k$}] (k) at (0,0);
        \coordinate (k1) at (270:2);
        \coordinate[label=right: \textcolor{dgreen}{$l$}] (l) at (135:2);
        \coordinate (k3) at (45:2);
        \coordinate[label=left: $l_2$] (l2) at ($(l) + (180:2)$);
        \coordinate[label=right: $l_3$] (l3) at ($(l) + (90:2)$);
        \draw[fill, red] (k) circle[radius=0.1];
        \draw[fill, dgreen] (l) circle[radius=0.1];
        
        \draw[color=red, decoration={markings}, postaction={decorate}] (k3) -- (k) node[below right, pos=0.5]{$G_{k_3 p_1}$};
        \draw[color=red, decoration={markings}, postaction={decorate}] (k1) -- (k) node[right, pos=0.5]{$G_{k_1 p_2}$};
        
        \draw[color=dgreen, decoration={markings}, postaction={decorate}] (l) -- (k) node[below left, pos=0.5]{$G_{k_2 l_1}$};
        \draw[decoration={markings, mark=at position 0.5 with {\arrow{Latex}}}, postaction={decorate}] (l2) -- (l);
        \draw[decoration={markings, mark=at position 0.5 with {\arrow{Latex}}}, postaction={decorate}] (l3) -- (l);
    \end{tikzpicture} } \normalsize
+ { \footnotesize  \begin{tikzpicture}[scale=0.5, baseline]
        \coordinate[label=right: \textcolor{red}{$k$}] (k) at (0,0);
        \coordinate (k1) at (270:2);
        \coordinate (k3) at (45:2);
        \coordinate (k2) at (135:2);
        \coordinate[label=left: \textcolor{dgreen}{$l$}] (l) at ($(k) -(0:2)$);
        \coordinate (l1) at ($(l) +(270:2)$);
        \coordinate[label=left: $l_2$] (l2) at ($(l) +(135:2)$);
        \coordinate[label=right: $l_3$] (l3) at ($(l) +(45:2)$);
        \draw[fill, red] (k) circle[radius=0.1];
        \draw[fill, dgreen] (l) circle[radius=0.1];

        \draw[color=red, decoration={markings}, postaction={decorate}] ($(k) + (90:0.5)$) circle (0.5) node[right=10]{$G_{k_2 k_3}$};
        \draw[color=red, decoration={markings}, postaction={decorate}] (k1) -- (k) node[right, pos=0.5]{$G_{k_1 p_2}$};
        
        \draw[color=dgreen, decoration={markings}, postaction={decorate}] (l1) -- (l) node[left, pos=0.5]{$G_{l_1 p_1}$};        
        \draw[decoration={markings, mark=at position 0.5 with {\arrow{Latex}}}, postaction={decorate}] (l2) -- (l);
        \draw[decoration={markings, mark=at position 0.5 with {\arrow{Latex}}}, postaction={decorate}] (l3) -- (l);
    \end{tikzpicture} } \normalsize \Bigg) \ .
\end{eqnarray*}

Finally, applying the second operator $\ii\,\hbar\,\BVL\,\sH$ gives the diagrams
\begin{eqnarray*}
& & \frac{3\cdot 2 \cdot 3 \cdot 2}{2\cdot 3 \cdot 1 \cdot 2} \, \Bigg(
    { \footnotesize \begin{tikzpicture}[scale=0.5, baseline]
        \coordinate[label=left: \textcolor{red}{$k$}] (k) at (0,0);
        \coordinate (k1) at (270:2);
        \coordinate (k3) at (45:2);
        \coordinate (k2) at (135:2);
        \coordinate[label=right: \textcolor{red}{$l$}] (l) at ($(k) +(0:2)$);
        \coordinate (l1) at ($(l) +(270:2)$);
        \coordinate (l2) at ($(l) +(135:2)$);
        \coordinate (l3) at ($(l) +(45:2)$);
        \draw[fill, red] (k) circle[radius=0.1];
        \draw[fill, red] (l) circle[radius=0.1];

        \draw[color=red,  postaction={decorate}] ($(k) + (90:0.5)$) circle (0.5) node[left=10]{$G_{k_2 k_3}$};
        \draw[color=red, postaction={decorate}] (k1) -- (k) node[left, pos=0.5]{$G_{k_1 p_1}$};
        
        \draw[color=red, postaction={decorate}] (l1) -- (l) node[right, pos=0.5]{$G_{l_1 p_2}$};
        \draw[color=dgreen,  postaction={decorate}] ($(l) + (90:0.5)$) circle (0.5) node[right=10]{$G_{l_2 l_3}$};
    \end{tikzpicture} } \normalsize
+  { \footnotesize  \begin{tikzpicture}[scale=0.5, baseline]
        \coordinate[label=left: \textcolor{red}{$k$}] (k) at (0,0);
        \coordinate (k1) at (270:2);
        \coordinate[label=left: \textcolor{red}{$l$}] (l) at (45:2);
        \coordinate (k2) at (135:2);
        \coordinate (l2) at ($(l) + (90:2)$);
        \coordinate (l3) at ($(l) + (0:2)$);
        \draw[fill, red] (k) circle[radius=0.1];
        \draw[fill, red] (l) circle[radius=0.1];
        
        \draw[color=red,  postaction={decorate}] (k2) -- (k) node[below left, pos=0.5]{$G_{k_2 p_2}$};
        \draw[color=red,  postaction={decorate}] (k1) -- (k) node[left, pos=0.5]{$G_{k_1 p_1}$};
        
        \draw[color=red,  postaction={decorate}] (l) -- (k) node[below right, pos=0.5]{$G_{k_3 l_1}$};
        \draw[color=dgreen,  postaction={decorate}] ($(l) + (45:0.5)$) circle (0.5) node[right=10]{$G_{l_2 l_3}$};
    \end{tikzpicture}
+   \begin{tikzpicture}[scale=0.5, baseline]
        \coordinate[label=left: \textcolor{red}{$k$}] (k) at (0,0);
        \coordinate (k1) at (270:2);
        \coordinate[label=right: \textcolor{red}{$l$}] (l) at (135:2);
        \coordinate (k3) at (45:2);
        \coordinate (l2) at ($(l) + (180:2)$);
        \coordinate (l3) at ($(l) + (90:2)$);
        \draw[fill, red] (k) circle[radius=0.1];
        \draw[fill, red] (l) circle[radius=0.1];
        
        \draw[color=red,  postaction={decorate}] (k3) -- (k) node[below right, pos=0.5]{$G_{k_3 p_2}$};
        \draw[color=red,  postaction={decorate}] (k1) -- (k) node[left, pos=0.5]{$G_{k_1 p_1}$};
        
        \draw[color=red,  postaction={decorate}] (l) -- (k) node[below left, pos=0.5]{$G_{k_2 l_1}$};
        \draw[color=dgreen,  postaction={decorate}] ($(l) + (135:0.5)$) circle (0.5) node[left=10]{$G_{l_2 l_3}$};
    \end{tikzpicture} } \normalsize \\
&& \hspace{2cm} + { \footnotesize \begin{tikzpicture}[scale=0.5, baseline]
        \coordinate[label=right: \textcolor{red}{$k$}] (k) at (0,0);
        \coordinate (k1) at (270:2);
        \coordinate[label=left: \textcolor{red}{$l$}] (l) at (45:2);
        \coordinate (k2) at (135:2);
        \coordinate (l2) at ($(l) + (90:2)$);
        \coordinate (l3) at ($(l) + (0:2)$);
        \draw[fill, red] (k) circle[radius=0.1];
        \draw[fill, red] (l) circle[radius=0.1];
        
        \draw[color=red,  postaction={decorate}] (k2) -- (k) node[below left, pos=0.5]{$G_{k_2 p_1}$};
        \draw[color=red,  postaction={decorate}] (k1) -- (k) node[right, pos=0.5]{$G_{k_1 p_2}$};
        
        \draw[color=red,  postaction={decorate}] (l) -- (k) node[below right, pos=0.5]{$G_{k_3 l_1}$};
        \draw[color=dgreen,  postaction={decorate}] ($(l) + (45:0.5)$) circle (0.5) node[right=10]{$G_{l_2 l_3}$};
    \end{tikzpicture} } \normalsize
+ {\footnotesize  \begin{tikzpicture}[scale=0.5, baseline]
        \coordinate[label=right: \textcolor{red}{$k$}] (k) at (0,0);
        \coordinate (k1) at (270:2);
        \coordinate[label=right: \textcolor{red}{$l$}] (l) at (135:2);
        \coordinate (k3) at (45:2);
        \coordinate (l2) at ($(l) + (180:2)$);
        \coordinate (l3) at ($(l) + (90:2)$);
        \draw[fill, red] (k) circle[radius=0.1];
        \draw[fill, red] (l) circle[radius=0.1];
        
        \draw[color=red,  postaction={decorate}] (k3) -- (k) node[below right, pos=0.5]{$G_{k_3 p_1}$};
        \draw[color=red,  postaction={decorate}] (k1) -- (k) node[right, pos=0.5]{$G_{k_1 p_2}$};
        
        \draw[color=red,  postaction={decorate}] (l) -- (k) node[below left, pos=0.5]{$G_{k_2 l_1}$};
        \draw[color=dgreen,  postaction={decorate}] ($(l) + (135:0.5)$) circle (0.5) node[left=10]{$G_{l_2 l_3}$};
    \end{tikzpicture} } \normalsize
+ { \footnotesize  \begin{tikzpicture}[scale=0.5, baseline]
        \coordinate[label=right: \textcolor{red}{$k$}] (k) at (0,0);
        \coordinate (k1) at (270:2);
        \coordinate (k3) at (45:2);
        \coordinate (k2) at (135:2);
        \coordinate[label=left: \textcolor{red}{$l$}] (l) at ($(k) -(0:2)$);
        \coordinate (l1) at ($(l) +(270:2)$);
        \coordinate (l2) at ($(l) +(135:2)$);
        \coordinate (l3) at ($(l) +(45:2)$);
        \draw[fill, red] (k) circle[radius=0.1];
        \draw[fill, red] (l) circle[radius=0.1];

        \draw[color=red,  postaction={decorate}] ($(k) + (90:0.5)$) circle (0.5) node[right=10]{$G_{k_2 k_3}$};
        \draw[color=red,  postaction={decorate}] (k1) -- (k) node[right, pos=0.5]{$G_{k_1 p_2}$};
        
        \draw[color=red,  postaction={decorate}] (l1) -- (l) node[left, pos=0.5]{$G_{l_1 p_1}$};        
        \draw[color=dgreen,  postaction={decorate}] ($(l) + (90:0.5)$) circle (0.5) node[left=10]{$G_{l_2 l_3}$};
    \end{tikzpicture} } \normalsize \Bigg) \ .
\end{eqnarray*}    
The final result can be read off from here and is given by
\begin{equation*}    
\tilde G_2^\star(p_1,p_2)^{\swone}_1 = 
6 \times {\footnotesize \begin{tikzpicture}[scale=0.5, baseline]
        \coordinate (k) at (-1.5,0);
        \coordinate (l) at (1.5,0);
        \coordinate[label=above: $p_1$] (p1) at (180:2.5);
        \coordinate[label=above: $p_2$] (p2) at (0:2.5);

        \draw[decoration={markings, mark=at position 0.5 with {\arrow{Latex[reversed]}} }, postaction={decorate}] (k) -- (p1);
        \draw[decoration={markings, mark=at position 0.5 with {\arrow{latex[reversed]}} }, postaction={decorate}] (l) -- (p2);
        \draw[postaction={decorate}] ($(k)+(0:0.5)$) circle (0.5);
        \draw[ postaction={decorate}] ($(l)-(0:0.5)$) circle (0.5);
    \end{tikzpicture} } \normalsize
\>+\> 12 \times { \footnotesize \begin{tikzpicture}[scale=0.5, baseline]
        \coordinate (k) at (0,0);
        \coordinate (l) at ($(k)+(90:1)$);
        \coordinate[label=above: $p_1$] (p1) at ($(k) +(180:2.5)$);
        \coordinate[label=above: $p_2$] (p2) at ($(k) +(0:2.5)$);

        \draw[decoration={markings, mark=at position 0.5 with {\arrow{Latex[reversed]}} }, postaction={decorate}] (k) -- (p1);
        \draw[decoration={markings, mark=at position 0.5 with {\arrow{Latex[reversed]}} }, postaction={decorate}] (k) -- (p2);
        \draw[ postaction={decorate}] (k) -- (l);
        \draw[postaction={decorate}] ($(l)+(90:0.5)$) circle (0.5);
    \end{tikzpicture} } \normalsize  \ .
\end{equation*}
Note that, after inserting the vertex contributions and using the corresponding delta-functions to perform integrations over internal momenta, all phase factors cancel, as expected.

The second term is calculated through analogous steps as
\begin{eqnarray*}
\tilde G_2^\star(p_1,p_2)^{\swone}_2 &=& \ii\,\hbar\,\BVL\,\sH\, \bigg( \ii\,\hbar\,\BVL\,\sH\, \Big( \Big\{\CS _{\rm int}, \sH\, \big\{\CS _{\rm int}, \sH \Bigg(
    { \footnotesize \begin{tikzpicture}[scale=0.5, baseline]
        \coordinate (p1) at (90:1);
        \coordinate[label=below: {$p_1$}] (p1in) at (270:1);
        \coordinate (p2) at ($(p1) +(0:2)$);
        \coordinate[label=below: {$p_2$}] (p2in) at ($(p1in) +(0:2)$);

        \draw[decoration={markings, mark=at position 0.5 with {\arrow{latex}}}, postaction={decorate}] (p1in) -- (p1);
        \draw[decoration={markings, mark=at position 0.5 with {\arrow{latex}}}, postaction={decorate}] (p2in) -- (p2);
    \end{tikzpicture} } \normalsize
\Bigg) \big\}_\star \Big\}_\star\, \Big) \bigg)\\[4pt]
&=& -\frac{3}{2}\,\ii\,\hbar\,\BVL\,\sH\, \bigg( \ii\,\hbar\,\BVL\,\sH\, \Big( \Big\{\CS _{\rm int}, \sH\, \Bigg(
   { \footnotesize \begin{tikzpicture}[scale=0.5, baseline]
        \coordinate[label=left: \textcolor{dgreen}{$k$}] (k) at (0,0);
        \coordinate (k1) at (270:2);
        \coordinate[label=right: {$k_3$}] (k3) at (45:2);
        \coordinate[label=left: {$k_2$}] (k2) at (135:2);
        \coordinate[label=below: {$p_2$}] (p2in) at ($(k1) +(0:2)$);
        \coordinate (p2) at ($(k) +(0:2)$);        
        \draw[fill, dgreen] (k) circle[radius=0.1];
        
        \draw[decoration={markings, mark=at position 0.5 with {\arrow{Latex}}}, postaction={decorate}] (k3) -- (k);
        \draw[decoration={markings, mark=at position 0.5 with {\arrow{Latex}}}, postaction={decorate}] (k2) -- (k);
        \draw[decoration={markings, mark=at position 0.5 with {\arrow{Latex}}}, postaction={decorate}] (p2in) -- (p2);

        \draw[color=dgreen,  postaction={decorate}] (k1) -- (k) node[left, pos=0.5]{$G_{k_1 p_1}$};
    \end{tikzpicture} } \normalsize
+  { \footnotesize \begin{tikzpicture}[scale=0.5, baseline]
        \coordinate[label=right: \textcolor{dgreen}{$k$}] (k) at (0,0);
        \coordinate (k1) at (270:2);
        \coordinate[label=right: {$k_3$}] (k3) at (45:2);
        \coordinate[label=left: {$k_2$}] (k2) at (135:2);
        \coordinate[label=below: {$p_1$}] (p1in) at ($(k1) +(180:2)$);
        \coordinate (p1) at ($(k) +(180:2)$);
        \draw[fill, dgreen] (k) circle[radius=0.1];
        
        \draw[decoration={markings, mark=at position 0.5 with {\arrow{Latex}}}, postaction={decorate}] (k3) -- (k);
        \draw[decoration={markings, mark=at position 0.5 with {\arrow{Latex}}}, postaction={decorate}] (k2) -- (k);
        \draw[decoration={markings, mark=at position 0.5 with {\arrow{Latex}}}, postaction={decorate}] (p1in) -- (p1);
    
        \draw[color=dgreen,  postaction={decorate}] (k1) -- (k) node[right, pos=0.5]{$G_{k_1 p_2}$};
    \end{tikzpicture} } \normalsize
\Bigg) \Big\}_\star\, \Big) \bigg) \ .
\end{eqnarray*}
 
Adding the second interaction vertex results in  
\begin{eqnarray*}
&&  \frac{3\cdot 3}{2\cdot 3}\,\ii\,\hbar\,\BVL\,\sH\, \bigg( \ii\,\hbar\,\BVL\,\sH\, \Bigg(
    \big(1+\e^{\,\ii\, (l_2 + l_3)\cdot \theta\, k_3 - \ii\, k_2\cdot \theta\, k_3}\big) \times { \footnotesize \begin{tikzpicture}[scale=0.5, baseline]
        \coordinate[label=left: \textcolor{red}{$k$}] (k) at (0,0);
        \coordinate (k1) at (270:2);
        \coordinate[label=right: \textcolor{dgreen}{$l$}] (l) at (135:2);
        \coordinate[label=right: {$k_3$}] (k3) at (45:2);
        \coordinate[label=below: {$p_2$}] (p2in) at ($(k1) +(0:2)$);
        \coordinate[label=left: $l_2$] (l2) at ($(l) + (180:2)$);
        \coordinate[label=right: $l_3$] (l3) at ($(l) + (90:2)$);
        \draw[fill, red] (k) circle[radius=0.1];
        \draw[fill, dgreen] (l) circle[radius=0.1];
        
        \draw[decoration={markings, mark=at position 0.5 with {\arrow{Latex}}}, postaction={decorate}] (k3) -- (k);
        \draw[color=red,  postaction={decorate}] (k1) -- (k) node[left, pos=0.5]{$G_{k_1 p_1}$};
        
        \draw[color=dgreen,  postaction={decorate}] (l) -- (k) node[below left, pos=0.5]{$G_{k_2 l_1}$};
        \draw[decoration={markings, mark=at position 0.5 with {\arrow{Latex}}}, postaction={decorate}] (l2) -- (l);
        \draw[decoration={markings, mark=at position 0.5 with {\arrow{Latex}}}, postaction={decorate}] (l3) -- (l);
        \draw[decoration={markings, mark=at position 0.5 with {\arrow{Latex}}}, postaction={decorate}] (p2in) -- (p2);
    \end{tikzpicture} } \normalsize \\
&& +\>   \big(1+\e^{\,\ii\, (l_2 + l_3)\cdot \theta\, k_3 - \ii\, k_2\cdot \theta\, k_3}\big) \times { \footnotesize \begin{tikzpicture}[scale=0.5, baseline]
        \coordinate[label=right: \textcolor{red}{$k$}] (k) at (0,0);
        \coordinate (k1) at (270:2);
        \coordinate[label=left: \textcolor{dgreen}{$l$}] (l) at (45:2);
        \coordinate[label=left: {$k_2$}]  (k2) at (135:2);
        \coordinate[label=left: $l_2$] (l2) at ($(l) + (90:2)$);
        \coordinate[label=right: $l_3$] (l3) at ($(l) + (0:2)$);
        \coordinate[label=below: {$p_1$}] (p1in) at ($(k1) +(180:2)$);
        \draw[fill, red] (k) circle[radius=0.1];
        \draw[fill, dgreen] (l) circle[radius=0.1];
        
        \draw[decoration={markings, mark=at position 0.5 with {\arrow{Latex}}}, postaction={decorate}] (k2) -- (k);
        \draw[color=red,  postaction={decorate}] (k1) -- (k) node[right, pos=0.5]{$G_{k_1 p_2}$};
        \draw[decoration={markings, mark=at position 0.5 with {\arrow{Latex}}}, postaction={decorate}] (p1in) -- (p1);
        
        \draw[color=dgreen,  postaction={decorate}] (l) -- (k) node[below right, pos=0.5]{$G_{k_3 l_1}$};
        \draw[decoration={markings, mark=at position 0.5 with {\arrow{Latex}}}, postaction={decorate}] (l2) -- (l);
        \draw[decoration={markings, mark=at position 0.5 with {\arrow{Latex}}}, postaction={decorate}] (l3) -- (l);
    \end{tikzpicture} } \normalsize
+  2 \times { \footnotesize \begin{tikzpicture}[scale=0.5, baseline]
        \coordinate[label=left: \textcolor{red}{$k$}] (k) at (0,0);
        \coordinate (k1) at (270:2);
        \coordinate[label=right: {$k_3$}] (k3) at (45:2);
        \coordinate[label=left: {$k_2$}] (k2) at (135:2);
        \coordinate (l1) at ($(k1) +(0:5)$);
        \coordinate[label=right: \textcolor{dgreen}{$l$}] (l) at ($(k) +(0:5)$);
        \coordinate[label=right: {$l_3$}] (l3) at ($(l) + (45:2)$);
        \coordinate[label=left: {$l_2$}] (l2) at ($(l) + (135:2)$);
        \draw[fill, red] (k) circle[radius=0.1];
        \draw[fill, dgreen] (l) circle[radius=0.1];
        
        \draw[decoration={markings, mark=at position 0.5 with {\arrow{Latex}}}, postaction={decorate}] (k3) -- (k);
        \draw[decoration={markings, mark=at position 0.5 with {\arrow{Latex}}}, postaction={decorate}] (k2) -- (k);
        \draw[decoration={markings, mark=at position 0.5 with {\arrow{Latex}}}, postaction={decorate}] (l3) -- (l);
        \draw[decoration={markings, mark=at position 0.5 with {\arrow{Latex}}}, postaction={decorate}] (l2) -- (l);

        \draw[color=red,  postaction={decorate}] (k1) -- (k) node[left, pos=0.5]{$G_{k_1 p_1}$};
        \draw[color=dgreen,  postaction={decorate}] (l1) -- (l) node[right, pos=0.5]{$G_{l_1 p_2}$};
    \end{tikzpicture} } \normalsize \Bigg) \bigg) \ .
\end{eqnarray*}

Now we have to contract the remaining lines by applying $\ii\,\hbar\,\BVL\,\sH$ twice, which amounts to calculating the remaining four-point functions using the braided Wick theorem. The final result is
\begin{equation*}    
\tilde G_2^\star(p_1,p_2)^{\swone}_2 = 18 \times { \footnotesize \begin{tikzpicture}[scale=0.5, baseline]
        \coordinate (k) at (-0.5,0);
        \coordinate (l) at (0.5,0);
        \draw[decoration={markings, mark=at position 0.5 with {\arrow{Latex[reversed]}} }, postaction={decorate}] (k) -- ($(k) + (180:2)$) node[above]{$p_1$};
        \draw[decoration={markings, mark=at position 0.5 with {\arrow{Latex[reversed]}} }, postaction={decorate}] (l) -- ($(l) + (0:2)$) node[above]{$p_2$};
        \draw[decoration={markings}, postaction={decorate}] ($(k)+(0:0.5)$) circle (0.5);    
    \end{tikzpicture} } \normalsize
+6 \times { \footnotesize \begin{tikzpicture}[scale=0.5, baseline]
        \coordinate (k) at (0,0);
        \coordinate (l) at ($(k)+(90:1)$);
        \coordinate[label=above: $p_1$] (p1) at ($(k) +(180:2.5)$);
        \coordinate[label=above: $p_2$] (p2) at ($(k) +(0:2.5)$);

        \draw[decoration={markings, mark=at position 0.5 with {\arrow{Latex[reversed]}} }, postaction={decorate}] (k) -- (p1);
        \draw[decoration={markings, mark=at position 0.5 with {\arrow{Latex[reversed]}} }, postaction={decorate}] (k) -- (p2);
        \draw[postaction={decorate}] (k) -- (l);
        \draw[postaction={decorate}] ($(l)+(90:0.5)$) circle (0.5);
    \end{tikzpicture} } \normalsize
+ 3 \times { \footnotesize \begin{tikzpicture}[scale=0.5, baseline]
        \coordinate (k) at (-1.5,0);
        \coordinate (l) at (1.5,0);
        \coordinate[label=above: $p_1$] (p1) at (180:2.5);
        \coordinate[label=above: $p_2$] (p2) at (0:2.5);

        \draw[decoration={markings, mark=at position 0.5 with {\arrow{Latex[reversed]}} }, postaction={decorate}] (k) -- (p1);
        \draw[decoration={markings, mark=at position 0.5 with {\arrow{latex[reversed]}} }, postaction={decorate}] (l) -- (p2);
        \draw[postaction={decorate}] ($(k)+(0:0.5)$) circle (0.5);
        \draw[ postaction={decorate}] ($(l)-(0:0.5)$) circle (0.5);
    \end{tikzpicture} } \normalsize \ ,
\end{equation*}
where again all phase factors cancel.

Adding the contributions $\tilde G_2^\star(p_1,p_2)^{\swone}_1$ and $\tilde G_2^\star(p_1,p_2)^{\swone}_2$ gives exactly (\ref{KorekcijaPhi3Diagrams}).

\subsection{Three-point function at one-loop}

The three-point function is defined similarly as
\begin{equation}\label{eq:3PointPhi3}
\tilde G_3^\star(p_1,p_2,p_3) = \sum_{m=1}^\infty \, \sP\,\big((-\ii\,\hbar\,\BVL\,\sH - \{\CS _{\rm int},-\}_\star\,\sH)^m\, (\tte^{p_1}\odot_\star\tte^{p_2}\odot_\star\tte^{p_3})\big)  \ .
\end{equation}

\subsubsection{Tree-level contribution}

The tree-level contribution is
\begin{align}
\begin{split}
\tilde G_3^\star(p_1,p_2, p_3)^{\swzero} &=  -(\ii\,\hbar\,\BVL\,\sH)^2\, \{\CS _{\rm int}, \sH\, (\tte^{p_1}\odot_\star\tte^{p_2}\odot_\star\tte^{p_3})\}_\star \\
& \quad \, - \ii\,\hbar\,\BVL\,\sH\,\big\{\CS_{\rm int},\sH\,\big(\ii\,\hbar\,\BVL\,\sH\,(\tte^{p_1}\odot_\star\tte^{p_2}\odot_\star\tte^{p_3})\big)\big\}_\star \ .\nn
\end{split}
\end{align}
The interaction vertex in the first term is evaluated as in \eqref{eq:SintH3e}, with the resulting four-point functions calculated using the braided Wick theorem, while the second term is evaluated as in \eqref{H31} and \eqref{DeltaSH1}.

Altogether one gets
\begin{align}
\begin{split}
\tilde G_3^\star(p_1,p_2, p_3)^{\swzero} &= - \lambda\,(\ii\, \hbar)^2\,\e^{\,\frac{\ii}{2}\,p_3 \cdot \theta\, p_2}  
 \, \frac{(2\pi)^d\, \delta(p_1+p_2+p_3)}{(p_1^2 - m^2)\,(p_2^2 - m^2)\,(p_3^2 - m^2)} \\
& \quad \, - \frac{\lambda}{2}\,(\ii\, \hbar)^2\, \int_k\, \frac{1}{k^2 - m^2} \, \Bigg( \frac{(2\pi)^d\, \delta(p_1)}{p_1^2 - m^2}\,\frac{(2\pi)^d\, \delta(p_2+p_3)}{p_2^2 - m^2} \\
& \hspace{5cm} +  \frac{(2\pi)^d\, \delta(p_2)}{p_2^2 - m^2}\,\frac{(2\pi)^d\, \delta(p_3+p_1)}{p_3^2 - m^2} \\
& \hspace{6cm} + \frac{(2\pi)^d\, \delta(p_3)}{p_3^2 - m^2}\,\frac{(2\pi)^d\, \delta(p_1+p_2)}{p_1^2 - m^2} \Bigg) \ .\label{G3Tree}
\end{split}
\end{align}
We see that the noncommutative contribution appears as a phase factor in external momenta of connected diagrams. The diagrammatic result is given by 
\begin{equation}
\tilde G_3^\star(p_1,p_2, p_3)^{\swzero} \> = \>  \mbox{$\sum\limits_\circlearrowright$}\,\Bigg(2\, \e^{\,\frac{\ii}{2}\,p_3 \cdot \theta\, p_2} \times { \footnotesize \begin{tikzpicture}[scale=0.25, baseline]
        \draw[decoration={markings, mark=at position 0.5 with {\arrow{Latex[reversed]}}}, postaction={decorate}] (0,0) -- (45:4);
        \draw[decoration={markings, mark=at position 0.5 with {\arrow{Latex[reversed]}}}, postaction={decorate}] (0,0) -- (135:4);
        \draw[decoration={markings, mark=at position 0.5 with {\arrow{Latex[reversed]}}}, postaction={decorate}] (0,0) -- (270:4);
        \draw (45:4.5) node[anchor=west] {$p_1$};
        \draw (135:4.5) node[anchor=east] {$p_2$};
        \draw (270:4.5) node[anchor=north] {$p_3$}; 
    \end{tikzpicture} } \normalsize \>+\> 3\times { \footnotesize \begin{tikzpicture}[scale=0.25, baseline]
        \draw[decoration={markings, mark=at position 0.5 with {\arrow{Latex[reversed]}}}, postaction={decorate}] (45:2) -- (45:4);
        \draw[postaction={decorate}] (45:1.5) circle (0.5);
        
        \draw[decoration={markings, mark=at position 0.5 with {\arrow{Latex[reversed]}}}, postaction={decorate}] (0,0) -- (135:4);
        \draw[decoration={markings, mark=at position 0.5 with {\arrow{Latex[reversed]}}}, postaction={decorate}] (0,0) -- (270:4);
        \draw (45:4.5) node[anchor=west] {$p_1$};
        \draw (135:4.5) node[anchor=east] {$p_2$};
        \draw (270:4.5) node[anchor=north] {$p_3$};
    \end{tikzpicture} } \normalsize \Bigg) \label{VertTree}
\end{equation}
with specific diagrams defined in Appendix~\ref{subsub:pictograms}, where $\mbox{$\sum_\circlearrowright$}\,$ indicates the sum over cyclic permutations of the external momenta $p_1$, $p_2$ and $p_3$.

\subsubsection{One-loop corrections}

The one-loop contribution to $\tilde G_3^\star(p_1,p_2, p_3)$ is 
\begin{equation}
\tilde G_3^\star(p_1,p_2,p_3)^{\swone} =  \sP\,\big((\ii\,\hbar\,\BVL\,\sH + \{\CS _{\rm int},-\}_\star\,\sH)^6\, (\tte^{p_1}\odot_\star\tte^{p_2}\odot_\star\tte^{p_3})\big) \ .\nn 
\end{equation}
Since $\ii\,\hbar\,\BVL\,\sH$ effectively reduces the number of antifields by two and $\{\CS _{\rm int},-\}_\star\,\sH$ adds one antifield, we can easily determine which combinations of $\BVL$ and $\CS _{\rm int}$ contribute. There are seven different contributions which we will label in an intuitive fashion as
\begin{align*}
\triangle_1 &= \BVL\, \CS _{\rm int} \, \BVL \, \CS _{\rm int} \, \CS _{\rm int} \BVL \ , \\[4pt]
\triangle_2 &= \BVL \, \BVL\,  \CS _{\rm int}\,  \CS _{\rm int}\,  \CS _{\rm int}\,  \BVL \ , \\[4pt]
\triangle_3 &= \BVL\,  \CS _{\rm int}\,  \BVL\,  \CS _{\rm int}\,  \BVL\,  \CS _{\rm int} \ , \\[4pt]
\triangle_4 &= \BVL \, \BVL \, \CS _{\rm int} \, \CS _{\rm int} \, \BVL\,  \CS _{\rm int} \ , \\[4pt]
\triangle_5 &= \BVL\,  \CS _{\rm int}\,  \BVL\,  \BVL \, \CS _{\rm int}\,  \CS _{\rm int} \ , \\[4pt]
\triangle_6 &= \BVL\,  \BVL \, \CS _{\rm int}\,  \BVL\,  \CS _{\rm int}\,  \CS _{\rm int} \ , \\[4pt]
\triangle_7 &= \BVL\,  \BVL\,  \BVL\,  \CS _{\rm int}\,  \CS _{\rm int}\,  \CS _{\rm int} \ .
\end{align*}

The first two combinations contribute only as disconnected correlation functions since the first operator acting is $\BVL$, which creates one propagator factor $ G_{p_ip_j}$ separated from the rest that yield the correction of $\tilde G_1^\star(p_k)$, for $i \neq j \neq k$. For the remaining combinations, we will give initial expressions coming from homological perturbation theory and the final results represented in terms of diagrams which are decoded in Appendix~\ref{subsub:pictograms}.

The contribution $\triangle_3 = \BVL\,  \CS _{\rm int}\,  \BVL\,  \CS _{\rm int}\,  \BVL \, \CS _{\rm int}$ is calculated via
\begin{align*}
&\tilde G_3^\star(p_1,p_2,p_3)^{\swone}_3 \\[4pt]
&\quad= \ii\,\hbar\,\BVL\,\sH\,  \Bigg( \bigg\{\CS _{\rm int}, \sH\,  \bigg(\ii\,\hbar\,\BVL\,\sH\,  \Big( \big\{\CS _{\rm int}, \sH\,  \big(\ii\,\hbar\,\BVL\,\sH\,  \{\CS _{\rm int}, \sH\, (\tte^{p_1}\odot_\star\tte^{p_2}\odot_\star\tte^{p_3}) \}_\star \big) \big\}_\star \Big)  \bigg) \bigg\}_\star \Bigg) \ .
\end{align*}
The result is given by
\begin{align*}
& \hspace{-2mm}\tilde G_3^\star(p_1,p_2,p_3)^{\swone}_3 \\[4pt]
& \hspace{-2mm} = \frac{3}{3} \cdot \frac{3}{2} \cdot \frac{3}{1} \times \mbox{$\sum\limits_\circlearrowright$}\,\Bigg(
\frac{16}{3}\, \e^{\,\frac{\ii}{2}\, p_3 \cdot \theta\, p_2} \times
           { \footnotesize \begin{tikzpicture}[scale=0.25, baseline]
                \draw[ postaction={decorate}] (0,0) -- (45:2);
                \draw[decoration={markings, mark=at position 0.5 with {\arrow{Latex[reversed]}}}, postaction={decorate}] (45:2) -- (45:4);
                \draw[postaction={decorate}] (45:2) -- +(135:1);
                \draw [ postaction={decorate}] (45:2) +(135:1.5) circle (0.5);                
                \draw[decoration={markings, mark=at position 0.5 with {\arrow{Latex[reversed]}}}, postaction={decorate}] (0,0) -- (135:4);
                \draw[decoration={markings, mark=at position 0.5 with {\arrow{Latex[reversed]}}}, postaction={decorate}] (0,0) -- (270:4);
                \draw (45:4.5) node[anchor=west] {$p_1$};
                \draw (135:4.5) node[anchor=east] {$p_2$};
                \draw (270:4.5) node[anchor=north] {$p_3$}; 
            \end{tikzpicture} } \normalsize
        + \frac{2}{3}\times { \footnotesize \begin{tikzpicture}[scale=0.25, baseline]
                \draw[decoration={markings, mark=at position 0.5 with {\arrow{Latex[reversed]}}}, postaction={decorate}] (45:2) -- (45:4);
                \draw[ postaction={decorate}] (45:2) -- +(135:1);
                \draw [ postaction={decorate}] (45:2) +(135:1.5) circle (0.5);
                \draw[postaction={decorate}] (45:2) -- +(315:1);
                \draw [ postaction={decorate}] (45:2) +(315:1.5) circle (0.5);
                \draw[decoration={markings, mark=at position 0.5 with {\arrow{Latex[reversed]}}}, postaction={decorate}] (0,0) -- (135:4);
                \draw[decoration={markings, mark=at position 0.5 with {\arrow{Latex[reversed]}}}, postaction={decorate}] (0,0) -- (270:4);
                \draw (45:4.5) node[anchor=west] {$p_1$};
                \draw (135:4.5) node[anchor=east] {$p_2$};
                \draw (270:4.5) node[anchor=north] {$p_3$};
            \end{tikzpicture} } \normalsize  \\ 
        & \hspace{3.75cm} +  \frac{4}{3}\times { \footnotesize \begin{tikzpicture}[scale=0.25, baseline]
                \draw[ postaction={decorate}] (45:2) -- (45:2.5);
                \draw[decoration={markings, mark=at position 0.5 with {\arrow{Latex[reversed]}}}, postaction={decorate}] (45:3.5) -- (45:4);
                \draw[ postaction={decorate}] (45:1.5) circle (0.5);
                \draw (45:3) circle (0.5);
                \draw[decoration={markings, mark=at position 0.5 with {\arrow{Latex[reversed]}}}, postaction={decorate}] (0,0) -- (135:4);
                \draw[decoration={markings, mark=at position 0.5 with {\arrow{Latex[reversed]}}}, postaction={decorate}] (0,0) -- (270:4);
                \draw (45:4.5) node[anchor=west] {$p_1$};
                \draw (135:4.5) node[anchor=east] {$p_2$};
                \draw (270:4.5) node[anchor=north] {$p_3$}; 
            \end{tikzpicture} } \normalsize +4 \times { \footnotesize  \begin{tikzpicture}[scale=0.25, baseline]
                \draw[decoration={markings, mark=at position 0.5 with {\arrow{Latex[reversed]}}}, postaction={decorate}] (45:2) -- (45:4);
                \draw[ postaction={decorate}] (45:1.5) circle (0.5);               
                \draw[decoration={markings, mark=at position 0.5 with {\arrow{Latex[reversed]}}}, postaction={decorate}] (0,0) -- (135:4);
                \draw[decoration={markings, mark=at position 0.5 with {\arrow{Latex[reversed]}}}, postaction={decorate}] (0,0) -- (270:4);
                \draw[ postaction={decorate}] (225:0) -- (225:2);
                \draw[postaction={decorate}] (225:2.5) circle (0.5);
                \draw (45:4.5) node[anchor=west] {$p_1$};
                \draw (135:4.5) node[anchor=east] {$p_2$};
                \draw (270:4.5) node[anchor=north] {$p_3$};
            \end{tikzpicture} } \normalsize
        + \frac{2}{3}\times { \footnotesize \begin{tikzpicture}[scale=0.25, baseline]
               \draw[decoration={markings, mark=at position 0.5 with {\arrow{Latex[reversed]}}}, postaction={decorate}] (45:2) -- (45:4);
                \draw (45:1.5) circle (0.5);           
                \draw[decoration={markings, mark=at position 0.5 with {\arrow{Latex[reversed]}}}, postaction={decorate}] (135:2) -- (135:4);
                \draw[postaction={decorate}] (135:1.5) circle (0.5);            
                \draw[decoration={markings, mark=at position 0.5 with {\arrow{Latex[reversed]}}}, postaction={decorate}] (270:2) -- (270:4);
                \draw[postaction={decorate}] (270:1.5) circle (0.5);               
                \draw (45:4.5) node[anchor=west] {$p_1$};
                \draw (135:4.5) node[anchor=east] {$p_2$};
                \draw (270:4.5) node[anchor=north] {$p_3$};
            \end{tikzpicture} } \normalsize \Bigg) \ .
\end{align*}

The fourth contribution $\triangle_4=\BVL \, \BVL\,  \CS _{\rm int}\,  \CS _{\rm int} \, \BVL\,  \CS _{\rm int}$ starts as
\begin{align*}
&\tilde G_3^\star(p_1,p_2,p_3)^{\swone}_4 \\[4pt]
& \qquad =\ii\,\hbar\,\BVL\,\sH\,  \bigg( \ii\,\hbar\,\BVL\,\sH\,  \Big( \Big\{\CS _{\rm int}, \sH\,  \big\{\CS _{\rm int}, \sH\,  \big(\ii\,\hbar\,\BVL\,\sH\,  \{\CS _{\rm int}, \sH\, (\tte^{p_1}\odot_\star\tte^{p_2}\odot_\star\tte^{p_3}) \}_\star \big) \big\}_\star \Big) \Big\}_\star \bigg)  \ ,
\end{align*}
and leads to the result
\begin{align*}
&  \tilde G_3^\star(p_1,p_2,p_3)^{\swone}_4 \\[4pt]
&  = \frac{3}{3} \cdot \frac{3}{2} \cdot \frac{3}{3} \times \mbox{$\sum\limits_\circlearrowright$}\,\Bigg(
8\,\e^{\,\frac{\ii}{2}\, p_3 \cdot \theta\, p_2} \times
        {\footnotesize    \begin{tikzpicture}[scale=0.25, baseline]
                \draw[ postaction={decorate}] (0,0) -- (45:2);
                \draw[decoration={markings, mark=at position 0.5 with {\arrow{Latex[reversed]}}}, postaction={decorate}] (45:2) -- (45:4);
                \draw[postaction={decorate}] (45:2) -- +(135:1);
                \draw [postaction={decorate}] (45:2) +(135:1.5) circle (0.5);
                \draw[decoration={markings, mark=at position 0.5 with {\arrow{Latex[reversed]}}}, postaction={decorate}] (0,0) -- (135:4);
                \draw[decoration={markings, mark=at position 0.5 with {\arrow{Latex[reversed]}}}, postaction={decorate}] (0,0) -- (270:4);
                \draw (45:4.5) node[anchor=west] {$p_1$};
                \draw (135:4.5) node[anchor=east] {$p_2$};
                \draw (270:4.5) node[anchor=north] {$p_3$}; 
            \end{tikzpicture} } \normalsize
+ 24\,\e^{\,\frac{\ii}{2}\, p_3 \cdot \theta\, p_2} \times
          {\footnotesize  \begin{tikzpicture}[scale=0.25, baseline]
                \draw[postaction={decorate}] (0,0) -- (45:1.5);
                \draw[decoration={markings, mark=at position 0.5 with {\arrow{Latex[reversed]}}}, postaction={decorate}] (45:2.5) -- (45:4);
                \draw[ postaction={decorate}] (45:2) circle (0.5);
                \draw[decoration={markings, mark=at position 0.5 with {\arrow{Latex[reversed]}}}, postaction={decorate}] (0,0) -- (135:4);
                \draw[decoration={markings, mark=at position 0.5 with {\arrow{Latex[reversed]}}}, postaction={decorate}] (0,0) -- (270:4);
                \draw (45:4.5) node[anchor=west] {$p_1$};
                \draw (135:4.5) node[anchor=east] {$p_2$};
                \draw (270:4.5) node[anchor=north] {$p_3$}; 
            \end{tikzpicture} } \normalsize \\
& \hspace{4cm} +  { \footnotesize      \begin{tikzpicture}[scale=0.25, baseline]
                \draw[decoration={markings, mark=at position 0.5 with {\arrow{Latex[reversed]}}}, postaction={decorate}] (45:2) -- (45:4);
                \draw[postaction={decorate}] (45:2) -- +(135:1);
                \draw [ postaction={decorate}] (45:2) +(135:1.5) circle (0.5);
                \draw[postaction={decorate}] (45:2) -- +(315:1);
                \draw [postaction={decorate}] (45:2) +(315:1.5) circle (0.5);
                \draw[decoration={markings, mark=at position 0.5 with {\arrow{Latex[reversed]}}}, postaction={decorate}] (0,0) -- (135:4);
                \draw[decoration={markings, mark=at position 0.5 with {\arrow{Latex[reversed]}}}, postaction={decorate}] (0,0) -- (270:4);
                \draw (45:4.5) node[anchor=west] {$p_1$};
                \draw (135:4.5) node[anchor=east] {$p_2$};
                \draw (270:4.5) node[anchor=north] {$p_3$};
            \end{tikzpicture} } \normalsize
+ 2 \times     { \footnotesize   \begin{tikzpicture}[scale=0.25, baseline]
                \draw[ postaction={decorate}] (45:2) -- (45:2.5);
                \draw[decoration={markings, mark=at position 0.5 with {\arrow{Latex[reversed]}}}, postaction={decorate}] (45:3.5) -- (45:4);
                \draw[ postaction={decorate}] (45:1.5) circle (0.5);
                \draw (45:3) circle (0.5);
                \draw[decoration={markings, mark=at position 0.5 with {\arrow{Latex[reversed]}}}, postaction={decorate}] (0,0) -- (135:4);
                \draw[decoration={markings, mark=at position 0.5 with {\arrow{Latex[reversed]}}}, postaction={decorate}] (0,0) -- (270:4);
                \draw (45:4.5) node[anchor=west] {$p_1$};
                \draw (135:4.5) node[anchor=east] {$p_2$};
                \draw (270:4.5) node[anchor=north] {$p_3$}; 
            \end{tikzpicture} } \normalsize
+ 6 \times { \footnotesize       \begin{tikzpicture}[scale=0.25, baseline]
                \draw[ postaction={decorate}] (45:1.5) circle (0.5);
                \draw[ postaction={decorate}] (45:1.5) +(45:1.5) arc [start angle=45, end angle=180, radius=0.75];
                \draw  (45:1.5) +(45:1.5) arc [start angle=45, end angle=-90, radius=0.75];
                \draw[decoration={markings, mark=at position 0.5 with {\arrow{Latex[reversed]}}}, postaction={decorate}] (45:3) -- (45:4);                    
                \draw[decoration={markings, mark=at position 0.5 with {\arrow{Latex[reversed]}}}, postaction={decorate}] (0,0) -- (135:4);
                \draw[decoration={markings, mark=at position 0.5 with {\arrow{Latex[reversed]}}}, postaction={decorate}] (0,0) -- (270:4);
                \draw (45:4.5) node[anchor=west] {$p_1$};
                \draw (135:4.5) node[anchor=east] {$p_2$};
                \draw (270:4.5) node[anchor=north] {$p_3$}; 
            \end{tikzpicture} } \normalsize \\
& \hspace{4cm}+ 6 \times { \footnotesize       \begin{tikzpicture}[scale=0.25, baseline]
                \draw[decoration={markings, mark=at position 0.5 with {\arrow{Latex[reversed]}}}, postaction={decorate}] (45:2) -- (45:4);
                \draw[ postaction={decorate}] (45:1.5) circle (0.5);
        \draw (0,0) circle (0.5);
                \draw[decoration={markings, mark=at position 0.5 with {\arrow{Latex[reversed]}}}, postaction={decorate}] (0,0) +(135:0.5) -- (135:4);
                \draw[decoration={markings, mark=at position 0.5 with {\arrow{Latex[reversed]}}}, postaction={decorate}] (0,0) +(270:0.5) -- (270:4);
                \draw (45:4.5) node[anchor=west] {$p_1$};
                \draw (135:4.5) node[anchor=east] {$p_2$};
                \draw (270:4.5) node[anchor=north] {$p_3$};
            \end{tikzpicture} } \normalsize
+ 6 \times { \footnotesize  \begin{tikzpicture}[scale=0.25, baseline]
                \draw[decoration={markings, mark=at position 0.5 with {\arrow{Latex[reversed]}}}, postaction={decorate}] (45:2) -- (45:4);
                \draw[ postaction={decorate}] (45:1.5) circle (0.5);
                \draw[decoration={markings, mark=at position 0.5 with {\arrow{Latex[reversed]}}}, postaction={decorate}] (0,0) -- (135:4);
                \draw[decoration={markings, mark=at position 0.5 with {\arrow{Latex[reversed]}}}, postaction={decorate}] (0,0) -- (270:4);
                \draw[ postaction={decorate}] (225:0) -- (225:2);
                \draw[postaction={decorate}] (225:2.5) circle (0.5);
                \draw (45:4.5) node[anchor=west] {$p_1$};
                \draw (135:4.5) node[anchor=east] {$p_2$};
                \draw (270:4.5) node[anchor=north] {$p_3$};
            \end{tikzpicture} } \normalsize
+     { \footnotesize     \begin{tikzpicture}[scale=0.25, baseline]
                \draw[decoration={markings, mark=at position 0.5 with {\arrow{Latex[reversed]}}}, postaction={decorate}] (45:2) -- (45:4);
                \draw (45:1.5) circle (0.5);
            \draw[decoration={markings, mark=at position 0.5 with {\arrow{Latex[reversed]}}}, postaction={decorate}] (135:2) -- (135:4);
                \draw[ postaction={decorate}] (135:1.5) circle (0.5);
            \draw[decoration={markings, mark=at position 0.5 with {\arrow{Latex[reversed]}}}, postaction={decorate}] (270:2) -- (270:4);
                \draw[postaction={decorate}] (270:1.5) circle (0.5);
                \draw (45:4.5) node[anchor=west] {$p_1$};
                \draw (135:4.5) node[anchor=east] {$p_2$};
                \draw (270:4.5) node[anchor=north] {$p_3$};
            \end{tikzpicture} } \normalsize \Bigg) \ .
\end{align*}

The fifth contribution $\triangle_5= \BVL\, \CS _{\rm int}\,  \BVL\,  \BVL\,  \CS _{\rm int} \,  \CS _{\rm int}$ is
\begin{align*}
&\tilde G_3^\star(p_1,p_2,p_3)^{\swone}_5 \\[4pt]
& \quad = \ii\,\hbar\,\BVL\,\sH\,  \Bigg( \bigg\{\CS _{\rm int}, \sH\,  \bigg( \ii\,\hbar\,\BVL\,\sH\,  \Big(\ii\,\hbar\,\BVL\,\sH\,  \big( \big\{\CS _{\rm int}, \sH\,   \{\CS _{\rm int}, \sH\, (\tte^{p_1}\odot_\star\tte^{p_2}\odot_\star\tte^{p_3}) \}_\star \big\}_\star \big) \Big) \bigg) \bigg\}_\star \Bigg) \ ,
\end{align*}
and it results in
\begin{align*}
&\tilde G_3^\star(p_1,p_2,p_3)^{\swone}_5  = \frac{3}{3} \cdot \frac{3}{4} \cdot \frac{3}{1} \times \mbox{$\sum\limits_\circlearrowright$}\,\Bigg(
\frac{224}{15}\,\e^{\,\frac{\ii}{2}\, p_3 \cdot \theta\, p_2}\times { \footnotesize
                        \begin{tikzpicture}[scale=0.25, baseline]
        \draw[postaction={decorate}] (0,0) -- (45:2);
        \draw[decoration={markings, mark=at position 0.5 with {\arrow{Latex[reversed]}}}, postaction={decorate}] (45:2) -- (45:4);
        \draw[ postaction={decorate}] (45:2) -- +(135:1);
        \draw [postaction={decorate}] (45:2) +(135:1.5) circle (0.5);
        \draw[decoration={markings, mark=at position 0.5 with {\arrow{Latex[reversed]}}}, postaction={decorate}] (0,0) -- (135:4);
        \draw[decoration={markings, mark=at position 0.5 with {\arrow{Latex[reversed]}}}, postaction={decorate}] (0,0) -- (270:4);
        \draw (45:4.5) node[anchor=west] {$p_1$};
        \draw (135:4.5) node[anchor=east] {$p_2$};
        \draw (270:4.5) node[anchor=north] {$p_3$}; 
    \end{tikzpicture} } \normalsize
        + \frac{16}{15}  \times { \footnotesize   \begin{tikzpicture}[scale=0.25, baseline]
            \draw[decoration={markings, mark=at position 0.5 with {\arrow{Latex[reversed]}}}, postaction={decorate}] (45:2) -- (45:4);
            \draw[postaction={decorate}] (45:2) -- +(135:1);
            \draw [ postaction={decorate}] (45:2) +(135:1.5) circle (0.5);
            \draw[ postaction={decorate}] (45:2) -- +(315:1);
            \draw [postaction={decorate}] (45:2) +(315:1.5) circle (0.5);
            \draw[decoration={markings, mark=at position 0.5 with {\arrow{Latex[reversed]}}}, postaction={decorate}] (0,0) -- (135:4);
            \draw[decoration={markings, mark=at position 0.5 with {\arrow{Latex[reversed]}}}, postaction={decorate}] (0,0) -- (270:4);
            \draw (45:4.5) node[anchor=west] {$p_1$};
            \draw (135:4.5) node[anchor=east] {$p_2$};
            \draw (270:4.5) node[anchor=north] {$p_3$};
        \end{tikzpicture} } \normalsize \\
&\hspace{7cm} + \frac{32}{15}  \times { \footnotesize \begin{tikzpicture}[scale=0.25, baseline]
            \draw[ postaction={decorate}] (45:2) -- (45:2.5);
            \draw[decoration={markings, mark=at position 0.5 with {\arrow{Latex[reversed]}}}, postaction={decorate}] (45:3.5) -- (45:4);
            \draw[ postaction={decorate}] (45:1.5) circle (0.5);
            \draw (45:3) circle (0.5);
            \draw[decoration={markings, mark=at position 0.5 with {\arrow{Latex[reversed]}}}, postaction={decorate}] (0,0) -- (135:4);
            \draw[decoration={markings, mark=at position 0.5 with {\arrow{Latex[reversed]}}}, postaction={decorate}] (0,0) -- (270:4);
            \draw (45:4.5) node[anchor=west] {$p_1$};
            \draw (135:4.5) node[anchor=east] {$p_2$};
            \draw (270:4.5) node[anchor=north] {$p_3$}; 
        \end{tikzpicture} } \normalsize
+ \frac{32}{5} \times { \footnotesize   \begin{tikzpicture}[scale=0.25, baseline]
            \node (b) at (0,0){};
            \draw[decoration={markings, mark=at position 0.5 with {\arrow{Latex[reversed]}}}, postaction={decorate}] (45:2) -- (45:4);
            \draw[ postaction={decorate}] (45:1.5) circle (0.5);
    \draw (0,0) circle (0.5);
    \draw[decoration={markings, mark=at position 0.5 with {\arrow{Latex[reversed]}}}, postaction={decorate}] (0,0) +(135:0.5) -- (135:4);
            \draw[decoration={markings, mark=at position 0.5 with {\arrow{Latex[reversed]}}}, postaction={decorate}] (0,0) +(270:0.5) -- (270:4);
            \draw (45:4.5) node[anchor=west] {$p_1$};
            \draw (135:4.5) node[anchor=east] {$p_2$};
            \draw (270:4.5) node[anchor=north] {$p_3$};
        \end{tikzpicture} } \normalsize \\
& \hspace{7cm} + \frac{32}{5} \times { \footnotesize   \begin{tikzpicture}[scale=0.25, baseline]
            \draw[decoration={markings, mark=at position 0.5 with {\arrow{Latex[reversed]}}}, postaction={decorate}] (45:2) -- (45:4);
            \draw[ postaction={decorate}] (45:1.5) circle (0.5);
            \draw[decoration={markings, mark=at position 0.5 with {\arrow{Latex[reversed]}}}, postaction={decorate}] (0,0) -- (135:4);
            \draw[decoration={markings, mark=at position 0.5 with {\arrow{Latex[reversed]}}}, postaction={decorate}] (0,0) -- (270:4);
            \draw[postaction={decorate}] (225:0) -- (225:2);
            \draw[postaction={decorate}] (225:2.5) circle (0.5);
            \draw (45:4.5) node[anchor=west] {$p_1$};
            \draw (135:4.5) node[anchor=east] {$p_2$};
            \draw (270:4.5) node[anchor=north] {$p_3$};
        \end{tikzpicture} } \normalsize
+ \frac{16}{15}  \times { \footnotesize   \begin{tikzpicture}[scale=0.25, baseline]
            \draw[decoration={markings, mark=at position 0.5 with {\arrow{Latex[reversed]}}}, postaction={decorate}] (45:2) -- (45:4);
            \draw (45:1.5) circle (0.5);
        \draw[decoration={markings, mark=at position 0.5 with {\arrow{Latex[reversed]}}}, postaction={decorate}] (135:2) -- (135:4);
            \draw[postaction={decorate}] (135:1.5) circle (0.5);
        \draw[decoration={markings, mark=at position 0.5 with {\arrow{Latex[reversed]}}}, postaction={decorate}] (270:2) -- (270:4);
            \draw[postaction={decorate}] (270:1.5) circle (0.5);
            \draw (45:4.5) node[anchor=west] {$p_1$};
            \draw (135:4.5) node[anchor=east] {$p_2$};
            \draw (270:4.5) node[anchor=north] {$p_3$};
        \end{tikzpicture} } \normalsize  \Bigg) \ .
\end{align*}

The sixth contribution $\triangle_6 = \BVL\, \BVL\, \CS _{\rm int}\, \BVL\, \CS _{\rm int} \, \CS _{\rm int}$ is given by
\begin{align*}
&\tilde G_3^\star(p_1,p_2,p_3)^{\swone}_6 \\[4pt]
& \qquad = \ii\,\hbar\,\BVL\,\sH\, \bigg( \ii\,\hbar\,\BVL\,\sH\, \Big( \Big\{\CS _{\rm int}, \sH\, \big(  \ii\,\hbar\,\BVL\,\sH\, \big\{\CS _{\rm int}, \sH\, \{\CS _{\rm int}, \sH\, (\tte^{p_1}\odot_\star\tte^{p_2}\odot_\star\tte^{p_3}) \}_\star \big\}_\star \big) \Big\}_\star \Big) \bigg) \ .
\end{align*}
It results in the diagrams
\begin{align*}
&\tilde G_3^\star(p_1,p_2,p_3)^{\swone}_6 \\[4pt]
&\quad = \frac{3}{3} \cdot \frac{3}{4} \cdot \frac{3}{3} \times \mbox{$\sum\limits_\circlearrowright$}\,\Bigg(
\frac{192}{5}\,\e^{\,\frac{\ii}{2}\, p_3 \cdot \theta\, p_2}\times { \footnotesize \begin{tikzpicture}[scale=0.25, baseline]
        \draw[postaction={decorate}] (0,0) circle (1);
        \draw[decoration={markings, mark=at position 0.5 with {\arrow{Latex[reversed]}}}, postaction={decorate}] (45:1) -- (45:4);
        \draw[decoration={markings, mark=at position 0.5 with {\arrow{Latex[reversed]}}}, postaction={decorate}] (135:1) -- (135:4);
        \draw[decoration={markings, mark=at position 0.5 with {\arrow{Latex[reversed]}}}, postaction={decorate}] (270:1) -- (270:4);
        \draw (45:4.5) node[anchor=west] {$p_1$};
        \draw (135:4.5) node[anchor=east] {$p_2$};
        \draw (270:4.5) node[anchor=north] {$p_3$}; 
    \end{tikzpicture} } \normalsize 
+    \frac{136}{5}\,\e^{\,\frac{\ii}{2}\, p_3 \cdot \theta\, p_2}   \times { \footnotesize     \begin{tikzpicture}[scale=0.25, baseline]
        \draw[postaction={decorate}] (0,0) -- (45:2);
        \draw[decoration={markings, mark=at position 0.5 with {\arrow{Latex[reversed]}}}, postaction={decorate}] (45:2) -- (45:4);
        \draw[ postaction={decorate}] (45:2) -- +(135:1);
        \draw [postaction={decorate}] (45:2) +(135:1.5) circle (0.5);
        \draw[decoration={markings, mark=at position 0.5 with {\arrow{Latex[reversed]}}}, postaction={decorate}] (0,0) -- (135:4);
        \draw[decoration={markings, mark=at position 0.5 with {\arrow{Latex[reversed]}}}, postaction={decorate}] (0,0) -- (270:4);
        \draw (45:4.5) node[anchor=west] {$p_1$};
        \draw (135:4.5) node[anchor=east] {$p_2$};
        \draw (270:4.5) node[anchor=north] {$p_3$}; 
    \end{tikzpicture}     } \normalsize
    \\
& \hspace{3.5cm}
+\frac{192}{5}\,\e^{\,\frac{\ii}{2}\, p_3 \cdot \theta\, p_2}  \times {\footnotesize  \begin{tikzpicture}[scale=0.25, baseline]
        \draw[ postaction={decorate}] (0,0) -- (45:1.5);
        \draw[decoration={markings, mark=at position 0.5 with {\arrow{Latex[reversed]}}}, postaction={decorate}] (45:2.5) -- (45:4);
        \draw[postaction={decorate}] (45:2) circle (0.5);
        \draw[decoration={markings, mark=at position 0.5 with {\arrow{Latex[reversed]}}}, postaction={decorate}] (0,0) -- (135:4);
        \draw[decoration={markings, mark=at position 0.5 with {\arrow{Latex[reversed]}}}, postaction={decorate}] (0,0) -- (270:4);
        \draw (45:4.5) node[anchor=west] {$p_1$};
        \draw (135:4.5) node[anchor=east] {$p_2$};
        \draw (270:4.5) node[anchor=north] {$p_3$}; 
    \end{tikzpicture} } \normalsize
+ \frac{8}{5}  \times { \footnotesize   \begin{tikzpicture}[scale=0.25, baseline]
        \draw[decoration={markings, mark=at position 0.5 with {\arrow{Latex[reversed]}}}, postaction={decorate}] (45:2) -- (45:4);
        \draw[ postaction={decorate}] (45:2) -- +(135:1);
        \draw [ postaction={decorate}] (45:2) +(135:1.5) circle (0.5);
        \draw[postaction={decorate}] (45:2) -- +(315:1);
        \draw [postaction={decorate}] (45:2) +(315:1.5) circle (0.5);
        \draw[decoration={markings, mark=at position 0.5 with {\arrow{Latex[reversed]}}}, postaction={decorate}] (0,0) -- (135:4);
        \draw[decoration={markings, mark=at position 0.5 with {\arrow{Latex[reversed]}}}, postaction={decorate}] (0,0) -- (270:4);
        \draw (45:4.5) node[anchor=west] {$p_1$};
        \draw (135:4.5) node[anchor=east] {$p_2$};
        \draw (270:4.5) node[anchor=north] {$p_3$};
    \end{tikzpicture} } \normalsize \\
& \hspace{3.5cm} + \frac{16}{5} \times { \footnotesize  \begin{tikzpicture}[scale=0.25, baseline]
        \draw[postaction={decorate}] (45:2) -- (45:2.5);
        \draw[decoration={markings, mark=at position 0.5 with {\arrow{Latex[reversed]}}}, postaction={decorate}] (45:3.5) -- (45:4);
        \draw[postaction={decorate}] (45:1.5) circle (0.5);
        \draw (45:3) circle (0.5);
        \draw[decoration={markings, mark=at position 0.5 with {\arrow{Latex[reversed]}}}, postaction={decorate}] (0,0) -- (135:4);
        \draw[decoration={markings, mark=at position 0.5 with {\arrow{Latex[reversed]}}}, postaction={decorate}] (0,0) -- (270:4);
        \draw (45:4.5) node[anchor=west] {$p_1$};
        \draw (135:4.5) node[anchor=east] {$p_2$};
        \draw (270:4.5) node[anchor=north] {$p_3$}; 
    \end{tikzpicture} } \normalsize + \frac{24}{5} \times { \footnotesize   \begin{tikzpicture}[scale=0.25, baseline]
        \draw[postaction={decorate}] (45:1.5) circle (0.5);
        \draw[postaction={decorate}] (45:1.5) +(45:1.5) arc [start angle=45, end angle=180, radius=0.75];
        \draw  (45:1.5) +(45:1.5) arc [start angle=45, end angle=-90, radius=0.75];
        \draw[decoration={markings, mark=at position 0.5 with {\arrow{Latex[reversed]}}}, postaction={decorate}] (45:3) -- (45:4);    
        \draw[decoration={markings, mark=at position 0.5 with {\arrow{Latex[reversed]}}}, postaction={decorate}] (0,0) -- (135:4);
        \draw[decoration={markings, mark=at position 0.5 with {\arrow{Latex[reversed]}}}, postaction={decorate}] (0,0) -- (270:4);
        \draw (45:4.5) node[anchor=west] {$p_1$};
        \draw (135:4.5) node[anchor=east] {$p_2$};
        \draw (270:4.5) node[anchor=north] {$p_3$}; 
    \end{tikzpicture} } \normalsize 
    +\frac{96}{5} \times { \footnotesize \begin{tikzpicture}[scale=0.25, baseline]
        \node (b) at (0,0){};
        \draw[decoration={markings, mark=at position 0.5 with {\arrow{Latex[reversed]}}}, postaction={decorate}] (45:2) -- (45:4);
        \draw[postaction={decorate}] (45:1.5) circle (0.5);
\draw (0,0) circle (0.5);
\draw[decoration={markings, mark=at position 0.5 with {\arrow{Latex[reversed]}}}, postaction={decorate}] (0,0) +(135:0.5) -- (135:4);
        \draw[decoration={markings, mark=at position 0.5 with {\arrow{Latex[reversed]}}}, postaction={decorate}] (0,0) +(270:0.5) -- (270:4);
        \draw (45:4.5) node[anchor=west] {$p_1$};
        \draw (135:4.5) node[anchor=east] {$p_2$};
        \draw (270:4.5) node[anchor=north] {$p_3$};  \end{tikzpicture} } \normalsize \\
& \hspace{3.5cm} + \frac{48}{5} \times { \footnotesize  \begin{tikzpicture}[scale=0.25, baseline]
        \draw[decoration={markings, mark=at position 0.5 with {\arrow{Latex[reversed]}}}, postaction={decorate}] (45:2) -- (45:4);
        \draw[ postaction={decorate}] (45:1.5) circle (0.5);
        \draw[decoration={markings, mark=at position 0.5 with {\arrow{Latex[reversed]}}}, postaction={decorate}] (0,0) -- (135:4);
        \draw[decoration={markings, mark=at position 0.5 with {\arrow{Latex[reversed]}}}, postaction={decorate}] (0,0) -- (270:4);
        \draw[postaction={decorate}] (225:0) -- (225:2);
        \draw[ postaction={decorate}] (225:2.5) circle (0.5);
        \draw (45:4.5) node[anchor=west] {$p_1$};
        \draw (135:4.5) node[anchor=east] {$p_2$};
        \draw (270:4.5) node[anchor=north] {$p_3$};
    \end{tikzpicture} } \normalsize
    + \frac{8}{5}  \times { \footnotesize  \begin{tikzpicture}[scale=0.25, baseline]
        \draw[decoration={markings, mark=at position 0.5 with {\arrow{Latex[reversed]}}}, postaction={decorate}] (45:2) -- (45:4);
        \draw (45:1.5) circle (0.5);
    \draw[decoration={markings, mark=at position 0.5 with {\arrow{Latex[reversed]}}}, postaction={decorate}] (135:2) -- (135:4);
        \draw[ postaction={decorate}] (135:1.5) circle (0.5);
    \draw[decoration={markings, mark=at position 0.5 with {\arrow{Latex[reversed]}}}, postaction={decorate}] (270:2) -- (270:4);
        \draw[ postaction={decorate}] (270:1.5) circle (0.5);
        \draw (45:4.5) node[anchor=west] {$p_1$};
        \draw (135:4.5) node[anchor=east] {$p_2$};
        \draw (270:4.5) node[anchor=north] {$p_3$};
    \end{tikzpicture} } \normalsize \Bigg) \ .
\end{align*}

Finally, the seventh contribution $\triangle_7=\BVL\, \BVL\,  \BVL \, \CS _{\rm int}\,  \CS _{\rm int}\,  \CS _{\rm int}$ is given by
\begin{align*}
&\tilde G_3^\star(p_1,p_2,p_3)^{\swone}_7  \\[4pt]
& \quad = \ii\,\hbar\,\BVL\,\sH\,  \Bigg( \ii\,\hbar\,\BVL\,\sH\,  \bigg( \ii\,\hbar\,\BVL\,\sH\,  \Big( \Big\{\CS _{\rm int}, \sH\,  \big\{\CS _{\rm int}, \sH\,  \{\CS _{\rm int}, \sH\,  (\tte^{p_1}\odot_\star\tte^{p_2}\odot_\star\tte^{p_3}) \}_\star \big\}_\star \Big\}_\star \Big) \bigg) \Bigg) \ .
\end{align*}
The result is given by
\begin{align*}
& \tilde G_3^\star(p_1,p_2,p_3)^{\swone}_7 \\[4pt]
& \quad = \frac{3 \cdot 3 \cdot 3 }{3 \cdot 4 \cdot 5} \times \mbox{$\sum\limits_\circlearrowright$}\, \Bigg( 96 \, \e^{\,\frac{\ii}{2}\,p_3 \cdot \theta\, p_2} \times { \footnotesize \begin{tikzpicture}[scale=0.25, baseline]
        \draw[ postaction={decorate}] (0,0) circle (1);
        \draw[decoration={markings, mark=at position 0.5 with {\arrow{Latex[reversed]}}}, postaction={decorate}] (45:1) -- (45:4);
        \draw[decoration={markings, mark=at position 0.5 with {\arrow{Latex[reversed]}}}, postaction={decorate}] (135:1) -- (135:4);
        \draw[decoration={markings, mark=at position 0.5 with {\arrow{Latex[reversed]}}}, postaction={decorate}] (270:1) -- (270:4);
        \draw (45:4.5) node[anchor=west] {$p_1$};
        \draw (135:4.5) node[anchor=east] {$p_2$};
        \draw (270:4.5) node[anchor=north] {$p_3$}; 
    \end{tikzpicture} } \normalsize
    + 40 \, \e^{\,\frac{\ii}{2}\,p_3 \cdot \theta\, p_2} \times { \footnotesize\begin{tikzpicture}[scale=0.25, baseline]
        \draw[postaction={decorate}] (0,0) -- (45:2);
        \draw[decoration={markings, mark=at position 0.5 with {\arrow{Latex[reversed]}}}, postaction={decorate}] (45:2) -- (45:4);
        \draw[postaction={decorate}] (45:2) -- +(135:1);
        \draw [postaction={decorate}] (45:2) +(135:1.5) circle (0.5);
        \draw[decoration={markings, mark=at position 0.5 with {\arrow{Latex[reversed]}}}, postaction={decorate}] (0,0) -- (135:4);
        \draw[decoration={markings, mark=at position 0.5 with {\arrow{Latex[reversed]}}}, postaction={decorate}] (0,0) -- (270:4);
        \draw (45:4.5) node[anchor=west] {$p_1$};
        \draw (135:4.5) node[anchor=east] {$p_2$};
        \draw (270:4.5) node[anchor=north] {$p_3$}; 
    \end{tikzpicture} } \normalsize \\
& \hspace{3.5cm} + 96 \,\e^{\,\frac{\ii}{2}\,p_3 \cdot \theta\, p_2}\times { \footnotesize \begin{tikzpicture}[scale=0.25, baseline]
        \draw[ postaction={decorate}] (0,0) -- (45:1.5);
        \draw[decoration={markings, mark=at position 0.5 with {\arrow{Latex[reversed]}}}, postaction={decorate}] (45:2.5) -- (45:4);
        \draw[  postaction={decorate}] (45:2) circle (0.5);
        \draw[decoration={markings, mark=at position 0.5 with {\arrow{Latex[reversed]}}}, postaction={decorate}] (0,0) -- (135:4);
        \draw[decoration={markings, mark=at position 0.5 with {\arrow{Latex[reversed]}}}, postaction={decorate}] (0,0) -- (270:4);
        \draw (45:4.5) node[anchor=west] {$p_1$};
        \draw (135:4.5) node[anchor=east] {$p_2$};
        \draw (270:4.5) node[anchor=north] {$p_3$}; 
    \end{tikzpicture} } \normalsize + 2 \times { \footnotesize \begin{tikzpicture}[scale=0.25, baseline]
        \draw[decoration={markings, mark=at position 0.5 with {\arrow{Latex[reversed]}}}, postaction={decorate}] (45:2) -- (45:4);
        \draw[postaction={decorate}] (45:2) -- +(135:1);
        \draw [ postaction={decorate}] (45:2) +(135:1.5) circle (0.5);
        \draw[postaction={decorate}] (45:2) -- +(315:1);
        \draw [postaction={decorate}] (45:2) +(315:1.5) circle (0.5);
        \draw[decoration={markings, mark=at position 0.5 with {\arrow{Latex[reversed]}}}, postaction={decorate}] (0,0) -- (135:4);
        \draw[decoration={markings, mark=at position 0.5 with {\arrow{Latex[reversed]}}}, postaction={decorate}] (0,0) -- (270:4);
        \draw (45:4.5) node[anchor=west] {$p_1$};
        \draw (135:4.5) node[anchor=east] {$p_2$};
        \draw (270:4.5) node[anchor=north] {$p_3$};
    \end{tikzpicture} } \normalsize + { \footnotesize
    4 \times \begin{tikzpicture}[scale=0.25, baseline]
        \draw[postaction={decorate}] (45:2) -- (45:2.5);
        \draw[decoration={markings, mark=at position 0.5 with {\arrow{Latex[reversed]}}}, postaction={decorate}] (45:3.5) -- (45:4);
        \draw[postaction={decorate}] (45:1.5) circle (0.5);
        \draw (45:3) circle (0.5);
        \draw[decoration={markings, mark=at position 0.5 with {\arrow{Latex[reversed]}}}, postaction={decorate}] (0,0) -- (135:4);
        \draw[decoration={markings, mark=at position 0.5 with {\arrow{Latex[reversed]}}}, postaction={decorate}] (0,0) -- (270:4);
        \draw (45:4.5) node[anchor=west] {$p_1$};
        \draw (135:4.5) node[anchor=east] {$p_2$};
        \draw (270:4.5) node[anchor=north] {$p_3$}; 
    \end{tikzpicture} } \normalsize \\
    & \hspace{3.5cm}
    + 12 \times { \footnotesize \begin{tikzpicture}[scale=0.25, baseline]
        \draw[ postaction={decorate}] (45:1.5) circle (0.5);
        \draw[postaction={decorate}] (45:1.5) +(45:1.5) arc [start angle=45, end angle=180, radius=0.75];
        \draw  (45:1.5) +(45:1.5) arc [start angle=45, end angle=-90, radius=0.75];
        \draw[decoration={markings, mark=at position 0.5 with {\arrow{Latex[reversed]}}}, postaction={decorate}] (45:3) -- (45:4);    
        \draw[decoration={markings, mark=at position 0.5 with {\arrow{Latex[reversed]}}}, postaction={decorate}] (0,0) -- (135:4);
        \draw[decoration={markings, mark=at position 0.5 with {\arrow{Latex[reversed]}}}, postaction={decorate}] (0,0) -- (270:4);
        \draw (45:4.5) node[anchor=west] {$p_1$};
        \draw (135:4.5) node[anchor=east] {$p_2$};
        \draw (270:4.5) node[anchor=north] {$p_3$}; 
    \end{tikzpicture} } \normalsize
    + 36\times { \footnotesize \begin{tikzpicture}[scale=0.25, baseline]
        \node (b) at (0,0){};
        \draw[decoration={markings, mark=at position 0.5 with {\arrow{Latex[reversed]}}}, postaction={decorate}] (45:2) -- (45:4);
        \draw[ postaction={decorate}] (45:1.5) circle (0.5);
\draw (0,0) circle (0.5);
\draw[decoration={markings, mark=at position 0.5 with {\arrow{Latex[reversed]}}}, postaction={decorate}] (0,0) +(135:0.5) -- (135:4);
        \draw[decoration={markings, mark=at position 0.5 with {\arrow{Latex[reversed]}}}, postaction={decorate}] (0,0) +(270:0.5) -- (270:4);
        \draw (45:4.5) node[anchor=west] {$p_1$};
        \draw (135:4.5) node[anchor=east] {$p_2$};
        \draw (270:4.5) node[anchor=north] {$p_3$};    
    \end{tikzpicture}   } \normalsize
    + 12 \times { \footnotesize \begin{tikzpicture}[scale=0.25, baseline]
        \draw[decoration={markings, mark=at position 0.5 with {\arrow{Latex[reversed]}}}, postaction={decorate}] (45:2) -- (45:4);
        \draw[postaction={decorate}] (45:1.5) circle (0.5);
        \draw[decoration={markings, mark=at position 0.5 with {\arrow{Latex[reversed]}}}, postaction={decorate}] (0,0) -- (135:4);
        \draw[decoration={markings, mark=at position 0.5 with {\arrow{Latex[reversed]}}}, postaction={decorate}] (0,0) -- (270:4);
        \draw[ postaction={decorate}] (225:0) -- (225:2);
        \draw[ postaction={decorate}] (225:2.5) circle (0.5);
        \draw (45:4.5) node[anchor=west] {$p_1$};
        \draw (135:4.5) node[anchor=east] {$p_2$};
        \draw (270:4.5) node[anchor=north] {$p_3$};
    \end{tikzpicture} } \normalsize \\
    & \hspace{3.5cm}
    + 2 \times { \footnotesize \begin{tikzpicture}[scale=0.25, baseline]
        \draw[decoration={markings, mark=at position 0.5 with {\arrow{Latex[reversed]}}}, postaction={decorate}] (45:2) -- (45:4);
        \draw (45:1.5) circle (0.5);
    \draw[decoration={markings, mark=at position 0.5 with {\arrow{Latex[reversed]}}}, postaction={decorate}] (135:2) -- (135:4);
        \draw[postaction={decorate}] (135:1.5) circle (0.5);
    \draw[decoration={markings, mark=at position 0.5 with {\arrow{Latex[reversed]}}}, postaction={decorate}] (270:2) -- (270:4);
        \draw[ postaction={decorate}] (270:1.5) circle (0.5);
        \draw (45:4.5) node[anchor=west] {$p_1$};
        \draw (135:4.5) node[anchor=east] {$p_2$};
        \draw (270:4.5) node[anchor=north] {$p_3$};
    \end{tikzpicture}  } \normalsize \Bigg) \ .
\end{align*}

Now that all calculations are completed, we can summarize our results. If we are interested in vertex corrections, we then look only at connected diagrams. We have already excluded the first two terms $\triangle_1$ and $\triangle_2$ whose contributions are solely disconnected. We will exclude all the other disconnected diagrams as well; in any case, we will discuss how to remove the tadpole contributions in Section~\ref{sub:moretadpole}. Finally the vertex corrections at one-loop level are in total given by
\begin{align*}
& \tilde G_3^\star(p_1,p_2,p_3)^{\swone}_{{\rm conn}} \\[4pt]
& \hspace{1cm} = \e^{\,\frac\ii2\,p_3\cdot\theta\,p_2} \  \mbox{$\sum\limits_\circlearrowright$}\,\Bigg(72   \times {\footnotesize \begin{tikzpicture}[scale=0.25, baseline]
        \draw[ postaction={decorate}] (0,0) circle (1);
        \draw[decoration={markings, mark=at position 0.5 with {\arrow{Latex[reversed]}}}, postaction={decorate}] (45:1) -- (45:4);
        \draw[decoration={markings, mark=at position 0.5 with {\arrow{Latex[reversed]}}}, postaction={decorate}] (135:1) -- (135:4);
        \draw[decoration={markings, mark=at position 0.5 with {\arrow{Latex[reversed]}}}, postaction={decorate}] (270:1) -- (270:4);
        \draw (45:4.5) node[anchor=west] {$p_1$};
        \draw (135:4.5) node[anchor=east] {$p_2$};
        \draw (270:4.5) node[anchor=north] {$p_3$}; 
    \end{tikzpicture} } \normalsize
+ 108 \times { \footnotesize \begin{tikzpicture}[scale=0.25, baseline]
        \draw[postaction={decorate}] (0,0) -- (45:2);
        \draw[decoration={markings, mark=at position 0.5 with {\arrow{Latex[reversed]}}}, postaction={decorate}] (45:2) -- (45:4);
        \draw[postaction={decorate}] (45:2) -- +(135:1);
        \draw [ postaction={decorate}] (45:2) +(135:1.5) circle (0.5);
        \draw[decoration={markings, mark=at position 0.5 with {\arrow{Latex[reversed]}}}, postaction={decorate}] (0,0) -- (135:4);
        \draw[decoration={markings, mark=at position 0.5 with {\arrow{Latex[reversed]}}}, postaction={decorate}] (0,0) -- (270:4);
        \draw (45:4.5) node[anchor=west] {$p_1$};
        \draw (135:4.5) node[anchor=east] {$p_2$};
        \draw (270:4.5) node[anchor=north] {$p_3$}; 
    \end{tikzpicture} } \normalsize
+ 108 \times { \footnotesize \begin{tikzpicture}[scale=0.25, baseline]
        \draw[postaction={decorate}] (0,0) -- (45:1.5);
        \draw[decoration={markings, mark=at position 0.5 with {\arrow{Latex[reversed]}}}, postaction={decorate}] (45:2.5) -- (45:4);
        \draw[ postaction={decorate}] (45:2) circle (0.5);
        \draw[decoration={markings, mark=at position 0.5 with {\arrow{Latex[reversed]}}}, postaction={decorate}] (0,0) -- (135:4);
        \draw[decoration={markings, mark=at position 0.5 with {\arrow{Latex[reversed]}}}, postaction={decorate}] (0,0) -- (270:4);
        \draw (45:4.5) node[anchor=west] {$p_1$};
        \draw (135:4.5) node[anchor=east] {$p_2$};
        \draw (270:4.5) node[anchor=north] {$p_3$}; 
    \end{tikzpicture}  \Bigg) } \normalsize \ .
\end{align*}
Note that the last two terms come with the same overall symmetry factor, as they should to correspond to the one-loop corrections to the connected two-point function  in \eqref{KorekcijaPhi3Diagrams}.

\subsection{Eliminating the tadpole diagrams with curvature}
\label{sub:moretadpole}

In the commutative theory, the issue of a non-trivial tadpole diagram is resolved by introducing a linear counterterm $Y\,\phi$ in the original action functional, see for example \cite{Srednicki}.
The constant $Y$ is determined so that it sets the vacuum expectation value of $\phi$ to zero. That is, the tadpole diagram is cancelled by a corresponding diagram coming from the linear counterterm. In braided $\lambda\,\phi^3$-theory, the divergent tadpole contribution can be removed in the following way.

For this, we consider the effective theory of Section~\ref{sub:sources} based on the {curved} braided $L_\infty$-algebra with curvature 
\begin{align*}
\ell_0^\star(1) = Y \ \in \ \FR \ ,
\end{align*}
where the ground field is regarded as the space of constant functions $\FR\subset V_2$. Then the action functional is modified to the curved Maurer--Cartan functional
\begin{equation}
S_Y(\phi) = S(\phi) + \langle\phi,\ell_0^\star(1)\rangle_\star   = \int\d^d x \ \frac{1}{2}\,\phi\,\big(-\square - m^2\big)\,\phi - \frac{\lambda}{3!}\,\phi\star\phi\star\phi + Y\,\phi \ . \label{eq:S0scalarY} 
\end{equation}
In the braided BV formalism, this modifies $\CS_{\rm int}$ by the addition of a linear interaction term $\CS_Y$ defined by
\begin{align}
\CS_Y = \langle\xi,\ell_0^{\star\,{\rm ext}}(1)\rangle_\star^{\rm ext} = Y\,\int_{k}\, (2\pi)^d \, \delta(k) \ \tte^k  \ .\label{SIntY} 
\end{align}

The contribution to the one-point function $\tilde G_1^\star(p)$ coming from (\ref{SIntY}) is easily calculated and  is given by
\begin{align}
\tilde G_{1}^\star(p)_Y = Y\,(2\pi)^d\,\frac{\delta(p)}{p^2 -m^2} \ .\nn
\end{align}
Therefore the full one-point function of braided $\lambda\,\phi^3$-theory at one-loop is
\begin{align}
\tilde G_1^\star(p)^{\swone} = \ii\,\hbar\, \Big(-\frac{\lambda}{2}\Big)\,(2\pi)^d\,\delta(p)\,\tilde \sgreen(p)\,\int_{k}\, \frac{1}{k^2 - m^2} + Y\,(2\pi)^d\,\frac{\delta(p)}{p^2 -m^2} \ , \nn 
\end{align}
which we represent diagrammatically by
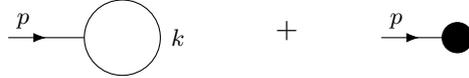
\begin{align*}\label{fig:Phi31Y}
{\footnotesize
\begin{tikzpicture}[baseline]
        \coordinate (k) at (0,0);
        \coordinate[label=above right: $p$] (p) at (180:1);
        \draw[decoration={markings, mark=at position 0.5 with {\arrow{Latex}} }, postaction={decorate}] (p) -- (k);
        \draw[ postaction={decorate}] ($(k)+(0:0.5)$) circle (0.5) node[right=15]{$k$};
    \end{tikzpicture}
\hspace{1cm} + \hspace{1cm}
\begin{tikzpicture}[baseline]
        \coordinate (k) at (0,0);
        \coordinate[label=above right: $p$] (p) at (180:1);
        \draw[decoration={markings, mark=at position 0.5 with {\arrow{Latex}} }, postaction={decorate}] (p) -- (k);
       \filldraw[color=black, fill=black]  circle (0.2) node[right=15] {};
    \end{tikzpicture} } \normalsize 
\end{align*}
Setting it to zero determines the constant $Y$ at one-loop as
\begin{equation}
Y^\swone = \ii\,\hbar\, \frac{\lambda}{2}\, \int_{k}\, \frac{1}{k^2 - m^2} \ , \label{Y1Loop}
\end{equation}
which we understand in terms of a cutoff regularization of the divergent integral.

One can proceed and calculate the contribution of $\CS _{Y}$ to the two-point function at one-loop in braided $\lambda\,\phi^3$-theory. Using the fixed value for $Y^\swone$ from (\ref{Y1Loop}), all tadpole contributions vanish. For example, the factors for the diagrams 
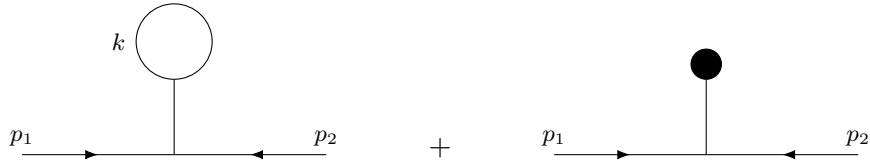
\begin{align*}
{\footnotesize
\begin{tikzpicture}[baseline]
        \coordinate (k) at (0,0);
        \coordinate (l) at ($(k)+(90:1)$);
        \coordinate[label=above: $p_1$] (p1) at ($(k) +(180:2)$);
        \coordinate[label=above: $p_2$] (p2) at ($(k) +(0:2)$);
        \draw[decoration={markings, mark=at position 0.5 with {\arrow{Latex}} }, postaction={decorate}] (p1) -- (k);
        \draw[decoration={markings, mark=at position 0.5 with {\arrow{Latex}} }, postaction={decorate}] (p2) -- (k);
        \draw[postaction={decorate}] (k) -- (l);
        \draw[postaction={decorate}] ($(l)+(90:0.5)$) circle (0.5) node[left=15]{$k$};
    \end{tikzpicture}
\hspace{1cm} + \hspace{1cm} 
    \begin{tikzpicture}[baseline]
        \coordinate (k) at (0,0);
        \coordinate (l) at ($(k)+(90:1)$);
        \coordinate[label=above: $p_1$] (p1) at ($(k) +(180:2)$);
        \coordinate[label=above: $p_2$] (p2) at ($(k) +(0:2)$);
        \draw[decoration={markings, mark=at position 0.5 with {\arrow{Latex}} }, postaction={decorate}] (p1) -- (k);
        \draw[decoration={markings, mark=at position 0.5 with {\arrow{Latex}} }, postaction={decorate}] (p2) -- (k);
        \draw[ postaction={decorate}] (k) -- (l);
        \filldraw[color=black, fill=black] ($(l)+(90:0.2)$) circle (0.2) node[left=15]{};
    \end{tikzpicture} } \normalsize
\end{align*}
ensure that the tadpole contributions exactly cancel.
Finally, we are left with 
\begin{align}
\tilde G_2^\star(p_1,p_2)^{\swone} = -\frac{(\hbar\,\lambda)^2}{2} \, \frac{(2\pi)^d\, \delta(p_1+p_2)}{(p_1^2 - m^2)\,(p_2^2 - m^2)} \ \int_k\, \frac{1}{\big(k^2-m^2\big)\,\big((p_1-k)^2 - m^2\big)} \ . \nn 
\end{align}

\subsection{A glimpse at two-loop corrections}
\label{sub:2loop}

The two-loop contribution to the two-point function is given by 
\begin{equation}\label{eq:2Point2Loop}
\tilde G_2^\star(p_1,p_2)^{\swtwo} =  \sP\,\big((-\ii\,\hbar\,\BVL\,\sH - \{\CS _{\rm int},-\}_\star\,\sH)^7\, (\tte^{p_1}\odot_\star\tte^{p_2})\big) \ ,
\end{equation}
with seven propagators and four vertices.
There are various terms resulting in disconnected or reducible diagrams. Here we would like to take a glimpse and see the form of the one-particle irreducible (1PI) contribution, i.e. whether it differs from the commutative case and what the difference is, regardless of its symmetry factor. For this purpose, we should only calculate the obvious term in \eqref{eq:2Point2Loop} containing the desired 1PI contribution
\begin{equation*}
\tilde G_2^\star(p_1,p_2)^{\swtwo}_{{\text{\tiny 1PI}}} = -(\ii\,\hbar\,\BVL\,\sH)^3\, \big( \{\CS _{\rm int}, \sH\, \{\CS _{\rm int}, \sH\, \{\CS _{\rm int}, \sH\, \{\CS _{\rm int},\sH\,(\tte^{p_1}\odot_\star\tte^{p_2}) \}_\star\}_\star\}_\star \}_\star \big) \ .
\end{equation*}

Starting from (\ref{eq:SintH3e}), adding one more vertex to \eqref{eq:SintHSintH3e} results in
\small
\begin{align}\label{S2_3vert}
\begin{split}
 & \{\CS _{\rm int}, \sH\, \{\CS _{\rm int}, \sH\, \{\CS _{\rm int},\sH(\tte^{p_1}\odot_\star\tte^{p_2}) \}_\star\}_\star\}_\star \\[4pt]
&  = -\frac{3}{4}\,\frac{3}{3}\,\frac{3}{2} \, \int_{k_1,k_2,k_3} \,  \int_{l_1,l_2,l_3} \, \int_{r_1,r_2,r_3} \!\!\!\!\!
\, V_3(k_1,k_2,k_3) \, V_3(l_1,l_2,l_3) \, V_3(r_1,r_2,r_3)\, \frac{(2\pi)^d}{k_1^2 -m^2}\frac{(2\pi)^d}{l_1^2 -m^2}\frac{(2\pi)^d}{r_1^2 -m^2} \\
& \quad \, \times \tte^{r_2}\odot_\star\tte^{r_3} \odot_\star \Big\{
\delta(k_1+p_1)\,\delta(l_1+k_2)\,\Big(  2\,\delta(r_1+l_2)\,\tte^{l_3}\odot_\star\tte^{k_3}\odot_\star\tte^{p_2} \\
& \hspace{7.5cm} + \e^{\,\ii\, r_1\cdot \theta\, l_1}\, \delta(r_1+k_3)\,\delta(l_1+k_2)\,\tte^{l_2}\odot_\star\tte^{l_3}\odot_\star\tte^{p_2}\Big) \\
& \hspace{2cm}+ \e^{-\ii\, k_1\cdot \theta\, p_1}\,\delta(k_1+p_2)\,\delta(l_1+k_2)\,\Big(  2\,\delta(r_1+l_2)\,\tte^{l_3}\odot_\star\tte^{k_3}\odot_\star\tte^{p_1} \\
& \hspace{7.5cm} + \e^{\,\ii\, r_1\cdot \theta\, l_1}\, \delta(r_1+k_3)\,\delta(l_1+k_2)\,\tte^{l_2}\odot_\star\tte^{l_3} \odot_\star\tte^{p_1}\Big) \\
& \hspace{2cm} + 4\,\delta(r_1+k_2)\,\Big( \e^{-\ii\, p_1\cdot \theta\, k_2}\,\delta(k_1+p_2)\,\delta(l_1+p_1) + \e^{\,\ii \, l_1\cdot \theta\, (k_2-p_1)}\,\delta(k_1+p_1)\,\delta(l_1+p_2) \Big)\\
& \hspace{11.5cm} \times \tte^{l_2}\odot_\star\tte^{l_3}\odot_\star\tte^{k_3} \Big\} \ .
\end{split}
\end{align}
\normalsize

Now we need to add the fourth vertex. The full result is very cumbersome and we will not present it here. However, to illustrate the terms appearing we present the partial contribution from the  term in the fourth line of (\ref{S2_3vert}), which reads
\small \begin{align}\label{S2_4vert}
\begin{split}
&\{\CS _{\rm int}, \sH\, \{\CS _{\rm int}, \sH\, \{\CS _{\rm int}, \sH\, \{\CS _{\rm int},\sH\,(\tte^{p_1}\odot_\star\tte^{p_2}) \}_\star\}_\star\}_\star \}_\star\big|_{\rm partial} \\[4pt]
& = \frac{3}{5}\,\frac{3}{4}\,\frac{3}{3}\,\frac{3}{2}\,\int_{k_1,k_2,k_3} \, \int_{l_1,l_2,l_3} \, \int_{r_1,r_2,r_3} \,\int_{s_1,s_2,s_3} \, 2\, V_3(k_1,k_2,k_3) \, V_3(l_1,l_2,l_3)\, V_3(r_1,r_2,r_3) \, V_3(s_1,s_2,s_3) \\
& \hspace{2cm} \times \frac{(2\pi)^d}{k_1^2 -m^2}\frac{(2\pi)^d}{l_1^2 -m^2}\frac{(2\pi)^d}{r_1^2 -m^2} \frac{(2\pi)^d}{s_1^2 -m^2} \,
\e^{\,\ii\, r_1\cdot \theta\, l_1} \, \delta(r_1+k_3)\,\delta(l_1+k_2)\, \delta(k_1+p_1) \\
& \hspace{3cm} \times \tte^{s_2} \odot_\star\tte^{s_3} \odot_\star \Big( 
2\,\delta(s_1+r_2)\,\tte^{r_3}\odot_\star \tte^{l_2}\odot_\star\tte^{l_3}\odot_\star\tte^{p_2} \\
& \hspace{6cm} 
+ 2\, \e^{\,\ii\, s_1\cdot \theta\, r_1}\,\delta(s_1+l_2)\,\tte^{r_2}\odot_\star \tte^{r_3}\odot_\star\tte^{l_3}\odot_\star\tte^{p_2}\\
& \hspace{7cm} + \e^{\,\ii\, s_1\cdot \theta\, (r_1+l_1)}\,\delta(s_1+p_2)\,\tte^{r_2}\odot_\star \tte^{r_3}\odot_\star\tte^{l_2}\odot_\star\tte^{l_3} \Big) \ .
\end{split}
\end{align} \normalsize

The six-point functions are calculated using the braided Wick theorem, see \eqref{BrWickG6}. Again, there are terms that contribute to disconnected and reducible diagrams that we will not write explicitly. As an example, we write here one term that contributes to 1PI diagrams through
\small
\begin{align}\label{S2_2loopfinal}
\begin{split}
&\hspace{-0.25cm}(\ii\,\hbar\,\BVL\,\sH)^3 \{\CS _{\rm int}, \sH \,\{\CS _{\rm int}, \sH\, \{\CS _{\rm int}, \sH \,\{\CS _{\rm int},\sH\,(\tte^{p_1}\odot_\star\tte^{p_2}) \}_\star\}_\star\}_\star \}_\star\big|_{\rm {partial}} \\[4pt]
&\hspace{-0.25cm} = -\frac{3}{5}\,\frac{3}{4}\,\frac{3}{3}\,\frac{3}{2} \, \int_{k_1,k_2,k_3} \,  \int_{l_1,l_2,l_3} \, \int_{r_1,r_2,r_3}\, \int_{s_1,s_2,s_3} \, 2\, V_3(k_1,k_2,k_3) \, V_3(l_1,l_2,l_3) \, V_3(r_1,r_2,r_3)\, V_3(s_1,s_2,s_3) \\
&  \times \frac{(2\pi)^d}{k_1^2 -m^2}\,\frac{(2\pi)^d}{l_1^2 -m^2}\,\frac{(2\pi)^d}{r_1^2 -m^2} \,\frac{(2\pi)^d}{s_1^2 -m^2}\, \e^{\,\ii\, r_1\cdot \theta\, l_1}\,\e^{\,\ii\, s_1\cdot \theta\, (r_1+l_1)}\,\delta(s_1+p_2)\, \delta(r_1+k_3)\,\delta(l_1+k_2)\, \delta(k_1+p_1)\\
& \hspace{6cm} \times \bcontraction{}{\tte^{s_2}}{}{\sfR_\alpha(\tte^{r_2})} \tte^{s_2}\,\sfR_\alpha(\tte^{r_2} )
\ \bcontraction{}{\sfR^\alpha(\tte^{s_3})}{}{\sfR_\beta(\tte^{l_2})} \sfR^\alpha(\tte^{s_3})\,\sfR_\beta(\tte^{l_2})
\ \bcontraction{}{\sfR^\beta(\tte^{r_3})}{}{\tte^{l_3}} \sfR^\beta(\tte^{r_3})\,\tte^{l_3} \ .
\end{split}
\end{align}
\normalsize
We introduced the symbolic notation
\begin{equation}
\bcontraction{}{\tte^{s}}{}{\tte^{r}} \tte^{s}\,\tte^{r} = -\ii\,\hbar\,\BVL\,\sH\, (\tte^{s}\odot_\star\tte^{r}) = \ii\,\hbar\, (2\pi)^d\,\frac{\delta(s+r)}{s^2-m^2} \ .\label{ContrExpBasis}
\end{equation}

Integrating over 10 out of 12 momenta in \eqref{S2_2loopfinal} and resolving the delta-functions, this term finally results in
\begin{align}\label{G2_2loop1PI}
\begin{split}
& \tilde G_2^\star(p_1,p_2)_{\text{\tiny 1PI}}^{\swtwo}\big|_{\rm partial} \\[4pt]
& \quad = (\ii\,\hbar)^3\,\Big(-\frac{\lambda}{3!}\Big)^4\, \frac{3}{5}\,\frac{3}{4}\,\frac{3}{3}\,\frac{3}{2}\, \frac{(2\pi)^d\,\delta(p_1+p_2)}{(p_1^2 -m^2)\,(p_2^2 -m^2)}  \\
& \hspace{1cm} \times \int_{l_1,l_2}\,\frac{1}{\big(l_1^2 -m^2\big)\,\big(l_2^2 -m^2\big)\,\big((l_1+l_2)^2 -m^2\big)\,\big((p_1+l_1)^2 -m^2\big) \, \big((p_2+l_2)^2 -m^2\big)} \ .
\end{split}
\end{align}
This partial contribution to the two-point function at two-loops in braided $\lambda\,\phi^3$-theory corresponds to the diagram
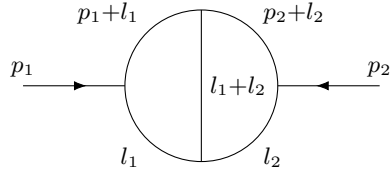
\begin{equation*} \label{fig:Phi77}
\begin{tikzpicture}[baseline, scale=0.67]
        \footnotesize
        \coordinate (k) at (-1.5,0);
        \coordinate (l) at (1.5,0);
        \draw[decoration={markings, mark=at position 0.5 with {\arrow{Latex[reversed]}} }, postaction={decorate}] (k) -- ($(k) + (180:2)$) node[above]{$p_1$};
        \draw[decoration={markings, mark=at position 0.5 with {\arrow{Latex[reversed]}} }, postaction={decorate}] (l) -- ($(l) + (0:2)$) node[above]{$p_2$};
        \draw[ postaction={decorate}] ($(k)+(0:1.5)$) circle (1.5) node[below left = 20]{$l_1$} node[below right = 20]{$l_2$} node[above left = 20]{$p_1 {+} l_1$} node[above right = 20]{$p_2 {+} l_2$};  
        \draw[ postaction={decorate}] ($(k) +(0:1.5) +(90:1.5)$) -- ($(k) +(0:1.5) +(270:1.5)$) node[right, pos=0.5]{$l_1 {+} l_2$}; \normalsize
    \end{tikzpicture}
    \end{equation*}
Again all noncommutative corrections vanish and the result is the same as in the commutative case.

\section{Braided Schwinger--Dyson equations}
\label{sec:SD}

In ordinary quantum field theories, the Schwinger--Dyson equations are consistency equations satisfied by full correlation functions which represent the quantum analogues of the classical equations of motion.
They are most readily derived in the path integral formalism with integration measure that is invariant under  infinitesimal variations of the fields. 
As our braided quantum field theories do not admit a path integral formulation, we will show how such consistency relations arise from the purely algebraic perspective of homological perturbation theory.
We adapt arguments originating in \cite{Okawa:2022sjf,Konosu2023,Konosu:2023rkm,Konosu:2024dpo} for quantum $A_\infty$-algebras to the setting of braided $L_\infty$-algebras. Our main goal is to better understand to what extent noncommutative corrections enter the correlation functions of braided scalar field theory through a detailed derivation and analysis of the braided analogues of the Schwinger--Dyson equations.

\subsection{Schwinger--Dyson equations as a recursion relation} 
\label{sub:general case BSDE}

Consider a braided scalar field theory in the setting of Section~\ref{sub:braidedphi3}.
Decomposing the perturbed projection as ${\tilde{\sP}} = \mathsf{P} + \mathsf{P}_{{{\mbf\delta}}}$, the braided homological perturbation lemma implies
\begin{equation*}
\begin{split}
    {\tilde{\sP}} &= \mathsf{P}\,\Big(\id_{\Sym_\RR(V[2])} + \big(\id_{\Sym_\RR(V[2])} - {{\mbf\delta}}\,\sH\big)^{-1}\,{{\mbf\delta}}\,\mathsf{H}\Big)
    = \mathsf{P}\,\big(\id_{\Sym_\RR(V[2])} - {{\mbf\delta}}\,\sH\big)^{-1} \ .
\end{split}
\end{equation*}
It follows that the full projector ${\tilde{\sP}}$ can be computed through the recursion relation (as also hinted at in \cite{Jurco:2019yfd})
\begin{equation}
\begin{split}\label{eq:recursion on Pfull}
    {\tilde{\sP}} = \mathsf{P} + {\tilde{\sP}} \, {{\mbf\delta}} \, \mathsf{H}\ .
\end{split}
\end{equation}

Acting on a basis of $\mathrm{Sym}_\mathcal{R}(V[2])$ of symmetric degree $n > 0$, the free projector $\sP$ vanishes by the trivial projection onto $\mathbbm{R}$ in (\ref{eq:trivial project}), giving
\begin{equation}\label{eq:scemidstep}
    {\tilde{\sP}} = {\tilde{\sP}} \, {{\mbf\delta}}\, \mathsf{H} \ .
\end{equation}
We show that this identity implies a braided version of the Schwinger--Dyson equations for the braided BV theory. 

As previously, we consider a basis for $\mathrm{Sym}_\mathcal{R}(V[2])$ in cohomological degree $0$, built from antifields $\phi_i^+\in V_2$, which in momentum space are the fields $\mathtt{e}^{p_i}$. 
We evaluate the recursion \eqref{eq:scemidstep} on
\begin{equation}
\begin{split}\label{eq:basis for sde}
    \mathtt{e}^{p_1} 
    \odot_\star \cdots \odot_\star 
    \mathtt{e}^{p_{n}} \ 
    \in  \ \mathrm{Sym}_\mathcal{R}(V[2])_0 \ .
\end{split}
\end{equation}
The left-hand side is simply the all-orders correlation function 
\begin{equation}
\begin{split}\label{eq:SDE_eomterm1st}
    \mathsf{P}_{\mbf\delta}
    (
    \mathtt{e}^{p_1} 
    \odot_\star \cdots \odot_\star 
    \mathtt{e}^{p_{n}}
    )
    &=  
    {\tilde{G}}^\star_{ n }(p_1, \ldots, p_{n}) \ .
\end{split}
\end{equation}
The right-hand side of \eqref{eq:scemidstep} requires a choice of deformation ${{\mbf\delta}}$.

\subsection{Free theory and the braided Wick theorem} 
\label{sub:free theory sde}

To derive the Schwinger--Dyson equations for the free braided scalar field theory, we use the deformation 
\begin{align*}
\mbf{\delta}_0 = -\mathrm{i}\, \hbar\, \mathsf{\Delta}_\BV \ .
\end{align*}
Because $\mathsf{\Delta}_{\BV}$ has symmetric degree $-2$ and we use the trivial projection $\sP$, correlation functions in the free theory are non-zero only for even numbers of field insertions. We use \eqref{eq:scemidstep} to show that they encode the braided Wick theorem proposed in \cite{DimitrijevicCiric:2023hua}.

Using \eqref{eq:SDE_eomterm1st} and \eqref{eq:scemidstep}, the free $2n$-point is given by
\begin{align*}
\tilde G_{2n}^{\star 0} (p_1,\dots,p_{2n}) := \mathsf{P}_{\mbf\delta_0}
    (
    \mathtt{e}^{p_1} 
    \odot_\star \cdots \odot_\star 
    \mathtt{e}^{p_{2n}}
    ) = \tilde\sP\,(-\ii\,\hbar\,\BVL\,\sH)\,(\mathtt{e}^{p_1} 
    \odot_\star \cdots \odot_\star 
    \mathtt{e}^{p_{2n}}
    ) \ .
\end{align*}
Using \eqref{eq:scalarpairing}, \eqref{eq:htwo},  \eqref{eq:sfH} and \eqref{eq:BVL} we obtain the  expansion
\begin{align}\label{eq:SDE_deltaterm}
\begin{split}
& -\ii\,\hbar\,\BVL\,\sH\,(\tte^{p_1}\odot_\star\cdots\odot_\star\tte^{p_{2n}}) \\[4pt]
& \hspace{2cm} = \frac1{2n} \, \sum_{i\neq j} \, \bcontraction{}{\phi_i}{}{\sfR_{\alpha_{i+1}}\cdots\sfR_{\alpha_{j-1}}(\phi_j)} \phi_i\,\sfR_{\alpha_{i+1}}\cdots\sfR_{\alpha_{j-1}}(\phi_j) \  \tte^{p_1}\odot_\star\cdots\odot_\star\tte^{p_{i-1}} \\ 
& \hspace{4.5cm}  \odot_\star\sfR^{\alpha_{i+1}}(\tte^{p_{i+1}})\odot_\star\cdots\odot_\star\sfR^{\alpha_{j-1}}(\tte^{p_{j-1}})\odot_\star \tte^{p_{j+1}}\odot_\star\cdots\odot_\star\tte^{p_{2n}} \ ,
\end{split}
\end{align}
for $n>1$, with the free two-point functions given by
\begin{equation}
\tilde G^{\star0}_2(p_i,p_j) = -\ii\,\hbar\,\BVL\,\sH\,( \tte^{p_i}\odot_\star \tte^{p_j}) = \ii\,\hbar\, \frac{(2\pi)^d\,\delta(p_i+p_j)}{p_i^2-m^2} =: \bcontraction{}{\phi_i}{}{\phi_j} \phi_i\,\phi_j =  \bcontraction{}{\phi_j}{}{\phi_i} \phi_j\,\phi_i \ , \label{BrWickG2}
\end{equation}
as in \eqref{G20}. The absence of braiding phase factors in the final equality is a consequence of momentum conservation.

Using \eqref{eq:Rekotimesep} to unravel the actions of the inverse $\RR$-matrices, and applying $\tilde\sP = \sP+\sP_{\mbf\delta_0}$ to the remaining products of antifields in \eqref{eq:SDE_deltaterm}, we arrive at the recursion relations
\begin{align}\label{eq:SDE_free}
\tilde G_{2n}^{\star 0} (p_1,\dots,p_{2n}) = \frac1{2n} \, \sum_{i\neq j} \, \e^{\,\ii\,p_j\cdot\theta\,(p_{i+1}+\cdots+p_{j})} \ \bcontraction{}{\phi_i}{}{\phi_j} \phi_i\,\phi_j \ \tilde G^{\star 0}_{2n-2}(p_1,\dots,\widehat{p_i},\dots,\widehat{p_j},\dots,p_{2n}) \ ,
\end{align}
for $n\geqslant1$, where we define $\tilde G^{\star 0}_0:=1$. The right-hand side is a sum of $2n\,(2n-1)$ terms involving $2n-2$-point correlation functions, 
where the hat notation means to omit the corresponding entry. For $n=1$ the formula \eqref{eq:SDE_free} reproduces the two-point function \eqref{BrWickG2}.
This is our form of the Schwinger--Dyson equations for the free braided scalar field theory.

Let us compare \eqref{eq:SDE_free} with the standard textbook form of the Schwinger--Dyson equation in the commutative case, which is also derived in \cite{Okawa:2022sjf,Konosu2023} from the perspective of $A_\infty$-algebras. We multiply through \eqref{eq:SDE_free} by $p_1^2-m^2$ to find 
\begin{align*}
&(p_1^2-m^2)\,\tilde G_{2n}^{\star 0} (p_1,\dots,p_{2n}) = \frac{\ii\,\hbar}{2n} \, \sum_{j=2}^n \, (2\pi)^d \ \delta(p_1+p_j) \ \e^{\,\ii\,p_j\cdot\theta\,(p_{2}+\cdots+p_j)} \ \tilde G^{\star 0}_{2n-2}(p_2,\dots,\widehat{p_j},\dots,p_{2n}) \\
& \hspace{2cm} + \frac1{2n} \, \sum_{i=2}^n \ \sum_{j\neq i} \, \e^{\,\ii\,p_j\cdot\theta\,(p_{i+1}+\cdots+p_j)} \ \bcontraction{}{\phi_i}{}{\phi_j} \phi_i\,\phi_j \ (p_1^2-m^2) \, \tilde G^{\star 0}_{2n-2}(p_1,\dots,\widehat{p_i},\dots,\widehat{p_j},\dots,p_{2n}) \ .
\end{align*}
Up to the overall factor $\frac1{2n}$, the first line coincides with the standard form of the Schwinger--Dyson equations, with the right-hand side representing a deformation of the usual {contact terms} with phase factors respecting the cyclic ordering of field insertions. 

The meaning of the remaining terms from the second line is that, rather than singling out a specific field insertion (here $\tte^{p_1}$), the (braided) $L_\infty$-algebra formulation averages the usual Schwinger--Dyson equations over all permutations of the field insertions, leaving an expression that is manifestly (braided) symmetric. This is in contrast to the $A_\infty$-algebra formulation of~\cite{Okawa:2022sjf,Konosu2023}, where the brackets do not have any symmetry properties like the $L_\infty$-brackets do. Indeed the classical algebra of observables for an $A_\infty$-structure is the unsymmetrized tensor algebra, and correspondingly the fattening of maps used in the homological perturbation lemma to compute correlators differ from those of an $L_\infty$-structure precisely by the symmetrization used in the latter~\cite{Berglund2014}. From an algebraic perspective, our form \eqref{eq:SDE_free} of the Schwinger--Dyson equations is much more elegant as it highlights the symmetries of the correlation functions, while it can be brought into the standard form (without the factor $\frac1{2n}$) by a combinatorial argument, as we discuss further below.

\subsubsection{Examples}
\label{subsub:Wickex}

We can illustrate the formula \eqref{eq:SDE_free} using the diagrammatic calculus of Section~\ref{sub:diagramcalculus}. In the following a large blue filled circle with $n$ emanating black lines represents the full $n$-point correlation function $\tilde G_n^{\star}(p_1,\dots,p_n)$.

\begin{example}[{\bf Four-point function}]
Let us derive the Schwinger--Dyson equation for $n=2$. We start from the element
\begin{eqnarray*}
    \tte^{p_1}\odot_\star \tte^{p_2}\odot_\star \tte^{p_3} \odot_\star \tte^{p_4}
=  { \footnotesize \begin{tikzpicture}[scale=0.5, baseline]
        \coordinate (k0) at (90:1);
        \coordinate (k1) at ($(k0) +(0:2)$);
        \coordinate (k2) at ($(k0) +(0:4)$);
        \coordinate (k3) at ($(k0) +(0:6)$);
        \coordinate[label=below: {$p_1$}] (k0in) at (270:1);
        \coordinate[label=below: {$p_2$}] (k1in) at ($(k0in) +(0:2)$);
        \coordinate[label=below: {$p_3$}] (k2in) at ($(k0in) +(0:4)$);
        \coordinate[label=below: {$p_4$}] (k3in) at ($(k0in) +(0:6)$);

        \draw[decoration={markings, mark=at position 0.5 with {\arrow{Latex}}}, postaction={decorate}] (k0in) -- (k0);
        \draw[decoration={markings, mark=at position 0.5 with {\arrow{Latex}}}, postaction={decorate}] (k1in) -- (k1);
        \draw[decoration={markings, mark=at position 0.5 with {\arrow{Latex}}}, postaction={decorate}] (k2in) -- (k2);
        \draw[decoration={markings, mark=at position 0.5 with {\arrow{Latex}}}, postaction={decorate}] (k3in) -- (k3);
    \end{tikzpicture} } \normalsize
\end{eqnarray*}
and apply both sides of \eqref{eq:scemidstep} to it. The left-hand side is the four-point function
\begin{eqnarray*}
\tilde{\sP}\, (\tte^{p_1} \odot_\star \tte^{p_2}\odot_\star \tte^{p_3} \odot_\star \tte^{p_4})  \ = \ \sP_{\mbf\delta_0} \Bigg( { \footnotesize
    \begin{tikzpicture}[scale=0.5, baseline]
        \coordinate (k0) at (90:1);
        \coordinate (k1) at ($(k0) +(0:2)$);
        \coordinate (k2) at ($(k0) +(0:4)$);
        \coordinate (k3) at ($(k0) +(0:6)$);
        \coordinate[label=below: {$p_1$}] (k0in) at (270:1);
        \coordinate[label=below: {$p_2$}] (k1in) at ($(k0in) +(0:2)$);
        \coordinate[label=below: {$p_3$}] (k2in) at ($(k0in) +(0:4)$);
        \coordinate[label=below: {$p_4$}] (k3in) at ($(k0in) +(0:6)$);

        \draw[decoration={markings, mark=at position 0.5 with {\arrow{Latex}}}, postaction={decorate}] (k0in) -- (k0);
        \draw[decoration={markings, mark=at position 0.5 with {\arrow{Latex}}}, postaction={decorate}] (k1in) -- (k1);
        \draw[decoration={markings, mark=at position 0.5 with {\arrow{Latex}}}, postaction={decorate}] (k2in) -- (k2);
        \draw[decoration={markings, mark=at position 0.5 with {\arrow{Latex}}}, postaction={decorate}] (k3in) -- (k3);
    \end{tikzpicture} \Bigg) } \normalsize \ = \ { \footnotesize \begin{tikzpicture}[scale=0.5, baseline]
    \coordinate[label=above left: {$p_1$}] (k0) at (135:2);
    \coordinate[label=above right: {$p_2$}] (k1) at (45:2);
    \coordinate[label=below right: {$p_3$}] (k2) at (315:2);
    \coordinate[label=below left: {$p_4$}] (k3) at (225:2);

    \draw (k0) -- (k2);
    \draw (k1) -- (k3);
    \draw[fill=blue!50] (0,0) circle[radius=0.5];
    \end{tikzpicture} } \normalsize
\end{eqnarray*}

For the right-hand side we calculate
\begin{eqnarray*}
 \tilde{\sP}\, \mbf\delta_0\, \sH\,(\tte^{p_1}\odot_\star \tte^{p_2}\odot_\star \tte^{p_3} \odot_\star \tte^{p_4}) &=&  \sP_{\mbf\delta_0}\, \mbf\delta_0\, \sH\, (\tte^{p_1}\odot_\star \tte^{p_2}\odot_\star \tte^{p_3} \odot_\star \tte^{p_4}) \\[4pt]
    &=& \sP\, (\mbf\delta_0\, \sH)^2\,(\tte^{p_1}\odot_\star \tte^{p_2}\odot_\star \tte^{p_3} \odot_\star \tte^{p_4})\\[4pt]
 &=& (\ii\,\hbar\,\BVL\,\sH)^2\, \Bigg( { \footnotesize
    \begin{tikzpicture}[scale=0.5, baseline]
        \coordinate (k0) at (90:1);
        \coordinate (k1) at ($(k0) +(0:2)$);
        \coordinate (k2) at ($(k0) +(0:4)$);
        \coordinate (k3) at ($(k0) +(0:6)$);
        \coordinate[label=below: {$p_1$}] (k0in) at (270:1);
        \coordinate[label=below: {$p_2$}] (k1in) at ($(k0in) +(0:2)$);
        \coordinate[label=below: {$p_3$}] (k2in) at ($(k0in) +(0:4)$);
        \coordinate[label=below: {$p_4$}] (k3in) at ($(k0in) +(0:6)$);

        \draw[decoration={markings, mark=at position 0.5 with {\arrow{Latex}}}, postaction={decorate}] (k0in) -- (k0);
        \draw[decoration={markings, mark=at position 0.5 with {\arrow{Latex}}}, postaction={decorate}] (k1in) -- (k1);
        \draw[decoration={markings, mark=at position 0.5 with {\arrow{Latex}}}, postaction={decorate}] (k2in) -- (k2);
        \draw[decoration={markings, mark=at position 0.5 with {\arrow{Latex}}}, postaction={decorate}] (k3in) -- (k3);
    \end{tikzpicture} } \normalsize \Bigg)\\[4pt]
&=&  { \footnotesize
    \begin{tikzpicture}[scale=0.5, baseline]
        \coordinate (k0) at (90:1);
        \coordinate (k1) at ($(k0) +(0:2)$);
        \coordinate (k0in) at (270:1);
        \coordinate (k1in) at ($(k0in) +(0:2)$);

        \draw[ color=dgreen] (k0in) -- (k0) node[below=10mm]{$G_{p_1 p_2}$};
        \draw[color=dgreen] (k1in) -- (k1) node[below=10mm]{$G_{p_3 p_4}$};
    \end{tikzpicture} } \normalsize
    + 
    \e^{\,\ii\, p_3 \cdot \theta\, p_2} \times { \footnotesize \begin{tikzpicture}[scale=0.5, baseline]
        \coordinate (k0) at (90:1);
        \coordinate (k1) at ($(k0) +(0:2)$);
        \coordinate (k0in) at (270:1);
        \coordinate (k1in) at ($(k0in) +(0:2)$);

        \draw[ color=dgreen] (k0in) -- (k0) node[below=10mm]{$G_{p_1 p_3}$};
        \draw[color=dgreen] (k1in) -- (k1) node[below=10mm]{$G_{p_2 p_4}$};
    \end{tikzpicture} } \normalsize
    + { \footnotesize
    \begin{tikzpicture}[scale=0.5, baseline]
        \coordinate (k0) at (90:1);
        \coordinate (k1) at ($(k0) +(0:2)$);
        \coordinate (k0in) at (270:1);
        \coordinate (k1in) at ($(k0in) +(0:2)$);

        \draw[ color=dgreen] (k0in) -- (k0) node[below=10mm]{$G_{p_1 p_4}$};
        \draw[color=dgreen] (k1in) -- (k1) node[below=10mm]{$G_{p_2 p_3}$};
    \end{tikzpicture} } \normalsize
    \\[4pt]
&=& { \footnotesize
    \begin{tikzpicture}[scale=0.5, baseline]
        \coordinate[label=right: {$p_2$}] (k0) at (90:1);
        \coordinate[label=right: {$p_3$}] (k1) at ($(k0) +(270:2)$);
        \coordinate[label=left: {$p_1$}] (k0in) at ($(k0)+(180:2)$);
        \coordinate[label=left: {$p_4$}] (k1in) at ($(k0in) +(270:2)$);

        \draw[] (k0in) -- (k0);
        \draw (k1in) -- (k1);
    \end{tikzpicture} } \normalsize
    \ + \ 
    \e^{\,\ii\, p_3 \cdot \theta\, p_2} \times { \footnotesize \begin{tikzpicture}[scale=0.5, baseline]
        \coordinate[label=right: {$p_3$}] (k0) at (90:1);
        \coordinate[label=right: {$p_2$}] (k1) at ($(k0) +(270:2)$);
        \coordinate[label=left: {$p_1$}] (k0in) at ($(k0)+(180:2)$);
        \coordinate[label=left: {$p_4$}] (k1in) at ($(k0in) +(270:2)$);

        \draw[] (k0in) -- (k0);
        \draw (k1in) -- (k1);
    \end{tikzpicture} } \normalsize
    \ + \ { \footnotesize
    \begin{tikzpicture}[scale=0.5, baseline]
        \coordinate[label=right: {$p_4$}] (k0) at (90:1);
        \coordinate[label=right: {$p_3$}] (k1) at ($(k0) +(270:2)$);
        \coordinate[label=left: {$p_1$}] (k0in) at ($(k0)+(180:2)$);
        \coordinate[label=left: {$p_2$}] (k1in) at ($(k0in) +(270:2)$);

        \draw[] (k0in) -- (k0);
        \draw (k1in) -- (k1);
    \end{tikzpicture} } \normalsize
\end{eqnarray*}
Compared to the right-hand side of \eqref{eq:SDE_free}, the 12 terms have combined to three terms and cancelled the overall factor of $\frac14$.

Putting everything together, we arrive at
\begin{eqnarray*}
  { \footnotesize  \begin{tikzpicture}[scale=0.5, baseline]
    \coordinate[label=above left: {$p_1$}] (k0) at (135:2);
    \coordinate[label=above right: {$p_2$}] (k1) at (45:2);
    \coordinate[label=below right: {$p_3$}] (k2) at (315:2);
    \coordinate[label=below left: {$p_4$}] (k3) at (225:2);

    \draw (k0) -- (k2);
    \draw (k1) -- (k3);
    \draw[fill=blue!50] (0,0) circle[radius=0.5];
    \end{tikzpicture} } \normalsize
    \ = \ { \footnotesize \begin{tikzpicture}[scale=0.5, baseline]
        \coordinate[label=right: {$p_2$}] (k0) at (90:1);
        \coordinate[label=right: {$p_3$}] (k1) at ($(k0) +(270:2)$);
        \coordinate[label=left: {$p_1$}] (k0in) at ($(k0)+(180:2)$);
        \coordinate[label=left: {$p_4$}] (k1in) at ($(k0in) +(270:2)$);

        \draw (k0in) -- (k0);
        \draw (k1in) -- (k1);
    \end{tikzpicture} } \normalsize
   \ + \ 
    \e^{\,\ii\, p_3 \cdot \theta\, p_2} \times { \footnotesize \begin{tikzpicture}[scale=0.5, baseline]
        \coordinate[label=right: {$p_3$}] (k0) at (90:1);
        \coordinate[label=right: {$p_2$}] (k1) at ($(k0) +(270:2)$);
        \coordinate[label=left: {$p_1$}] (k0in) at ($(k0)+(180:2)$);
        \coordinate[label=left: {$p_4$}] (k1in) at ($(k0in) +(270:2)$);

        \draw (k0in) -- (k0);
        \draw (k1in) -- (k1);
    \end{tikzpicture} } \normalsize
    \ + \ { \footnotesize
    \begin{tikzpicture}[scale=0.5, baseline]
        \coordinate[label=right: {$p_4$}] (k0) at (90:1);
        \coordinate[label=right: {$p_3$}] (k1) at ($(k0) +(270:2)$);
        \coordinate[label=left: {$p_1$}] (k0in) at ($(k0)+(180:2)$);
        \coordinate[label=left: {$p_2$}] (k1in) at ($(k0in) +(270:2)$);

        \draw (k0in) -- (k0);
        \draw (k1in) -- (k1);
    \end{tikzpicture} } \normalsize
\end{eqnarray*}
which translates into the analytic expression
\begin{align}\label{eq:4pointfree}
\tilde G_4^{\star 0}(p_1,p_2,p_3,p_4) = \bcontraction{}{\phi_1}{}{\phi_2} \phi_1\,\phi_2 \ \bcontraction{}{\phi_3}{}{\phi_4} \phi_3\,\phi_4 + \e^{\,\ii\, p_3 \cdot \theta\, p_2} \ \bcontraction{}{\phi_1}{}{\phi_3} \phi_1\,\phi_3 \ \bcontraction{}{\phi_2}{}{\phi_4} \phi_2\,\phi_4 + \bcontraction{}{\phi_1}{}{\phi_4} \phi_1\,\phi_4 \  \bcontraction{}{\phi_2}{}{\phi_3} \phi_2\,\phi_3 \ .
\end{align}
This is the braided Schwinger--Dyson equation for the four-point function in the free braided scalar field theory, which is just the braided Wick theorem in this case~\cite{DimitrijevicCiric:2023hua}.
\end{example}

\begin{example}[{\bf Six-point function}]
One can compute an explicit form for the six-point function from the Schwinger--Dyson equation \eqref{eq:SDE_free} for $n=3$ by substituting the four-point functions \eqref{eq:4pointfree} into the 30 terms appearing on the right-hand side; this is used in the calculations of Section~\ref{sub:2loop}. The final result is the braided Wick theorem for six antifields in momentum space proposed by~\cite{DimitrijevicCiric:2023hua}:
{ \small
\begin{align}
\begin{split}
\tilde{G}_6^{\star 0}(p_1,\dots,p_6) &= \bcontraction{}{\phi_1}{}{\phi_2} \phi_1\,\phi_2 \ \bcontraction{}{\phi_3}{}{\phi_4} \phi_3\,\phi_4 \ \bcontraction{}{\phi_5}{}{\phi_6} \phi_5\,\phi_6 +
\e^{\,\ii\,p_5\cdot\theta\,p_4} \ \bcontraction{}{\phi_1}{}{\phi_2} \phi_1\,\phi_2 \ \bcontraction{}{\phi_3}{}{\phi_5} \phi_3\,\phi_5 \ \bcontraction{}{\phi_4}{}{\phi_6} \phi_4\,\phi_6 +
\bcontraction{}{\phi_1}{}{\phi_2} \phi_1\,\phi_2 \ \bcontraction{}{\phi_3}{}{\phi_6} \phi_3\,\phi_6 \ \bcontraction{}{\phi_4}{}{\phi_5} \phi_4\,\phi_5  \\ 
&\quad\,+ \e^{\,\ii\,p_3\cdot\theta\,p_2} \ \bcontraction{}{\phi_1}{}{\phi_3} \phi_1\,\phi_3 \ \bcontraction{}{\phi_2}{}{\phi_4} \phi_2\,\phi_4 \ \bcontraction{}{\phi_5}{}{\phi_6} \phi_5\,\phi_6 +
\e^{\,\ii\,p_3\cdot\theta\,p_2}\,\e^{\,\ii\,p_5\cdot\theta\,p_4} \ \bcontraction{}{\phi_1}{}{\phi_3} \phi_1\,\phi_3 \ \bcontraction{}{\phi_2}{}{\phi_5} \phi_2\,\phi_5 \ \bcontraction{}{\phi_4}{}{\phi_6} \phi_4\,\phi_6 \label{BrWickG6}\\
&\quad\,+ \bcontraction{}{\phi_1}{}{\phi_6} \phi_1\,\phi_6 \ \bcontraction{}{\phi_2}{}{\phi_5} \phi_2\,\phi_5 \ \bcontraction{}{\phi_3}{}{\phi_4} \phi_3\,\phi_4 + \e^{\,\ii\,p_3\cdot\theta\,p_2} \ \bcontraction{}{\phi_1}{}{\phi_3} \phi_1\,\phi_3 \ \bcontraction{}{\phi_2}{}{\phi_6} \phi_2\,\phi_6 \ \bcontraction{}{\phi_4}{}{\phi_5} \phi_4\,\phi_5 +
\bcontraction{}{\phi_1}{}{\phi_4} \phi_1\,\phi_4 \ \bcontraction{}{\phi_2}{}{\phi_3} \phi_2\,\phi_3 \ \bcontraction{}{\phi_5}{}{\phi_6} \phi_5\,\phi_6  \\
&\quad\,+ \e^{\,\ii\,p_4\cdot\theta\,p_2}\,\e^{\,\ii\,p_4\cdot\theta\,p_3}\,\e^{\,\ii\,p_5\cdot\theta\,p_3} \  \bcontraction{}{\phi_1}{}{\phi_4} \phi_1\,\phi_4 \ \bcontraction{}{\phi_2}{}{\phi_5} \phi_2\,\phi_5 \ \bcontraction{}{\phi_3}{}{\phi_6} \phi_3\,\phi_6 +
\e^{\,\ii\,p_5\cdot\theta\,p_4} \ \bcontraction{}{\phi_1}{}{\phi_5} \phi_1\,\phi_5 \ \bcontraction{}{\phi_2}{}{\phi_3} \phi_2\,\phi_3 \ \bcontraction{}{\phi_4}{}{\phi_6} \phi_4\,\phi_6  \\ 
&\quad\,+ \e^{\,\ii\,p_4\cdot\theta\,(p_2+p_3)} \ \bcontraction{}{\phi_1}{}{\phi_4} \phi_1\,\phi_4 \ \bcontraction{}{\phi_2}{}{\phi_6} \phi_2\,\phi_6 \ \bcontraction{}{\phi_3}{}{\phi_5} \phi_3\,\phi_5 +
\e^{\,\ii\,(p_5+p_4)\cdot\theta\,p_3} \ \bcontraction{}{\phi_1}{}{\phi_5} \phi_1\,\phi_5 \ \bcontraction{}{\phi_2}{}{\phi_4} \phi_2\,\phi_4 \ \bcontraction{}{\phi_3}{}{\phi_6} \phi_3\,\phi_6 \\
&\quad\,+ \e^{\,\ii\,p_5\cdot\theta\,p_2} \ \bcontraction{}{\phi_1}{}{\phi_5} \phi_1\,\phi_5 \ \bcontraction{}{\phi_2}{}{\phi_6} \phi_2\,\phi_6 \ \bcontraction{}{\phi_3}{}{\phi_4} \phi_3\,\phi_4 + \bcontraction{}{\phi_1}{}{\phi_6} \phi_1\,\phi_6 \ \bcontraction{}{\phi_2}{}{\phi_3} \phi_2\,\phi_3 \ \bcontraction{}{\phi_4}{}{\phi_5} \phi_4\,\phi_5 +
\e^{\,\ii\,p_4\cdot\theta\,p_3} \ \bcontraction{}{\phi_1}{}{\phi_6} \phi_1\,\phi_6 \ \bcontraction{}{\phi_2}{}{\phi_4} \phi_2\,\phi_4 \ \bcontraction{}{\phi_4}{}{\phi_5} \phi_3\,\phi_5 \ . 
\end{split}
\end{align}
} \normalsize
\end{example}

\subsubsection{Braided Wick expansion}

We will now extablish that, generally, iterating \eqref{eq:SDE_free} reproduces the braided Wick theorem, adapted to momentum space from \cite{DimitrijevicCiric:2023hua}, which determines the free $2n$-point function as
\begin{equation}
\begin{split}\label{eq:braidedwick}
\tilde G^{\star 0}_{2n}(p_1,\dots,p_{2n})  = \frac1{n!\, 2^{n}} \    \sum_{\sigma \in S_{2n}} \, \e^{-\frac\ii2\,\sum\limits_{i<j}\,p_i\cdot\theta\,p_j} \ 
\prod_{k=1}^n\, \bcontraction{}{\phi_{\sigma(2k-1)}}{}{\phi_{\sigma(2k)}} \phi_{\sigma(2k-1)}\,\phi_{\sigma(2k)} \ ,
\end{split}
\end{equation}
where $S_{2n}$ is the symmetric group of degree $2n$. 
The momentum-dependent phase factor implicitly depends on the permutations, since for each $\sigma\in S_{2n}$ it is accompanied by a product of delta-functions $\delta(p_{\sigma(2k-1)}+p_{\sigma(2k)})$ from the individual two-point functions \eqref{BrWickG2} for $k=1,\dots,n$, which trivialises many terms. This expresses the braiding that arises in going from ordering $1\cdots 2n$ to ordering $\sigma(1)\cdots\sigma(2n)$ of the field insertions.

The general formula \eqref{eq:braidedwick} was written down in \cite{DimitrijevicCiric:2023hua} using heuristic arguments, and checked explicitly for $n=2$ using the braided version of the homological perturbation lemma. Indeed, we have already checked the formula \eqref{eq:braidedwick} for $n=1,2,3$ in Section~\ref{subsub:Wickex} using the Schwinger--Dyson equations \eqref{eq:SDE_free}. We can now provide a rigorous proof of the braided Wick expansion by showing that the recursive equation \eqref{eq:SDE_free} has solution of the form~\eqref{eq:braidedwick}.

Starting from \eqref{eq:braidedwick}, we observe that the sum over products of contractions can be rearranged. The sum over all permutations of the product of $n$ contractions contains every possible pair of contracted fields $\phi_a$ and $\phi_b$ with their momenta labelled respectively as $p_a,p_b \in \{p_1, \dots, p_{2n}\}$, where $a \neq b$. Although contraction is invariant under permutation between contracted fields, they are labelled by permutations which do care about the order. Therefore we fix the permutation so that $\sigma(2k_0-1) = a$ and $\sigma(2k_0) = b$, for some $k_0 \in \{1, \dots, n\}$, and write \eqref{eq:SDE_free} as
\begin{equation*}
    \tilde G^{\star 0}_{2n}(p_1,\dots,p_{2n})  = \frac1{n!\, 2^{n}} \  \sum_{\sigma \in S_{2n}} \, \e^{-\frac\ii2\,\sum\limits_{i<j}\,p_i\cdot\theta\,p_j} \ \bcontraction{}{\phi_{\sigma(2k_0-1)}}{}{\phi_{\sigma(2k_0)}} \phi_{\sigma(2k_0-1)}\,\phi_{\sigma(2k_0)} \ \prod_{\substack{k=1\\k\neq k_0}}^n \, \bcontraction{}{\phi_{\sigma(2k-1)}}{}{\phi_{\sigma(2k)}} \phi_{\sigma(2k-1)}\,\phi_{\sigma(2k)} \ .
\end{equation*}

We can identify permutations $\sigma \in S_{2n}$ and $\tau \in S_{2n-2}$ such that $\sigma(2k-1) = \tau(2l-1)$ and $\sigma(2k) = \tau(2l)$ for \smash{$k \in \{1, \dots, \widehat{k}_0, \dots, n\}$} and $l \in \{1, \dots, n-1\}$. Whereas $\sigma \in S_{2n}$ permutes fields with every possible momentum in $\{p_1, \dots, p_{2n}\}$, the permutation $\tau \in S_{2n-2}$ only permutes fields that do not  carry either momentum $p_a$ or $p_b$, i.e. it permutes fields with every momenta in $\{p_1, \dots, \widehat{p}_a, \dots, \widehat{p}_b, \dots, p_{2n}\}$.

Summing over $\sigma \in S_{2n}$ is then equivalent to summing over $\tau \in S_{2n-2}$ and over all possible unordered pairs $a\neq b$, or twice the sum over ordered pairs  $a < b$:
\begin{equation*}
    \tilde G^{\star 0}_{2n}(p_1,\dots,p_{2n})  = \frac1{n!\, 2^{n}} \ 2 \times\, \sum_{a < b} \, \bcontraction{}{\phi_{a}}{}{\phi_{b}} \phi_{a}\,\phi_{b} \ \sum_{\tau \in S_{2n-2}} \ \e^{-\frac\ii2\,\sum\limits_{i<j}\,p_i\cdot\theta\,p_j} \ \prod_{l=1}^{n-1} \, \bcontraction{}{\phi_{\tau(2l-1)}}{}{\phi_{\tau(2l)}} \phi_{\tau(2l-1)}\,\phi_{\tau(2l)} \ .
\end{equation*}
We recognize the last sum as part of $\tilde G^{\star 0}_{2n-2}(p_1, \dots \widehat{p}_a,\dots, \widehat{p}_b, \dots, p_{2n})$ according to \eqref{eq:braidedwick}. In order to recreate the whole correlation function \smash{$\tilde G^{\star 0}_{2n-2}(p_1, \dots \widehat{p}_a,\dots, \widehat{p}_b, \dots, p_{2n})$}, we have to work out the phase factor in the external momenta, keeping in mind the ordering $(p_1, \dots, p_a, \dots, p_b, \dots, p_{2n})$.

We extract terms containing $p_a$ or $p_b$ from the sum over external momenta to get
\begin{align*}
    \sum\limits_{i<j}\,p_i\cdot\theta\,p_j &=  - p_a\cdot\theta\,p_b + \sum\limits_{i<a}\,p_i\cdot\theta\,p_a +  \sum\limits_{a<i}\,p_a\cdot\theta\,p_i + \sum\limits_{i<b}\,p_i\cdot\theta\,p_b + \sum\limits_{b<i}\,p_b\cdot\theta\,p_i \\
    & \qquad \, +  \sum_{\substack{i'<j' \\ i',j' \in \{1, \dots, \widehat{a}, \dots, \widehat{b}, \dots, n\}}} \,p_{i'}\cdot\theta\,p_{j'} \ .
\end{align*}
The first term had to be subtracted since the second and third sums both contain the same term $p_a\cdot\theta\,p_b$. This yields
\begin{align*}
    \tilde G^{\star 0}_{2n}(p_1,\dots,p_{2n})  &= \frac1{n} \ \sum_{a < b} \, \e^{-\frac\ii2\,(- p_a\cdot\theta\,p_b + \sum\limits_{i<a}\,p_i\cdot\theta\,p_a +  \sum\limits_{a<i}\,p_a\cdot\theta\,p_i + \sum\limits_{i<b}\,p_i\cdot\theta\,p_b + \sum\limits_{b<i}\,p_b\cdot\theta\,p_i)} \ \bcontraction{}{\phi_{a}}{}{\phi_{b}} \phi_{a}\,\phi_{b} \\
    &\qquad \, \times \frac{1}{(n-1)!\,2^{n-1}} \ \sum_{\tau \in S_{2n-2}} \, \e^{-\frac\ii2\,\sum\limits_{i'<j'} \,p_{i'}\cdot\theta\,p_{j'}} \ \prod_{l=1}^{n-1} \, \bcontraction{}{\phi_{\tau(2l-1)}}{}{\phi_{\tau(2l)}} \phi_{\tau(2l-1)}\,\phi_{\tau(2l)} \ .
\end{align*}
In the second line we identify the correlation function $\tilde G^{\star 0}_{2n-2}(p_1, \dots \widehat{p}_a, \dots, \widehat{p}_b, \dots, p_{2n})$.

Whenever we contract fields carrying momenta $p_a$ and $p_b$, the result contains the conserving delta-distribution $\delta(p_a+p_b)$. Substituting $p_a = - p_b$ in the remaining terms from the initial sum over external momenta we get
\begin{align*}
 & - p_a\cdot\theta\,p_b + \sum\limits_{i<a}\,p_i\cdot\theta\,p_a +  \sum\limits_{a<i}\,p_a\cdot\theta\,p_i + \sum\limits_{i<b}\,p_i\cdot\theta\,p_b + \sum\limits_{b<i}\,p_b\cdot\theta\,p_i \\[4pt]
    & \hspace{1cm} = \sum\limits_{i<a}\,p_i\cdot\theta\,p_a +  \sum\limits_{a<i<b}\,p_a\cdot\theta\,p_i + \sum\limits_{b<i}\,p_a\cdot\theta\,p_i + \sum\limits_{i<a}\,p_i\cdot\theta\,p_b + \sum\limits_{a<i<b}\,p_i\cdot\theta\,p_b + \sum\limits_{b<i}\,p_b\cdot\theta\,p_i \\[4pt]
    & \hspace{1cm} = -\sum\limits_{i<a}\,p_i\cdot\theta\,p_b -  \sum\limits_{a<i<b}\,p_b\cdot\theta\,p_i - \sum\limits_{b<i}\,p_b\cdot\theta\,p_i + \sum\limits_{i<a}\,p_i\cdot\theta\,p_b + \sum\limits_{a<i<b}\,p_i\cdot\theta\,p_b + \sum\limits_{b<i}\,p_b\cdot\theta\,p_i \\[4pt]
    & \hspace{1cm} = -2\, \sum\limits_{a<i<b}\,p_b\cdot\theta\,p_i = -2\, p_b \cdot\theta\,(p_{a+1} + \dots + p_{b-1}) \ .
\end{align*}

Putting everything together, we arrive at the recursion relation
\begin{equation*}
    \tilde G^{\star 0}_{2n}(p_1,\dots,p_{2n})  = \frac1{n} \ \sum_{a < b}\, \e^{\,\ii\,p_b \cdot\theta(p_{a+1} + \dots + p_{b-1})} \  \bcontraction{}{\phi_{a}}{}{\phi_{b}} \phi_{a}\,\phi_{b} \ \tilde G^{\star 0}_{2n-2}(p_1,\dots,\widehat{p}_a, \dots, \widehat{p}_b, \dots, p_{2n}) \ .
\end{equation*}
Since the order of the fields is immaterial, this is the same as the Schwinger--Dyson equation \eqref{eq:SDE_free}, which completes the proof of the braided Wick theorem \eqref{eq:braidedwick}.

\subsection{Interacting theory} 
\label{sub:interacting theory sde}

Consider now the deformation to the quantum differential which defines the interacting quantum field theory. It is given as usual by the small perturbation
\begin{align*}
\mbf{\delta} = -\mathrm{i}\, \hbar\, \mathsf{\Delta}_{\textrm{\tiny BV}}\, - \left\{ \mathcal{S}_{\rm int}, - \right\} \ .
\end{align*}
In this case applying the recursion \eqref{eq:scemidstep} to \eqref{eq:basis for sde} yields
\begin{equation}\label{eq:SDrecursion}
    \tilde G_n^\star(p_1,\dots,p_n) = \tilde{\mathsf{P}}\,( -\mathrm{i}\, \hbar\, \mathsf{\Delta}_{\textrm{\tiny BV}}\, \mathsf{H} )\,(\tte^{p_1}\odot_\star\cdots\odot_\star\tte^{p_n})
    - \mathsf{\tilde{P}}\, \{\mathcal{S}_{\rm int}, \mathsf{H}\,(\tte^{p_1}\odot_\star\cdots\odot_\star\tte^{p_n}) \}_\star  \ ,
\end{equation}
for $n\geqslant1$.

For the term involving the BV differential, we adapt our derivation of \eqref{eq:SDE_free} to the interacting theory. Using the obvious analogue of \eqref{eq:SDE_deltaterm} for $n$ field insertions, and applying $\tilde\sP = \sP + \sP_{\mbf\delta}$ to the remaining products of antifields, the first term on the right-hand side of \eqref{eq:SDrecursion} expands into
\begin{align*}
\tilde{\mathsf{P}}\,( -\mathrm{i}\, \hbar\, \mathsf{\Delta}_{\textrm{\tiny BV}}\, \mathsf{H} )\,(\tte^{p_1}\odot_\star\cdots\odot_\star\tte^{p_n}) = \frac1{n} \, \sum_{i\neq j} \, \e^{\,\ii\,p_j\cdot\theta\,(p_{i+1}+\cdots+p_{j})} \ \bcontraction{}{\phi_i}{}{\phi_j} \phi_i\,\phi_j \ \tilde G^{\star}_{n-2}(p_1,\dots,\widehat{p_i},\dots,\widehat{p_j},\dots,p_{n})
\end{align*}
for $n\geqslant 2$, where again we define $\tilde G_0^\star:=1$. 

For the term involving the interaction functional, we observe that the operator  $\{\mathcal{S}_{\rm int},-\}_\star\,\sH$ acts as strict derivation, due to the momentum conserving delta-functions in \eqref{eq:Vint3} which trivialise any braiding phase factors, and the antibrackets can only pair fields with antifields. Hence
\begin{equation*} 
\begin{split}
    \left\{ \mathcal{S}_{\rm int} , \mathsf{H} \,
    (
    \tte^{p_1}
    \odot_\star \ldots 
    \odot_\star 
    \mathtt{e}^{p_n}) 
    \right\}_\star
    &
    = \frac{1}{n} \,
    \sum_{i=1}^n \,
    \tte^{p_1}
    \odot_\star \cdots \odot_\star 
    \left\{ \mathcal{S}_{\rm int},\mathsf{h}(  \mathtt{e}^{p_i}) \right\}_\star
    \odot_\star \cdots\odot_\star 
    \mathtt{e}^{p_n} \ .
\end{split}
\end{equation*}
Substituting the explicit form of the interaction functional \eqref{IntVert3}, the remaining antibrackets can be evaluated using the braided derivation property to get
\begin{align*}
\big\{\sfh(\tte^{p_i}),\tte^{k_1}\odot_\star\cdots\odot_\star\tte^{k_r}\big\}_\star = \sum_{j=1}^r\,\e^{-\ii\,p_i\cdot(k_1+\cdots+k_{j-1})} \ \langle\sfh(\tte^{p_i}),\tte^{k_j}\rangle_\star \ \tte^{k_1}\odot_\star\cdots\odot_\star \widehat{\tte^{k_j}}\odot_\star\cdots\odot_\star\tte^{k_r} \ ,
\end{align*}
for $r\geqslant3$, where $k_0:=0$ and the inner products are given by
\begin{align*}
\langle\sfh(\tte^{p_i}), \tte^{k_j}\rangle_\star = -\frac{(2\pi)^d \, \delta(k_j+p_i)}{p_i^2-m^2} \ .
\end{align*}
By relabelling the internal momenta $k_j$ and using the braided symmetry of the vertices \eqref{eq:V3braidedsym}, we can eliminate all noncommutative phase factors and render each term in the sum over $j$ equal to the term with $j=r$. 

Applying the projection $\mathsf{\tilde{P}}$, the recursion relation \eqref{eq:SDrecursion} takes the form
\begin{align}\label{eq:fullSDE}
\begin{split}
\tilde G_n^\star(p_1,\dots,p_n) &= \frac1{n} \, \sum_{i\neq j} \, \e^{\,\ii\,p_j\cdot\theta\,(p_{i+1}+\cdots+p_{j-1})} \ \bcontraction{}{\phi_i}{}{\phi_j} \phi_i\,\phi_j \ \tilde G^{\star}_{n-2}(p_1,\dots,\widehat{p_i},\dots,\widehat{p_j},\dots,p_{n}) \\
 & \qquad \, + \sum_{r\geqslant3}  \, \frac rn \ \sum_{i=1}^n \, \frac1{p_i^2-m^2} \ \int_{k_1,\dots,k_{r-1}} \, V_r(k_1,\dots,k_{r-1},-p_i) \\
 & \hspace{4cm} \times \tilde G_{n+r-2}^\star(p_1,\dots,p_{i-1}, k_1,\dots,k_{r-1},p_{i+1},\dots,p_n) \ ,
\end{split}
\end{align}
for $n\geqslant1$, where we define $\tilde G_{-1}^\star := 0$, and the contraction of scalar fields in the first line is the free propagator \eqref{BrWickG2}.
This is our final form of the Schwinger--Dyson equations for a general braided scalar field theory.  The expression \eqref{eq:fullSDE} determines the full $n$-point function recursively in terms of correlation functions of both lower multiplicity $n-2$ and higher multiplicities $n+r-2$ for $r\geqslant3$.

\begin{example}[{\bf Two-point function in braided $\mbf{\lambda\,\phi^3}$-theory}]   
We derive the Schwinger--Dyson  equations for $n=2$ in braided $\phi^3$-theory. We use our diagrammatic calculus, starting from the element  
\begin{eqnarray*}
   \tte^{p_1}\odot_\star \tte^{p_2} = { \footnotesize \begin{tikzpicture}[scale=0.5, baseline]
        \coordinate (k0) at (90:1);
        \coordinate (k1) at ($(k0) +(0:2)$);
        \coordinate[label=below: {$p_1$}] (k0in) at (270:1);
        \coordinate[label=below: {$p_2$}] (k1in) at ($(k0in) +(0:2)$);

        \draw[decoration={markings, mark=at position 0.5 with {\arrow{Latex}}}, postaction={decorate}] (k0in) -- (k0);
        \draw[decoration={markings, mark=at position 0.5 with {\arrow{Latex}}}, postaction={decorate}] (k1in) -- (k1);
    \end{tikzpicture} } \normalsize
\end{eqnarray*}
and apply both sides of \eqref{eq:scemidstep} to it. The left-hand side is the full two-point function
\begin{eqnarray*}
\tilde{\sP}\, (\tte^{p_1} \odot_\star \tte^{p_2})  \ = \ \sP_{\mbf\delta}\, \Bigg( { \footnotesize
    \begin{tikzpicture}[scale=0.5, baseline]
        \coordinate (k0) at (90:1);
        \coordinate (k1) at ($(k0) +(0:2)$);
        \coordinate[label=below: {$p_1$}] (k0in) at (270:1);
        \coordinate[label=below: {$p_2$}] (k1in) at ($(k0in) +(0:2)$);

        \draw[decoration={markings, mark=at position 0.5 with {\arrow{Latex}}}, postaction={decorate}] (k0in) -- (k0);
        \draw[decoration={markings, mark=at position 0.5 with {\arrow{Latex}}}, postaction={decorate}] (k1in) -- (k1);
    \end{tikzpicture} } \normalsize \Bigg)
 \ = \ { \footnotesize \begin{tikzpicture}[scale=0.5, baseline]
        \coordinate[label=right: {$p_2$}] (k0) at (0:2);
        \coordinate[label=left: {$p_1$}] (k0in) at ($(k0)+(180:4)$);
    
        \draw (k0in) -- (k0);
        \draw[fill=blue!50] (0,0) circle[radius=0.5];
\end{tikzpicture} } \normalsize
\end{eqnarray*}

Since $\mbf\delta = -\ii\,\hbar\,\BVL - \{\CS_{\rm int},-\}_\star$ is a sum of two terms, we first act with $\mbf\delta\, \sH$ to write
\begin{eqnarray*}
    \mbf\delta\, \sH\, ( \tte^{p_1} \odot_\star \tte^{p_2} ) &=& -\ii\,\hbar\,\BVL \, \sH\, \Bigg( { \footnotesize
    \begin{tikzpicture}[scale=0.5, baseline]
        \coordinate (k0) at (90:1);
        \coordinate (k1) at ($(k0) +(0:2)$);
        \coordinate[label=below: {$p_1$}] (k0in) at (270:1);
        \coordinate[label=below: {$p_2$}] (k1in) at ($(k0in) +(0:2)$);

        \draw[decoration={markings, mark=at position 0.5 with {\arrow{Latex}}}, postaction={decorate}] (k0in) -- (k0);
        \draw[decoration={markings, mark=at position 0.5 with {\arrow{Latex}}}, postaction={decorate}] (k1in) -- (k1);
    \end{tikzpicture} } \normalsize \Bigg)
    - \Bigg\{\CS _{\rm int}\,,\,
    \sH\, \Bigg( { \footnotesize \begin{tikzpicture}[scale=0.5, baseline]
        \coordinate (k0) at (90:1);
        \coordinate (k1) at ($(k0) +(0:2)$);
        \coordinate[label=below: {$p_1$}] (k0in) at (270:1);
        \coordinate[label=below: {$p_2$}] (k1in) at ($(k0in) +(0:2)$);

        \draw[decoration={markings, mark=at position 0.5 with {\arrow{Latex}}}, postaction={decorate}] (k0in) -- (k0);
        \draw[decoration={markings, mark=at position 0.5 with {\arrow{Latex}}}, postaction={decorate}] (k1in) -- (k1);
    \end{tikzpicture} } \normalsize \Bigg) \Bigg\}_\star \\[4pt]
    &=& { \footnotesize
    \begin{tikzpicture}[scale=0.5, baseline]
        \coordinate (k0) at (0,0);
        \coordinate (k0in) at (270:2);

        \draw[ color=dgreen] (k0in) -- (k0) node[below=10mm]{$G_{p_1 p_2}$};
    \end{tikzpicture}
+  \ \frac{3}{2} \times \begin{tikzpicture}[scale=0.5, baseline]
        \coordinate[label=left: \textcolor{dgreen}{$k$}] (p) at (0,0);
        \coordinate (p3) at (270:2);
        \coordinate[label=right: {$k_2$}] (p2) at (45:2);
        \coordinate[label=left: {$k_1$}] (p1) at (135:2);
        \coordinate[label=below: {$p_2$}] (k1in) at ($(p3) +(0:2)$);
        \coordinate (k1) at ($(p) +(0:2)$);        
        \draw[fill, dgreen] (p) circle[radius=0.1];
        \draw[decoration={markings, mark=at position 0.5 with {\arrow{Latex}}}, postaction={decorate}] (p2) -- (p);
        \draw[decoration={markings, mark=at position 0.5 with {\arrow{Latex}}}, postaction={decorate}] (p1) -- (p);
        \draw[decoration={markings, mark=at position 0.5 with {\arrow{Latex}}}, postaction={decorate}] (k1in) -- (k1);

        \draw[ color=dgreen] (p3) -- (p) node[below=10mm]{$G_{p_1 k_3}$};
    \end{tikzpicture}
+  \ \frac{3}{2} \times \begin{tikzpicture}[scale=0.5, baseline]
        \coordinate[label=right: \textcolor{dgreen}{$k$}] (p) at (0,0);
        \coordinate (p3) at (270:2);
        \coordinate[label=right: {$k_2$}] (p2) at (45:2);
        \coordinate[label=left: {$k_1$}] (p1) at (135:2);
        \coordinate[label=below: {$p_1$}] (k0in) at ($(p3) +(180:2)$);
        \coordinate (k0) at ($(p) +(180:2)$);        
        \draw[fill, dgreen] (p) circle[radius=0.1];
        \draw[decoration={markings, mark=at position 0.5 with {\arrow{Latex}}}, postaction={decorate}] (p2) -- (p);
        \draw[decoration={markings, mark=at position 0.5 with {\arrow{Latex}}}, postaction={decorate}] (p1) -- (p);
        \draw[decoration={markings, mark=at position 0.5 with {\arrow{Latex}}}, postaction={decorate}] (k0in) -- (k0);

        \draw[color=dgreen] (p3) -- (p) node[below=10mm]{$G_{p_2 k_3}$};
    \end{tikzpicture} } \normalsize
\end{eqnarray*}

We now calculate the right-hand side of the Schwinger--Dyson equations to get
\begin{eqnarray*}
\tilde{\sP}\, \mbf\delta\, \sH\, (\tte^{p_1}\odot_\star \tte^{p_2}) &=& (\sP + \sP_{\mbf\delta})\,\Bigg( { \footnotesize
    \begin{tikzpicture}[scale=0.5, baseline]
        \coordinate (k0) at (0,0);
        \coordinate (k0in) at (270:2);

        \draw[ color=dgreen] (k0in) -- (k0) node[below=10mm]{$G_{p_1 p_2}$};
    \end{tikzpicture} } \normalsize
+ \  \frac{3}{2} \times { \footnotesize \begin{tikzpicture}[scale=0.5, baseline]
        \coordinate[label=left: \textcolor{dgreen}{$k$}] (p) at (0,0);
        \coordinate (p3) at (270:2);
        \coordinate[label=right: {$k_2$}] (p2) at (45:2);
        \coordinate[label=left: {$k_1$}] (p1) at (135:2);
        \coordinate[label=below: {$p_2$}] (k1in) at ($(p3) +(0:2)$);
        \coordinate (k1) at ($(p) +(0:2)$);        
        \draw[fill, dgreen] (p) circle[radius=0.1];
        \draw[decoration={markings, mark=at position 0.5 with {\arrow{Latex}}}, postaction={decorate}] (p2) -- (p);
        \draw[decoration={markings, mark=at position 0.5 with {\arrow{Latex}}}, postaction={decorate}] (p1) -- (p);
        \draw[decoration={markings, mark=at position 0.5 with {\arrow{Latex}}}, postaction={decorate}] (k1in) -- (k1);

        \draw[ color=dgreen] (p3) -- (p) node[below=10mm]{$G_{p_1 k_3}$};
    \end{tikzpicture} } \normalsize
+ \  \frac{3}{2} \times { \footnotesize \begin{tikzpicture}[scale=0.5, baseline]
        \coordinate[label=right: \textcolor{dgreen}{$k$}] (p) at (0,0);
        \coordinate (p3) at (270:2);
        \coordinate[label=right: {$k_2$}] (p2) at (45:2);
        \coordinate[label=left: {$k_1$}] (p1) at (135:2);
        \coordinate[label=below: {$p_1$}] (k0in) at ($(p3) +(180:2)$);
        \coordinate (k0) at ($(p) +(180:2)$);        
        \draw[fill, dgreen] (p) circle[radius=0.1];
        \draw[decoration={markings, mark=at position 0.5 with {\arrow{Latex}}}, postaction={decorate}] (p2) -- (p);
        \draw[decoration={markings, mark=at position 0.5 with {\arrow{Latex}}}, postaction={decorate}] (p1) -- (p);
        \draw[decoration={markings, mark=at position 0.5 with {\arrow{Latex}}}, postaction={decorate}] (k0in) -- (k0);

        \draw[color=dgreen] (p3) -- (p) node[below=10mm]{$G_{p_2 k_3}$};
    \end{tikzpicture} } \normalsize
    \Bigg)\\[4pt]
&=& { \footnotesize
    \begin{tikzpicture}[scale=0.5, baseline]
        \coordinate (k0) at (90:1);
        \coordinate (k0in) at (270:1);

        \draw[ color=dgreen] (k0in) -- (k0) node[below=10mm]{$G_{p_1 p_2}$};
    \end{tikzpicture} } \normalsize
+ \  \frac{3}{2} \times { \footnotesize \begin{tikzpicture}[scale=0.5, baseline]
        \coordinate[label=left: \textcolor{dgreen}{$k$}] (k0) at (90:1);
        \coordinate (k0in) at (270:1);

        \draw[fill, dgreen] (k0) circle[radius=0.1];
        \draw[ color=dgreen] (k0in) -- (k0) node[below=10mm]{$G_{p_1 k_3}$};
    \end{tikzpicture} } \normalsize \times \sP_{\mbf\delta}\, \Bigg( { \footnotesize \begin{tikzpicture}[scale=0.5, baseline]
        \coordinate (p1) at (90:1);
        \coordinate (p2) at ($(p1) +(0:2)$);
        \coordinate (k1) at ($(p1) +(0:4)$);
        \coordinate[label=below: {$k_1$}] (p1in) at (270:1);
        \coordinate[label=below: {$k_2$}] (p2in) at ($(p1in) +(0:2)$);
        \coordinate[label=below: {$p_2$}] (k1in) at ($(p1in) +(0:4)$);

        \draw[decoration={markings, mark=at position 0.5 with {\arrow{Latex}}}, postaction={decorate}] (p1in) -- (p1);
        \draw[decoration={markings, mark=at position 0.5 with {\arrow{Latex}}}, postaction={decorate}] (p2in) -- (p2);
        \draw[decoration={markings, mark=at position 0.5 with {\arrow{Latex}}}, postaction={decorate}] (k1in) -- (k1);
    \end{tikzpicture} } \normalsize \Bigg) \\
&&  \qquad\, + \ \frac{3}{2} \times \sP_{\mbf\delta}\, \Bigg( { \footnotesize \begin{tikzpicture}[scale=0.5, baseline]
        \coordinate (k0) at (90:1);
        \coordinate (p1) at ($(k0) +(0:2)$);
        \coordinate (p2) at ($(k0) +(0:4)$);
        \coordinate[label=below: {$p_1$}] (k0in) at (270:1);
        \coordinate[label=below: {$k_1$}] (p1in) at ($(k0in) +(0:2)$);
        \coordinate[label=below: {$k_2$}] (p2in) at ($(k0in) +(0:4)$);

        \draw[decoration={markings, mark=at position 0.5 with {\arrow{Latex}}}, postaction={decorate}] (k0in) -- (k0);
        \draw[decoration={markings, mark=at position 0.5 with {\arrow{Latex}}}, postaction={decorate}] (p1in) -- (p1);
        \draw[decoration={markings, mark=at position 0.5 with {\arrow{Latex}}}, postaction={decorate}] (p2in) -- (p2);
    \end{tikzpicture} \Bigg) } \normalsize \times { \footnotesize \begin{tikzpicture}[scale=0.5, baseline]
        \coordinate[label=right: \textcolor{dgreen}{$k$}] (k1) at (90:1);
        \coordinate (k1in) at (270:1);

        \draw[fill, dgreen] (k1) circle[radius=0.1];
        \draw[color=dgreen] (k1in) -- (k1) node[below=10mm]{$G_{p_2 k_3}$};
    \end{tikzpicture} } \normalsize \\[4pt]
&=& { \footnotesize \begin{tikzpicture}[scale=0.5, baseline]
        \coordinate[label=right: {$p_2$}] (k0) at (0:1);
        \coordinate[label=left: {$p_1$}] (k0in) at ($(k0)+(180:2)$);

        \draw[] (k0in) -- (k0);
    \end{tikzpicture} } \normalsize
+ \  \frac{3}{2} \times { \footnotesize \begin{tikzpicture}[scale=0.5, baseline]
    \coordinate (p) at (0:0.5);
    \coordinate (k) at (180:0.5);

    \draw[decoration={markings, mark=at position 0.35 with {\arrow{Latex}}}, postaction={decorate}] ($(k)+(180:2)$) node[left]{$p_1$} -- (k);
    \draw (k) -- ($(p)+(0:2)$) node[right, pos=1]{$p_2$};
    \draw[fill=white] (k) circle[radius=1];
    \draw[fill=blue!50] (p) circle[radius=0.5];
    \end{tikzpicture} } \normalsize
+ \  \frac{3}{2} \times { \footnotesize \begin{tikzpicture}[scale=0.5, baseline]
    \coordinate (p) at (0:0.5);
    \coordinate (k) at (180:0.5);

    \draw ($(k)+(180:2)$) node[left]{$p_1$} -- (k);
    \draw[decoration={markings, mark=at position 0.9 with {\arrow{Latex[reversed]}}}, postaction={decorate}] (k) -- ($(p)+(0:2)$) node[right, pos=1]{$p_2$};
    \draw[fill=white] (p) circle[radius=1];
    \draw[fill=blue!50] (k) circle[radius=0.5];
    \end{tikzpicture} } \normalsize
\end{eqnarray*}

Finally we obtain the $n=2$ Schwinger--Dyson equation
\begin{eqnarray*}
  { \footnotesize  \begin{tikzpicture}[scale=0.5, baseline]
        \coordinate[label=right: {$p_2$}] (k0) at (0:2);
        \coordinate[label=left: {$p_1$}] (k0in) at ($(k0)+(180:4)$);
    
        \draw (k0in) -- (k0);
        \draw[fill=blue!50] (0,0) circle[radius=0.5];
\end{tikzpicture} } \normalsize
 \ = \ { \footnotesize  \begin{tikzpicture}[scale=0.5, baseline]
        \coordinate[label=right: {$p_2$}] (k0) at (0:1);
        \coordinate[label=left: {$p_1$}] (k0in) at ($(k0)+(180:2)$);

        \draw (k0in) -- (k0);
    \end{tikzpicture} } \normalsize
 \ + \  \frac{3}{2} \times { \footnotesize \begin{tikzpicture}[scale=0.5, baseline]
    \coordinate (p) at (0:0.5);
    \coordinate (k) at (180:0.5);

    \draw[decoration={markings, mark=at position 0.35 with {\arrow{Latex}}}, postaction={decorate}] ($(k)+(180:2)$) node[left]{$p_1$} -- (k);
    \draw (k) -- ($(p)+(0:2)$) node[right, pos=1]{$p_2$};
    \draw[fill=white] (k) circle[radius=1];
    \draw[fill=blue!50] (p) circle[radius=0.5];
    \end{tikzpicture} } \normalsize
 \ + \  \frac{3}{2} \times { \footnotesize \begin{tikzpicture}[scale=0.5, baseline]
    \coordinate (p) at (0:0.5);
    \coordinate (k) at (180:0.5);

    \draw ($(k)+(180:2)$) node[left]{$p_1$} -- (k);
    \draw[decoration={markings, mark=at position 0.9 with {\arrow{Latex[reversed]}}}, postaction={decorate}] (k) -- ($(p)+(0:2)$) node[right, pos=1]{$p_2$};
    \draw[fill=white] (p) circle[radius=1];
    \draw[fill=blue!50] (k) circle[radius=0.5];
    \end{tikzpicture} } \normalsize
\end{eqnarray*}
which unravels into the analytic expression
\begin{align}\label{eq:sde-scalar-3}
\tilde G_2^\star(p_1,p_2) &= \bcontraction{}{\phi_1}{}{\phi_2} \phi_1\,\phi_2 + \frac\lambda4 \, \int_k\ \frac{\e^{\,\frac\ii2\,k\cdot\theta\,p_1}}{p_1^2-m^2} \ \tilde G_3^\star(k,p_1-k,p_2) + \frac\lambda4 \, \int_k\ \frac{\e^{\,\frac\ii2\,k\cdot\theta\,p_2}}{p_2^2-m^2} \ \tilde G_3^\star(p_1,k,p_2-k) \ .
\end{align}
Thus the full propagator is determined recursively from the three-point vertex.
After multiplying \eqref{eq:sde-scalar-3} by $p_1^2-m^2=p_2^2-m^2$, the first term is identified as the contact term and the commutative limit $\theta=0$ is the usual Schwinger--Dyson equation for $\lambda\,\phi^3$-theory.  
\end{example}

\section{Discussion and outlook}
\label{sec:discussion}

In this paper we studied the simplest class of examples of a braided noncommutative field theory: a real massive scalar field $\phi$  in $d$ dimensions with polynomial interactions and the Moyal--Weyl twist deformation. We used the framework of braided  $L_\infty$-algebras and homological perturbation theory. 

We studied the example of braided $\lambda\,\phi^3$-theory in detail. We derived explicit expressions for correlation functions up to two-loops and multiplicity three. Similarly to the commutative theory, the vacuum expectation value of the field $\phi$ is different from zero. This manifests itself in a non-zero divergent tadpole diagram. We removed the tadpole diagram by extension to the effective theory based on a curved $L_\infty$-algebra, with a suitable central curvature that retains an interacting theory based on the same cochain complex of free fields as the flat case.

In addition to computational techniques of homotopical algebra, a novel graphical method for calculating correlation functions in homological perturbation theory was presented. This method significantly simplifies long calculations in higher order correlation functions and at higher loops. Both the analytic and diagrammatic methods lead to the same results: The two-point function up to two-loop order remains equal the corresponding commutative two-point function, while the three-point function has a noncommutative correction in the form of a phase factor in external momenta. From these results we conclude that, up to two-loops and multiplicity three, no UV/IR-mixing is present in braided $\lambda\,\phi^3$-theory. This is consistent with results reported in \cite{Oeckl:2000eg,Balachandran:2005pn,Bu:2006ha,Fiore:2007vg} through somewhat adhoc and less systematic computations than our rigorous formalism, but different from the standard noncommutative scalar field theory based on Feynman perturbation theory~\cite{Minwalla:1999px,Szabo:2001kg}. 

We derived an algebraic version of the Schwinger--Dyson equations in (\ref{eq:fullSDE}) for arbitrary braided scalar field theories. Noncommutative corrections again appear as phase factors in the corresponding external momenta. Unlike the standard commutative Schwinger--Dyson equations that follow from the path integral formalism and single out a specific field insertion, the braided $L_\infty$-algebra formulation averages the usual Schwinger--Dyson equations over all permutations of the field insertions, leaving an expression that is manifestly braided symmetric. In the case of a free scalar field theory, we used the braided Schwinger--Dyson equations to give a rigorous proof of the braided Wick theorem proposed in~\cite{DimitrijevicCiric:2023hua}. 

From the results obtained in the present paper we can conclude that the simple model of braided $\lambda\,\phi^3$-theory, up to two-loop order and multiplicity three, is free from UV/IR-mixing; this was also confirmed for the one-loop two-point function for braided $\phi^4$-theory in~\cite{DimitrijevicCiric:2023hua}. Nevertheless, the correlation functions of braided scalar field theories satisfy a braided noncommutative deformation of the usual Schwinger--Dyson equations. Our works~\cite{DimitrijevicCiric:2023hua, SQEDInPreparation} suggest that braided abelian gauge theories are also free from UV/IR-mixing. We plan to extend our work to braided non-abelian gauge theories and investigate their correlation functions as well as UV/IR-mixing. 

In addition to correlation functions, it is also very important to study scattering amplitudes for a given braided field theory, as these encode the physical processes in the model. We will report in the future on the construction of amplitudes in braided field theories by using homological perturbation theory to develop a braided version of the perturbiner expansion. Quantum perturbiner expansions for both scattering amplitudes and correlation functions at arbitrary loop order are defined in~\cite{Lee:2022aiu} using Schwinger--Dyson equations. Given the availability of a braided version of the latter, as demonstrated in the present paper, this could prove to be a promising direction for constructing recursion relations for loop-level amplitudes and correlators of braided quantum field theories.

\subsection*{Acknowledgements}

We thank Larisa Jonke for discussions. This article is based upon work from COST Actions CaLISTA CA21109 and THEORY-CHALLENGES CA22113 supported by COST (European Cooperation in Science and Technology). Part of this work was carried out during the Mini-Symposium on Geometry and Physics at the Rudjer Bo\v{s}kovi{\'c} Institute in Zagreb, Croatia; {\sc Dj.B., M.D.C.} and {\sc R.J.S.} are grateful to the organisors for hospitality and for providing a stimulating environment. The work of {\sc Dj.B., M.D.C.} and  {\sc V.R.} is supported by Project 451-03-66/2024-03/200162 of the Serbian Ministry of Science,
Technological Development and Innovation. The work of M.D.C and R.J.S. was partially supported by the Croatian Science Foundation Project IP-2019-04-4168. The work of {\sc G.T.} is supported by the Doctoral Training Partnership Award ST/T506114/1 from the UK Science and Technology Facilities Council. 

\appendix

\renewcommand{\theequation}{\Alph{section}.\arabic{equation}}
\setcounter{equation}{0}
\section{Braided Batalin--Vilkovisky quantization}
\label{app:BV}

In this appendix we briefly discuss braided $L_\infty$-algebras of field theory and quantization via the procedure of homotopy transfer. Further details can be found in \cite{DimitrijevicCiric:2021jea,Nguyen:2021rsa,Giotopoulos:2021ieg,DimitrijevicCiric:2023hua}. Here we generalize the treatment to include curving, adapting the analogous formalism for curved $A_\infty$-algebras from~\cite{Masuda:2020tfa,Okawa:2022sjf}. We recommend the review~\cite{Kraft} for the relevant mathematical background on the classical theory, which we extend to the braided setting in this paper.

Throughout this paper, degree shifting of a $\mathbbm{Z}$-graded vector space $V = \bigoplus_{k\in\RZ}\,V_k$ is performed under the convention $(V[l])_k = V_{k+l}$. 
Consequently, linear maps of degree $l$ between graded vector spaces $V$ and $W$ are written as $V \longrightarrow W[l]$, unless specified otherwise.
We assume the Einstein summation convention over repeated upper and lower indices.

\subsection{Flat and curved braided \texorpdfstring{$L_\infty$}\ -algebras}

Homotopy Lie algebras are vast generalisations of differential graded Lie algebras, where the antisymmetry of the binary bracket operation holds strictly, while the Jacobi identity only holds weakly, that is, up to coherent homotopies of higher operations. A braided homotopy Lie algebra is a homotopy Lie algebra in a symmetric monoidal category with a non-trivial braiding isomorphism, rather than the usual category of vector spaces which has trivial braiding. In this paper we work in the representation category of a triangular Hopf algebra where our braiding appears concretely through an associated $\RR$-matrix via the Drinfel'd twist formalism, as discussed at the beginning of Section~\ref{sub:braidedphi3}. However, the discussion which follows also encompasses more general cases which do not necessarily arise from deformations of classical structures. See~\cite{DimitrijevicCiric:2021jea,Nguyen:2021rsa,Giotopoulos:2021ieg} for further details.

A \textit{braided $L_\infty$-algebra} (over $\FR$) is a $\mathbbm{Z}$-graded real vector space {$V=\bigoplus_{k\in \mathbbm{Z}}\, V_{k}$} with a sequence of graded braided antisymmetric multilinear maps $\ell_n^\star$ of degree $2-n$ for $n\geqslant 0$, called \textit{$n$-brackets}:
\begin{align}
    \ell^\star_{n}:  V^{\otimes n} \longrightarrow &\, V[2-n] \ , \quad  v_1\otimes \cdots\otimes v_n \longmapsto 
    \ell^\star_{n} 
    (v_1,\dots,v_n)  \ , \nn\\[4pt]
    \ell^\star_{n} (\dots, v,v',\dots) &\,=\>  -(-1)^{|v|\,|v'|} \ \ell^\star_{n} \big(\dots, \sfR_\alpha (v'),\sfR^\alpha( v), \ldots\big) \ ,\nn
\end{align}
where $|v|$ denotes the degree of a homogeneous element $v\in V$. 

The $n$-brackets are required to fulfil \emph{braided homotopy Jacobi identities} $ \mathcal{J}^\star_n(v_1, \ldots, v_n) = 0$ for all $n\geqslant0$ and $v_i\in V$, where
\begin{equation}\label{eq:BraidedJacobi_curved}
\begin{split}
    \mathcal{J}^\star_n :=
    \sum_{i=0}^n \, (-1)^{i\,(n-i)}  \ 
    \ell_{n+1-i}^\star \, \circ \, \big( \ell_i^\star \otimes \mathrm{id}_V^{\,\otimes\, n-i}\big) \ \circ \ 
    \sum_{\sigma \in \mathrm{Sh}(i;n)}\, \mathrm{sgn}(\sigma) \ \sigma_\RR \ .
    \end{split}
\end{equation}
The shuffle group $\mathrm{Sh}(i;n) \subset S_n$ represents how two decks of cards of size $(i, n-i)$ are shuffled together once, with $\mathrm{Sh}(0; n)$ consisting solely of the identity. The permutation $\sigma_\RR$ acts by braided transposition, and the sign of a permutation ${\rm sgn}(\sigma)$ is given by the Koszul convention, taking into account the gradings of the elements of $V$.

A  braided $L_\infty$-algebra is \textit{flat} if $\ell_0^\star=0$. Then the braided homotopy Jacobi identities have sums running over $i=1,\dots,n$, and the first three relations are given by
\begin{align}
&\underline{n=1}:  \quad \ell^\star_{1}\big(\ell^\star_{1}(v)\big) = 0 \ ,\label{Cochain}\\[4pt]
&\underline{n=2}:  \quad \ell^\star_{1}\big(\ell^\star_{2}(v_1,v_2)\big) = \ell^\star_{2}\big(\ell^\star_{1}(v_1),v_2\big) + (-1)^{|v_1|} \ 
\ell^\star_{2}\big(v_1, \ell^\star_{1}(v_2)\big)\ , \label{Leibniz}\\[4pt]
&\underline{n=3}: \quad  \ell_{1}^\star\big(\ell^\star_{3}(v_1,v_2,v_3)\big) =  - \ell^\star_{3}\big(\ell^\star_{1}(v_1),v_2,v_3\big) - (-1)^{|v_1|} \ 
\ell^\star_{3}\big(v_1, \ell^\star_{1}(v_2), v_3\big) \nn\\
& \hspace{5cm}- (-1)^{|v_1|+|v_2|} \ \ell^\star_{3}\big(v_1,v_2, \ell^\star_{1}(v_3)\big) -\ell^\star_{2}\big(\ell^\star_{2}(v_1,v_2),v_3\big)\nn\\
&\hspace{5cm} -(-1)^{(|v_1|+|v_2|)\,|v_3|} \  \ell^\star_{2}\big(\ell^\star_{2}(\sfR_\alpha\,\sfR_\beta( v_3),\sfR^\beta (v_1)), \sfR^\alpha (v_2)\big) \label{eq:homotopyJacobi} \\
&\hspace{5cm} -(-1)^{(|v_2|+|v_3|)\,|v_1|} \ \ell^\star_{2}\big(\ell^\star_{2}(\sfR_\alpha( v_2), \sfR_\beta( v_3)), \sfR^\alpha\,\sfR^\beta (v_1)\big) \ . \nn
\end{align}
In particular, a flat braided $L_\infty$-algebra has an underlying cochain complex $(V, \ell^\star_1)$, and the $2$-bracket $\ell_2^\star:V\otimes V\longrightarrow V$ is a cochain map.
We recognise the third relation as the usual braided graded Jacobi identity up to the coboundary of the $3$-bracket $\ell_3^\star$. 

A braided $L_\infty$-algebra is \textit{curved} if the lowest map $\ell_0^\star : V^{\otimes 0} \simeq \mathbbm{R} \longrightarrow V[2]$ is non-zero.
By linearity this map is uniquely fixed by its image at $1 \in \mathbbm{R}$, called the \emph{curvature}.
The first three braided homotopy Jacobi identities are now given by 
\begin{align*}
& \hspace{-9mm} \underline{n=0}: \quad \ell^\star_1\big( \ell^\star_0(1)\big) = 0 \ , \\[4pt]
& \hspace{-9mm} \underline{n=1}: \quad \ell_1^\star\big(\ell_1^\star(v)\big) = -\ell_2^\star\big( \ell_0^\star(1), v\big) \ , \\[4pt]
& \hspace{-9mm} \underline{n=2}: \quad \ell_{1}^\star\big(\ell^\star_{2}(v_1,v_2)\big)  + \ell_3^\star(\ell^\star_0(1), v_1, v_2)
    = \ell^\star_{2}\big(\ell^\star_{1}(v_1),v_2\big) + (-1)^{|v_1|} \ 
    \ell^\star_{2}\big(v_1, \ell^\star_{1}(v_2)\big) \ .
\end{align*}
In particular, the curvature $\ell^\star_0(1)$ is `constant' in the sense that it is $\ell^\star_1$-closed. The curvature controls the deviation of the $1$-bracket $\ell_1^\star$ from a differential and a strict graded derivation of the $2$-bracket~$\ell_2^\star$.

A {braided} $L_\infty$-algebra is \emph{cyclic} if it additionally comes with a graded braided symmetric and non-degenerate bilinear map $\langle-,-\rangle_\star:V\otimes V\longrightarrow\mathbbm{R}[-3]$ satisfying
\begin{align}\label{Pairing}
\begin{split}
    &\langle v_0\,,\,\ell^\star_n(v_1,\dots,v_{n-1},v_n)\rangle_\star \\[4pt]
    & \hspace{3cm} = 
    \pm
     \ \langle\, \sfR_{\alpha_0}\,\sfR_{\alpha_1}\cdots\sfR_{\alpha_{n-1}}(v_n)\,,\, 
    \ell_n^\star(
    \sfR^{\alpha_0}(v_0),\sfR^{\alpha_1}(v_1),\dots,\sfR^{\alpha_{n-1}}(v_{n-1}))\,
    \rangle_\star \ , 
\end{split}
\end{align}
for all $n\geqslant 1$ and $v_i\in V$. 
Here and in the following we write $\pm$ for the Koszul signs arising from the gradings  when permuting elements involved in all operations. 

\subsection{Batalin--Vilkovisky formalism}

As discussed in \cite{Hohm:2017pnh,Jurco:2018sby}, the data of any classical perturbative field theory are completely encoded in a corresponding $L_\infty$-algebra, through its Maurer--Cartan theory. The underlying graded vector space $V=\bigoplus_{k\in\RZ}\,V_k$ contains the gauge and higher gauge parameters for $k\leqslant0$, $V_1$ contains the physical fields, $V_2$ contains the equations of motion, while $V_k$ for $k\geqslant3$ contain the second Noether and higher second Noether identities. Lagrangian field theories further involve a cyclic structure which dually pairs fields in degrees $k\leqslant1$ with their \emph{antifields} in degrees $3-k\geqslant2$.

From the data of a classical Lagrangian field theory, the Batalin--Vilkovisky (BV) formalism constructs the \emph{derived} space of classical observables of the theory. Starting from a braided  $L_\infty$-algebra $(V,\{\ell^\star_n\})$ with a cyclic structure $\langle-,-\rangle_\star$ of degree~$-3$, this is the braided commutative algebra $(\Sym_\RR(V[1])^*, \odot_\star)$, viewed as the space of braided symmetric polynomials on the vector space $V[1]$. Using the non-degenerate canonical symplectic pairing, we identify the dual $V^*\simeq V[1]$ and
\begin{align}\nn
    \Sym_\RR(V[1])^*\simeq \Sym_\RR(V[2]) \ .
\end{align}

This data is supplemented by the antibracket, which is  the graded braided Poisson structure 
\begin{align*}
\{-,-\}_\star:\Sym_\RR (V[2])\otimes \Sym_\RR (V[2])\longrightarrow \Sym_\RR (V[2])[1] \ ,
\end{align*}
defined on generators $v_1,v_2\in V[2]$ by
\begin{align*}
\{v_1,v_2\}_\star = \langle v_1,v_2\rangle_\star \ ,
\end{align*}
and extended as a  graded braided derivation on $\Sym_\RR (V[2])$ in each of its slots; for example
\begin{align}\label{eq:der}
\{a_1,a_2\odot_\star a_3\}_\star = \{ a_1, a_2\}_\star\odot_\star a_3 \pm \sfR_\alpha (a_2) \odot_\star\{ \sfR^\alpha (a_1),a_3\}_\star \ ,
\end{align}
for $a_1, a_2, a_3\in \Sym_\RR V[2]$.

Finally, the BV formalism constructs a differential $$Q_\BV: \mathrm{Sym}_\mathcal{R} V[2] \longrightarrow (\mathrm{Sym}_\mathcal{R} V[2])[1]$$ which is a strict graded derivation of the antibracket $\{-,-\}_\star$, while at the same time a braided graded derivation of the braided graded commutative product $\odot_\star$. The cohomology of $Q_\BV$ in degree~$0$ simultaneously encodes the quotients of the space of physical fields by the equations of motion and by the action of gauge transformations. The derived space of classical observables is thus fully specified by the triple $\big(\Sym_\RR(V[2]),Q_\BV, \{-,-\}_\star\big)$, which has the structure of a braided $P_0$-algebra~\cite{Nguyen:2021rsa}.

Moving between the descriptions in terms of braided $L_\infty$-algebras and braided $P_0$-algebras is facilitated with the introduction of \emph{contracted coordinate functions}~\cite{Jurco:2018sby}
$$\xi := \tte^k\otimes \tte_k  \ \in  \ \Sym_\RR(V[2])\otimes V$$ of degree $1$, for a basis $\{\tte_k\}\subset V$ with corresponding dual basis $\{\tte^k\}\subset V^*\simeq V[3]$ satisfying $\langle\tte^k,\tte_l\rangle_\star = \delta^k_l$ for all $k,l$. 
The braided $L_\infty$-structure on $V$ naturally extends to the tensor product $\Sym_\RR(V[2])\otimes V$ through the extended brackets
\begin{align*}
\begin{split}
\ell_0^{\star\,{\rm ext}}(1) &= 1\otimes\ell_0^\star(1) \ , \\[4pt]
    \ell_1^{\star\,\rm ext}(a_1\otimes v_1) &= \pm\,a_1\otimes\ell^\star_1(v_1) \ , \\[4pt]
    \ell_n^{\star\, {\rm ext}}(a_1\otimes v_1,\dots,a_n\otimes v_n) &= \pm\,\big(a_1\odot_\star\sfR_{\alpha^1_1}(a_2)\odot_\star\cdots\odot_\star \sfR_{\alpha_{n-1}^1}\cdots\sfR_{\alpha_1^{n-1}}(a_n)\big) \\
    & \hspace{2cm} \otimes\,\ell_n^\star\big(\sfR^{\alpha_{n-1}^1}\cdots\sfR^{\alpha_1^1}(v_1),\dots,\sfR^{\alpha_1^{n-1}}(v_{n-1}),v_n\big) \ ,
\end{split}
\end{align*}
for all $n\geqslant2$, $a_1,\dots,a_n\in\Sym_\RR V[2]$ and $v_1,\dots,v_n\in V$.
Similarly, the cyclic structure is extended via the $\Sym_\RR V[2]$-valued pairing
\begin{align*}
\langle a_1\otimes v_1,a_2\otimes v_2\rangle^{\,\rm ext}_\star = \pm\, \big(a_1\odot_\star\sfR_\alpha(a_2)\big) \, \langle\sfR^\alpha(v_1),v_2\rangle_\star \ .
\end{align*}

The BV functional is obtained from the contracted coordinate functions as the analogue of the curved braided Maurer--Cartan functional
\begin{align*}
    \CS_\BV = \sum_{n\geqslant 0} \, \frac{(-1)^{n\choose 2}}{(n+1)!} \, \langle\xi,\ell_n^{\star\,{\rm ext}}(\xi^{\otimes n})\rangle_\star^{\rm ext} \ \in \ (\Sym_\RR V[2])_0 \ ,
\end{align*}
where \smash{${0\choose 2} :=0$}.
Using the antibracket, this generates the BV differential on the space of observables as
\begin{align*} 
    Q_\BV = \{\CS_\BV,-\}_\star \ ,
\end{align*}
whose nilpotency $(Q_\BV)^2=0$ is equivalent to the {classical master equation}
\begin{align*}
    \{\CS_\BV,\CS_\BV\}_\star = 0 \ .
\end{align*}
This is equivalent to the braided homotopy Jacobi identities of the cyclic braided $L_\infty$-algebra $(V,\{\ell_n^\star\},\langle-,-\rangle_\star)$, while compatibility with the antibracket $\{-,-\}_\star$ is a consequence of cyclicity of the inner product $\langle-,-\rangle_\star$.

\subsection{Homological perturbation lemma}

Let us now explain how quantisation works in the framework of the BV formalism. If the curvature $\ell_0^\star(1)$ is non-zero, then $\ell_1^\star$ is not a differential in general, unless $\ell_0^\star(1)$ is \emph{central}:
\begin{align*}
\ell_{n+1}^\star\big(\ell_0^\star(1),v_1,\dots,v_n\big) = 0 \ ,
\end{align*}
for all $n\geqslant1$ and $v_i\in V$. This is a necessary and sufficient condition guaranteeing $(\ell_1^{\star})^{2}=0$. In this case we may study the field theory as a perturbation, which includes the constant curvature, around the solutions to its linearized equations of motion. 

The underlying cochain complex $(V, \ell_1^\star)$ of the classical theory is homotopy equivalent to its minimal model $(H^\bullet(V), 0)$ through a diagram
\begin{equation}\label{SDR1}
\begin{tikzcd}
    \arrow[out=200,in=160,loop,looseness=3,"\sfh"] (V,\ell^\star_1) \ar[r,shift right=1ex,swap,"\sfp"] & \ar[l,shift right=1ex,swap,"\iota"] \big(H^\bullet(V),0\big) 
\end{tikzcd} \ .
\end{equation}
A homotopy equivalence is a set of maps $\mathsf{h} : V \longrightarrow V[-1]$ such that the degree $0$ cochain maps $\mathsf{p} : V \longrightarrow H^\bullet(V)$ and $\iota: H^\bullet(V)\longrightarrow V $ decompose the endomorphism $\iota\circ \mathsf{p}$ into a Hodge--Kodaira relation
\begin{equation}\label{eq:hodge-kodaira}
    \iota \circ \mathsf{p} = \mathrm{id}_{{V}} + \ell^\star_1 \circ\mathsf{h} + \mathsf{h} \circ\ell^\star_1 \ .
\end{equation}
This can always be refined to an increasingly stronger set of conditions:
\begin{myitemize}
    \item $\sfp \circ \iota = \id_{H^\bullet(V)}$, making the data a {deformation retract}. 
    \item $\sfh^2=0$,  $\sfh\circ \iota=0$ and $\sfp\circ \sfh=0$, making the data a {strong deformation retract}.
\end{myitemize}
In the Drinfel'd twist formalism, the kinetic term is unchanged in the braided theory, and so the cochain complexes of the deformed and undeformed theories are isomorphic.
In other words, the braided and unbraided free field theories are classically equivalent.

The strong deformation retract structure extends to the braided $P_0$-algebra of classical observables $\big(\Sym_\RR V[2],\ell^{\star}_1, \{-,-\}_\star\big)$ of the \textit{free} theory, where here $\ell_1^{\star}$ is the linear part of the BV differential $Q_\BV$ extended from the differential of the $L_\infty$-structure as a strict graded derivation of symmetric degree~$0$.
Using the symmetric tensor trick (reviewed in e.g.~\cite{Berglund2014}), we obtain a strong deformation retract on the space of free classical observables.
This leads to the diagram
\begin{equation}\label{SDR2}
\begin{tikzcd}
    \arrow[loop left, distance=2em, start anchor={[yshift=-1ex]west}, end anchor={[yshift=1ex]west}]{}{\mathsf{H}} 
    \big(\Sym_\RR(V[2]),\ell_1^{\star}\big) \ar[r,shift right=1ex,swap,"\sP"] & \ar[l,shift right=1ex,swap,"\sI"] \big(\Sym_\RR (H^\bullet(V[2])),0\big) 
\end{tikzcd} \ ,
\end{equation}
where the maps $\sP$, $\sI$ and $\sH$ are extensions of the corresponding maps in \eqref{SDR1} to braided symmetric algebras using the symmetric tensor trick.

To incorporate interactions and quantize the theory, we use the homological perturbation lemma, see e.g.~\cite{2004Crainic,Doubek:2017naz}, adapted to the braided setting in~\cite{Nguyen:2021rsa}. 
The differential $\ell_1^{\star}$ is perturbed to the differential $Q_{\mbf\delta} = \ell_1^{\star} + \mbf\delta$, where $\mbf\delta$ is a small perturbation in the sense that the map \smash{$\mathrm{id}_{\Sym_\RR(V[2])} - \mbf\delta\, \mathsf{H}$} is invertible.
This leads to a new strong deformation retract with new maps $\tilde{\sP}$, $\tilde{\sI}$ and $\tilde{\sH}$ fitting into the diagram
\begin{equation}\label{SDR3}
\begin{tikzcd}
\arrow[loop left, distance=2.2em, start anchor={[yshift=-1ex]west}, end anchor={[yshift=1ex]west}]{}{\mathsf{\tilde{H}}} 
\big(\Sym_\RR(V[2]),Q_{\mbf\delta}\big) \ar[r,shift right=1ex,swap,"\tilde{\sP}"] & \ar[l,shift right=1ex,swap,"\tilde{\sI}"] 
\big( \Sym_\RR (H^\bullet (V[2] )\,)\, ,\, \tilde{\mbf\delta}\,\big) 
\end{tikzcd}\ .
\end{equation}
In this paper we only need the explicit form of the deformed projection map, which is given by $\tilde{\sP}= \sP + \sP_{\mbf\delta}$ with
\begin{align} \label{eq:BQFTnpoint}
    \sP_{\mbf\delta} = \sP\,\big(\id_{\Sym_\RR(V[2])} - {\mbf\delta}\,\sH\big)^{-1} \, \mbf\delta \, \sH \ .
\end{align}
In the classical setting, it was shown in \cite{Doubek:2017naz} that $\sP_{\mbf\delta}$ computes the path integral. 

\subsection{Braided Laplacian}

The free braided quantum field theory is defined by the perturbation $\mbf\delta_0 = - \ii\,\hbar\,\BVL$ with the BV Laplacian $$\BVL:\Sym_\RR ( V[2])\longrightarrow \Sym_\RR  (V[2])[1] \ .$$
The BV Laplacian satisfies the two key properties $(\BVL)^2=0$ and $\BVL\circ\ell_1^\star = -\ell_1^\star\circ\BVL$ which guarantee that $Q_{\mbf\delta_0}=\ell^\star_1 - \ii\,\hbar\,\BVL$ is a differential, $(Q_{\mbf\delta_0})^2=0$. 
Its failure to derive the braided graded commutative product is measured by the antibracket through
\begin{align}\label{eq:BVLantibracket}
    \BVL(a_1\odot_\star a_2) = \BVL(a_1)\odot_\star a_2 + (-1)^{\vert a_1\vert}\, 
    a_1\odot_\star\BVL(a_2) +  \{a_1,a_2\}_\star \ ,
\end{align}
for all $a_1,a_2\in \Sym_\RR (V[2])$. Equivalently, $\BVL$ is a strict graded derivation of the antibracket.

Explicitly, denoting the identity by $1 \in (\mathrm{Sym}_{\mathcal{R}} V[2])_0$, the BV Laplacian is defined by its images on generators through
\begin{align}\label{eq:BVL}
    \BVL(1)=0 & \qquad , \qquad \BVL(v_1)=0 \qquad , \qquad  \BVL(v_1\odot_\star v_2) = \langle v_1, v_2\rangle_\star  \ , \nn\\[4pt]
    \BVL( v_1\odot_\star\cdots\odot_\star v_n) & = \sum_{i<j} \, \pm \, \langle  v_i,\sfR_{\alpha_{i+1}}\cdots\sfR_{\alpha_{j-1}}( v_j)\rangle_\star \   v_1\odot_\star\cdots\odot_\star v_{i-1}\\ 
    & \hspace{1.2cm} \odot_\star\sfR^{\alpha_{i+1}}( v_{i+1})\odot_\star\cdots\odot_\star\sfR^{\alpha_{j-1}}( v_{j-1})\odot_\star  v_{j+1}\odot_\star\cdots\odot_\star v_n \ ,\nn
\end{align}
for all $ v_1,\dots, v_n\in V[2]$. 

The interacting braided quantum field theory is defined by taking $\mbf\delta = -\ii\,\hbar\,\BVL - \{\CS _{\rm int},-\}_\star$, where $\CS_{\rm int}$ consists of the curved and interacting parts of the BV functional $\CS_\BV$:
\begin{align*}
\CS_{\rm int} = \langle\xi,\ell_0^{\star\,{\rm ext}}(1)\rangle_\star^{\rm ext}  + \sum_{n\geqslant 2} \, \frac{(-1)^{n\choose 2}}{(n+1)!} \, \langle\xi,\ell_n^{\star\,{\rm ext}}(\xi^{\otimes n})\rangle_\star^{\rm ext} \ .
\end{align*}
The classical master equation together with  $\BVL(\CS_\BV)=0$ ensure that $Q_{\mbf\delta}$ is a differential.
Then \eqref{eq:BQFTnpoint} is a formal power series in $\hbar$ and in the coupling constants appearing in $\CS_{\rm int}$, representing the perturbative expansion.

\section{Glossary of diagrams}
\label{subsub:pictograms}

In this appendix we present the analytic expressions for the diagrams that appear in the loop calculations of Section~\ref{sec:phi3corrs} for braided $\lambda\,\phi^3$-theory using our graphical calculus. The explicit formulas up to two loops and multiplicity three are given as follows.

\bigskip\bigskip

\centerline{\underline{Multiplicity one}}
\vspace{-5mm}
\begin{eqnarray*}
  { \footnotesize  \begin{tikzpicture}[scale=0.5, baseline]
        \coordinate (k) at (0,0);
        \coordinate[label=above left: $p$] (p) at (180:3);

        \draw[decoration={markings, mark=at position 0.5 with {\arrow{Latex}} }, postaction={decorate}] (p) -- (k);
        \draw[decoration={markings }, postaction={decorate}] ($(k)+(0:1)$) circle (1);
    \end{tikzpicture} } \normalsize &=& -\frac{\lambda}{3!} \, \ii\,\hbar \,  (2 \pi)^d\, \delta(p)\, \tilde \sgreen(p)\, \int_k\, \tilde \sgreen(k) 
    \end{eqnarray*}
    
    \bigskip\bigskip
    
    \centerline{\underline{Multiplicity two}}
    \vspace{-5mm}
\begin{eqnarray*}
  {\footnotesize  \begin{tikzpicture}[scale=0.5, baseline]
        \coordinate[label=above:$p_1$] (p1) at (180:2.5);
        \coordinate[label=above:$p_2$] (p2) at (0:2.5);

        \draw[decoration={markings }, postaction={decorate}] (p1) -- (p2);
    \end{tikzpicture} } \normalsize &=& (2 \pi)^d\, \delta(p_1 + p_2)\, \tilde \sgreen(p_1) \end{eqnarray*}
\begin{eqnarray*}   {\footnotesize \begin{tikzpicture}[scale=0.5, baseline]
        \coordinate (k) at (-0.5,0);
        \coordinate (l) at (0.5,0);
        \draw[decoration={markings, mark=at position 0.5 with {\arrow{Latex[reversed]}} }, postaction={decorate}] (k) -- ($(k) + (180:2)$) node[above]{$p_1$};
        \draw[decoration={markings, mark=at position 0.5 with {\arrow{Latex[reversed]}} }, postaction={decorate}] (l) -- ($(l) + (0:2)$) node[above]{$p_2$};
        \draw ($(k)+(0:0.5)$) circle (0.5);    
    \end{tikzpicture} } \normalsize &=& \Big( -\frac{\lambda}{3!} \Big)^2 \, (\ii\,\hbar)^2 \,  (2 \pi)^d\, \delta(p_1 + p_2)\, \tilde \sgreen(p_1)\, \tilde \sgreen(p_2)\, \int_k\, \tilde \sgreen(k)\, \tilde \sgreen(p_1-k) \end{eqnarray*} 
 \begin{eqnarray*}  { \footnotesize \begin{tikzpicture}[scale=0.5, baseline]
        \coordinate (k) at (0,0);
        \coordinate (l) at ($(k)+(90:1)$);
        \coordinate[label=above: $p_1$] (p1) at ($(k) +(180:2.5)$);
        \coordinate[label=above: $p_2$] (p2) at ($(k) +(0:2.5)$);

        \draw[decoration={markings, mark=at position 0.5 with {\arrow{Latex[reversed]}} }, postaction={decorate}] (k) -- (p1);
        \draw[decoration={markings, mark=at position 0.5 with {\arrow{Latex[reversed]}} }, postaction={decorate}] (k) -- (p2);
        \draw (k) -- (l);
        \draw ($(l)+(90:0.5)$) circle (0.5);
    \end{tikzpicture} } \normalsize &=& \Big( -\frac{\lambda}{3!} \Big) ^2\, (\ii\, \hbar)^2\, (2 \pi)^d\, \delta(p_1 + p_2)\, \tilde \sgreen(p_1)\, \tilde \sgreen(p_2)\, \int_k\, \tilde \sgreen(k)\, \tilde \sgreen(0) \end{eqnarray*}
    \begin{eqnarray*} {\footnotesize
    \begin{tikzpicture}[scale=0.5, baseline]
        \coordinate (k) at (-1.5,0);
        \coordinate (l) at (1.5,0);
        \coordinate[label=above: $p_1$] (p1) at (180:2.5);
        \coordinate[label=above: $p_2$] (p2) at (0:2.5);

        \draw[decoration={markings, mark=at position 0.5 with {\arrow{Latex[reversed]}} }, postaction={decorate}] (k) -- (p1);
        \draw[decoration={markings, mark=at position 0.5 with {\arrow{latex[reversed]}} }, postaction={decorate}] (l) -- (p2);
        \draw ($(k)+(0:0.5)$) circle (0.5);
        \draw ($(l)-(0:0.5)$) circle (0.5);
    \end{tikzpicture} } \normalsize &=& \Big( -\frac{\lambda}{3!} \Big) ^2 \, (\ii \,\hbar)^2\,  (2 \pi)^d\, \delta(p_1)\, \tilde \sgreen(p_1)\, \int_k\, \tilde \sgreen(k) \ (2\pi)^d\, \delta(p_2)\, \tilde \sgreen(p_2)\, \int_l\, \tilde \sgreen(l) \end{eqnarray*}
    \begin{eqnarray*} { \footnotesize
     \begin{tikzpicture}[scale=0.5, baseline]
        \coordinate (k) at (-1,0);
        \coordinate (l) at (1,0);
        \draw[decoration={markings, mark=at position 0.5 with {\arrow{Latex[reversed]}} }, postaction={decorate}] (k) -- ($(k) + (180:1.5)$) node[above]{$p_1$};
        \draw[decoration={markings, mark=at position 0.5 with {\arrow{Latex[reversed]}} }, postaction={decorate}] (l) -- ($(l) + (0:1.5)$) node[above]{$p_2$};
        \draw ($(k)+(0:1)$) circle (1);  
        \draw ($(k) +(0:1) +(90:1)$) -- ($(k) +(0:1) +(270:1)$);
    \end{tikzpicture} } \normalsize &=& \Big(-\frac{\lambda}{3!}\Big)^4\,(\ii\, \hbar)^3\,(2 \pi)^d\, \delta(p_1 + p_2)\, \tilde \sgreen(p_1)\, \tilde \sgreen(p_2)\, \\
    &&\hspace{4cm} \times \int_{k,l}\, \tilde \sgreen(k)\, \tilde \sgreen(l)\, \tilde \sgreen(k+l)\, \tilde \sgreen(p_1 + k)\, \tilde \sgreen(p_2 + l)
\end{eqnarray*}

\bigskip\bigskip

\centerline{\underline{Multiplicity three}}
\vspace{-5mm}
\begin{eqnarray*} {\footnotesize
\begin{tikzpicture}[scale=0.25, baseline]
        \draw[decoration={markings, mark=at position 0.5 with {\arrow{Latex[reversed]}}}, postaction={decorate}] (0,0) -- (45:4);
        \draw[decoration={markings, mark=at position 0.5 with {\arrow{Latex[reversed]}}}, postaction={decorate}] (0,0) -- (135:4);
        \draw[decoration={markings, mark=at position 0.5 with {\arrow{Latex[reversed]}}}, postaction={decorate}] (0,0) -- (270:4);
        \draw (45:4.5) node[anchor=west] {$p_1$};
        \draw (135:4.5) node[anchor=east] {$p_2$};
        \draw (270:4.5) node[anchor=north] {$p_3$}; 
    \end{tikzpicture} } \normalsize &=& -\frac{\lambda}{3!}\,(\ii\, \hbar)^2 \, (2 \pi)^d\, \delta(p_1 + p_2 + p_3)\, \tilde \sgreen(p_1)\, \tilde \sgreen(p_2)\, \tilde \sgreen(p_3) \end{eqnarray*}
    \begin{eqnarray*} { \footnotesize
    \begin{tikzpicture}[scale=0.25, baseline]
        \draw[decoration={markings, mark=at position 0.5 with {\arrow{Latex[reversed]}}}, postaction={decorate}] (45:2) -- (45:4);
        \draw(45:1.5) circle (0.5);
        
        \draw[decoration={markings, mark=at position 0.5 with {\arrow{Latex[reversed]}}}, postaction={decorate}] (0,0) -- (135:4);
        \draw[decoration={markings, mark=at position 0.5 with {\arrow{Latex[reversed]}}}, postaction={decorate}] (0,0) -- (270:4);
        \draw (45:4.5) node[anchor=west] {$p_1$};
        \draw (135:4.5) node[anchor=east] {$p_2$};
        \draw (270:4.5) node[anchor=north] {$p_3$};
    \end{tikzpicture} } \normalsize &=& -\frac{\lambda}{3!}\,(\ii\, \hbar)^2\,  (2 \pi)^d\, \delta(p_2 + p_3)\, \tilde \sgreen(p_2) \ (2\pi)^d\, \delta(p_1)\, \tilde \sgreen(p_1)\, \int_k\, \tilde \sgreen(k) \end{eqnarray*}
    \begin{eqnarray*} { \footnotesize
    \begin{tikzpicture}[scale=0.25, baseline]
        \draw (0,0) circle (1);
        \draw[decoration={markings, mark=at position 0.5 with {\arrow{Latex[reversed]}}}, postaction={decorate}] (45:1) -- (45:4);
        \draw[decoration={markings, mark=at position 0.5 with {\arrow{Latex[reversed]}}}, postaction={decorate}] (135:1) -- (135:4);
        \draw[decoration={markings, mark=at position 0.5 with {\arrow{Latex[reversed]}}}, postaction={decorate}] (270:1) -- (270:4);
        \draw (45:4.5) node[anchor=west] {$p_1$};
        \draw (135:4.5) node[anchor=east] {$p_2$};
        \draw (270:4.5) node[anchor=north] {$p_3$}; 
    \end{tikzpicture} } \normalsize &=& \Big(-\frac{\lambda}{3!}\Big)^3\,(\ii\, \hbar)^3\, (2 \pi)^d\, \delta(p_1 + p_2 + p_3)\, \tilde \sgreen(p_1)\, \tilde \sgreen(p_2)\, \tilde \sgreen(p_3) \\[-30pt]
    && \hspace{4cm} \times \int_k\, \tilde \sgreen(k)\, \tilde \sgreen(p_2 + k)\, \tilde \sgreen(p_3 - k) \end{eqnarray*}
    \begin{eqnarray*} { \footnotesize
        \begin{tikzpicture}[scale=0.25, baseline]
        \draw (0,0) -- (45:2);
        \draw[decoration={markings, mark=at position 0.5 with {\arrow{Latex[reversed]}}}, postaction={decorate}] (45:2) -- (45:4);
        \draw (45:2) -- +(135:1);
        \draw (45:2) +(135:1.5) circle (0.5);
        
        \draw[decoration={markings, mark=at position 0.5 with {\arrow{Latex[reversed]}}}, postaction={decorate}] (0,0) -- (135:4);
        \draw[decoration={markings, mark=at position 0.5 with {\arrow{Latex[reversed]}}}, postaction={decorate}] (0,0) -- (270:4);
        \draw (45:4.5) node[anchor=west] {$p_1$};
        \draw (135:4.5) node[anchor=east] {$p_2$};
        \draw (270:4.5) node[anchor=north] {$p_3$}; 
    \end{tikzpicture} } \normalsize &=& \Big(-\frac{\lambda}{3!}\Big)^3\,(\ii\, \hbar)^3 \, (2 \pi)^d\, \delta(p_1 + p_2 + p_3)\, \tilde \sgreen(p_1)^2\, \tilde \sgreen(p_2)\, \tilde \sgreen(p_3)\,  \int_k\, \tilde \sgreen(k)\, \tilde \sgreen(0) \end{eqnarray*}
    \begin{eqnarray*} { \footnotesize
    \begin{tikzpicture}[scale=0.25, baseline]
        \draw (0,0) -- (45:1.5);
        \draw[decoration={markings, mark=at position 0.5 with {\arrow{Latex[reversed]}}}, postaction={decorate}] (45:2.5) -- (45:4);
        \draw (45:2) circle (0.5);
        
        \draw[decoration={markings, mark=at position 0.5 with {\arrow{Latex[reversed]}}}, postaction={decorate}] (0,0) -- (135:4);
        \draw[decoration={markings, mark=at position 0.5 with {\arrow{Latex[reversed]}}}, postaction={decorate}] (0,0) -- (270:4);
        \draw (45:4.5) node[anchor=west] {$p_1$};
        \draw (135:4.5) node[anchor=east] {$p_2$};
        \draw (270:4.5) node[anchor=north] {$p_3$}; 
    \end{tikzpicture} } \normalsize &=& \Big(-\frac{\lambda}{3!}\Big)^3\,(\ii \, \hbar)^3\, (2 \pi)^d\, \delta(p_1 + p_2 + p_3)\, \tilde \sgreen(p_1)^2\, \tilde \sgreen(p_2)\, \tilde \sgreen(p_3)\,  \int_k\, \tilde \sgreen(k)\, \tilde \sgreen(p_1 + k) 
\end{eqnarray*}
\begin{eqnarray*} { \footnotesize
\begin{tikzpicture}[scale=0.25, baseline]
        \draw[decoration={markings, mark=at position 0.5 with {\arrow{Latex[reversed]}}}, postaction={decorate}] (45:2) -- (45:4);
        \draw (45:2) -- +(135:1);
        \draw (45:2) +(135:1.5) circle (0.5);
        \draw (45:2) -- +(315:1);
        \draw (45:2) +(315:1.5) circle (0.5);
        
        \draw[decoration={markings, mark=at position 0.5 with {\arrow{Latex[reversed]}}}, postaction={decorate}] (0,0) -- (135:4);
        \draw[decoration={markings, mark=at position 0.5 with {\arrow{Latex[reversed]}}}, postaction={decorate}] (0,0) -- (270:4);
        \draw (45:4.5) node[anchor=west] {$p_1$};
        \draw (135:4.5) node[anchor=east] {$p_2$};
        \draw (270:4.5) node[anchor=north] {$p_3$};
    \end{tikzpicture} } \normalsize &=& \Big(-\frac{\lambda}{3!}\Big)^3\,(\ii\, \hbar)^3\,  (2 \pi)^d\, \delta(p_2 + p_3)\, \tilde \sgreen(p_2) \ (2\pi)^d\, \delta(p_1)\, \tilde \sgreen(p_1)\, \left(\int_k\, \tilde \sgreen(k)\, \tilde \sgreen(0) \right)^2 \end{eqnarray*}
    \begin{eqnarray*} { \footnotesize
    \begin{tikzpicture}[scale=0.25, baseline]
        \draw (45:2) -- (45:2.5);
        \draw[decoration={markings, mark=at position 0.5 with {\arrow{Latex[reversed]}}}, postaction={decorate}] (45:3.5) -- (45:4);
        \draw (45:1.5) circle (0.5);
        \draw (45:3) circle (0.5);
        \draw[decoration={markings, mark=at position 0.5 with {\arrow{Latex[reversed]}}}, postaction={decorate}] (0,0) -- (135:4);
        \draw[decoration={markings, mark=at position 0.5 with {\arrow{Latex[reversed]}}}, postaction={decorate}] (0,0) -- (270:4);
        \draw (45:4.5) node[anchor=west] {$p_1$};
        \draw (135:4.5) node[anchor=east] {$p_2$};
        \draw (270:4.5) node[anchor=north] {$p_3$}; 
    \end{tikzpicture} } \normalsize &=& \Big(-\frac{\lambda}{3!}\Big)^3\,(\ii\, \hbar)^3\,  (2 \pi)^d\, \delta(p_2 + p_3)\, \tilde \sgreen(p_2) \\[-30pt]
    && \hspace{4cm} \times \, (2\pi)^d\, \delta(p_1)\, \tilde \sgreen(p_1)^2  \int_{k}\, \tilde \sgreen(k)^2 \ \int_l\,\tilde\sgreen(l)  \end{eqnarray*}
    \begin{eqnarray*} { \footnotesize
\begin{tikzpicture}[scale=0.25, baseline]
        \draw (45:1.5) circle (0.5);
        \draw (45:1.5) +(45:1.5) arc [start angle=45, end angle=180, radius=0.75];
        \draw  (45:1.5) +(45:1.5) arc [start angle=45, end angle=-90, radius=0.75];
        \draw[decoration={markings, mark=at position 0.5 with {\arrow{Latex[reversed]}}}, postaction={decorate}] (45:3) -- (45:4);    
        
        \draw[decoration={markings, mark=at position 0.5 with {\arrow{Latex[reversed]}}}, postaction={decorate}] (0,0) -- (135:4);
        \draw[decoration={markings, mark=at position 0.5 with {\arrow{Latex[reversed]}}}, postaction={decorate}] (0,0) -- (270:4);
        \draw (45:4.5) node[anchor=west] {$p_1$};
        \draw (135:4.5) node[anchor=east] {$p_2$};
        \draw (270:4.5) node[anchor=north] {$p_3$}; 
    \end{tikzpicture} } \normalsize &=& \Big(-\frac{\lambda}{3!}\Big)^3\,(\ii\, \hbar)^3\, (2 \pi)^d\, \delta(p_2 + p_3)\, \tilde \sgreen(p_2) \\[-30pt]
    && \hspace{4cm} \times \, (2\pi)^d\, \delta(p_1)\, \tilde \sgreen(p_1)  \int_{k,l}\, \tilde \sgreen(k)^2\, \tilde \sgreen(l)\, \tilde \sgreen(k + l)
\end{eqnarray*}
\begin{eqnarray*} { \footnotesize
    \begin{tikzpicture}[scale=0.25, baseline]
        \node (b) at (0,0){};
        \draw[decoration={markings, mark=at position 0.5 with {\arrow{Latex[reversed]}}}, postaction={decorate}] (45:2) -- (45:4);
        \draw (45:1.5) circle (0.5);

        \draw (0,0) circle (0.5);

        \draw[decoration={markings, mark=at position 0.5 with {\arrow{Latex[reversed]}}}, postaction={decorate}] (0,0) +(135:0.5) -- (135:4);
        \draw[decoration={markings, mark=at position 0.5 with {\arrow{Latex[reversed]}}}, postaction={decorate}] (0,0) +(270:0.5) -- (270:4);
        \draw (45:4.5) node[anchor=west] {$p_1$};
        \draw (135:4.5) node[anchor=east] {$p_2$};
        \draw (270:4.5) node[anchor=north] {$p_3$};    
    \end{tikzpicture} } \normalsize &=& \Big(-\frac{\lambda}{3!}\Big)^3\,(\ii\, \hbar)^3\, (2 \pi)^d\, \delta(p_2 + p_3)\, \tilde \sgreen(p_2)\, \tilde \sgreen(p_3)\,\int_k\, \tilde \sgreen(k) \, \tilde \sgreen(p_2 + k) \\[-30pt]
    && \hspace{4cm} \times \, (2\pi)^d\,\delta(p_1)\, \tilde \sgreen(p_1)\, \int_l\, \tilde \sgreen(l) \end{eqnarray*}
    \begin{eqnarray*} {\footnotesize
    \begin{tikzpicture}[scale=0.25, baseline]
        \draw[decoration={markings, mark=at position 0.5 with {\arrow{Latex[reversed]}}}, postaction={decorate}] (45:2) -- (45:4);
        \draw (45:1.5) circle (0.5);
        
        \draw[decoration={markings, mark=at position 0.5 with {\arrow{Latex[reversed]}}}, postaction={decorate}] (0,0) -- (135:4);
        \draw[decoration={markings, mark=at position 0.5 with {\arrow{Latex[reversed]}}}, postaction={decorate}] (0,0) -- (270:4);
        \draw (225:0) -- (225:2);
        \draw (225:2.5) circle (0.5);
        \draw (45:4.5) node[anchor=west] {$p_1$};
        \draw (135:4.5) node[anchor=east] {$p_2$};
        \draw (270:4.5) node[anchor=north] {$p_3$};
    \end{tikzpicture} } \normalsize &=& \Big(-\frac{\lambda}{3!}\Big)^3\,(\ii\, \hbar)^3\,  (2 \pi)^d\, \delta(p_2 + p_3)\, \tilde \sgreen(p_2)\, \tilde \sgreen(p_3)\, \int_k\, \tilde \sgreen(k)\, \tilde \sgreen(0) \\[-30pt]
    && \hspace{4cm} \times\,(2\pi)^d\,\delta(p_1)\, \tilde \sgreen(p_1)\, \int_l\, \tilde \sgreen(l) \end{eqnarray*}
    \begin{eqnarray*} { \footnotesize
    \begin{tikzpicture}[scale=0.25, baseline]
        \draw[decoration={markings, mark=at position 0.5 with {\arrow{Latex[reversed]}}}, postaction={decorate}] (45:2) -- (45:4);
        \draw (45:1.5) circle (0.5);
    
        \draw[decoration={markings, mark=at position 0.5 with {\arrow{Latex[reversed]}}}, postaction={decorate}] (135:2) -- (135:4);
        \draw (135:1.5) circle (0.5);
    
        \draw[decoration={markings, mark=at position 0.5 with {\arrow{Latex[reversed]}}}, postaction={decorate}] (270:2) -- (270:4);
        \draw (270:1.5) circle (0.5);
        
        \draw (45:4.5) node[anchor=west] {$p_1$};
        \draw (135:4.5) node[anchor=east] {$p_2$};
        \draw (270:4.5) node[anchor=north] {$p_3$};
    \end{tikzpicture} } \normalsize &=& \Big(-\frac{\lambda}{3!}\Big)^3\,(\ii\, \hbar)^3\, (2 \pi)^d\, \delta(p_1)\, \tilde \sgreen(p_1)\, \int_k\, \tilde \sgreen(k) \\[-30pt]
&& \hspace{3cm} \times \, (2\pi)^d\,  \delta(p_2)\, \tilde \sgreen(p_2)\, \int_l\, \tilde \sgreen(l) \ (2\pi)^d\,\delta(p_3)\, \tilde \sgreen(p_3)\, \int_q\, \tilde \sgreen(q) 
\end{eqnarray*}

\bibliographystyle{ourstyle}  
\bibliography{bsqed.bib}

\end{document}